\documentclass{aa}

\usepackage{graphicx}
\usepackage{txfonts}
\usepackage{xcolor,varwidth}
\usepackage{lscape}
\usepackage{booktabs}
\usepackage{multirow}
\usepackage{amsmath}
\usepackage{longtable}
\usepackage{chemmacros}

\newcommand{\pht}{\phantom{^\dagger}}
\newcommand{\phs}{\phantom{<}}
\newcommand{\phns}{\phantom{0<}}
\newcommand{\phn}{\phantom{0}}
\newcommand{\phnn}{\phantom{0}\phantom{0}}
\newcommand{\hii}{\ion{H}{ii}}

\usepackage{natbib}
\bibpunct{(}{)}{;}{a}{}{,}
\usepackage{hyperref}
\hypersetup{
  colorlinks = true, 
  urlcolor   = blue, 
  linkcolor  = blue, 
  citecolor  = blue 
}

\begin{document} 

   \title{A search for radio jets from massive young stellar objects}

   \subtitle{Association of radio jets with H$_2$O and CH$_3$OH masers}

   \author{\"U.~Kavak\inst{1, 2, 3, 4}
          \and \'A.~S\'anchez-Monge\inst{1}
          \and A.~L\'opez-Sepulcre\inst{5,6}
          \and R.~Cesaroni\inst{7}
          \and F.~F.~S.~van der Tak\inst{3,4} 
          \and L.~Moscadelli\inst{7}
          \and M.~T.~Beltr\'an\inst{7}
          \and P.~Schilke\inst{1}
          }
   \institute{I.\ Physikalisches Institut, Universit\"at zu Köln, Zülpicher Str.\ 77, 50937 Köln, Germany \\
              \email{kavak@astro.rug.nl}
         \and
             Institute of Graduate Studies in Science, Program of Astronomy and Space Sciences, Istanbul University, Istanbul, Turkey
         \and
             Kapteyn Astronomical Institute, Landleven 12, 9747AD, Groningen, The Netherlands
         \and 
             SRON Netherlands Institute for Space Research, Landleven 12, 9747AD, Groningen, The Netherlands
         \and
             Universit\'e Grenoble Alpes, CNRS, IPAG, 38000 Grenoble, France
         \and
             Institut de Radioastronomie Milim\'etrique (IRAM), 300 rue de la Piscine, F-38406 Saint-Martin-d'H\`eres, France
         \and
             INAF, Osservatorio Astrofisico di Arcetri, Largo Enrico Fermi 5, I-50125, Florence, Italy
             }

   \date{Received February 4, 2020; accepted ??}


  \abstract
   {Recent theoretical and observational studies debate the similarities between the formation process of high-mass ($>8$~$M_\odot$) and low-mass stars. The formation of low-mass star formation is directly associated with the presence of disks and jets. Theoretical models predict that stars with masses up to 140~$M_\odot$ can be formed through disk-mediated accretion in disk-jet systems. According to this scenario, radio jets are expected to be common in high-mass star-forming regions.}
   {We aim to increase the number of known radio jets in high-mass star forming regions by searching for radio jet candidates at radio continuum wavelengths.}
   {We have used the Karl G.\ Jansky Very Large Array (VLA) to observe 18 high-mass star-forming regions in the C~band (6~cm, $\approx$1\farcs0 resolution) and K~band (1.3~cm, $\approx$0\farcs3 resolution). We have searched for radio jet candidates by studying the association of radio continuum sources with shock activity signposts (e.g., molecular outflows, extended green objects, and maser emission). Our VLA observations also target the 22~GHz H$_2$O and 6.7~GHz CH$_3$OH maser lines.}
   {We have identified 146 radio continuum sources, with 40 of them being located within the field of view of both images (C and K~band maps). For these sources we have derived their spectral index, which is consistent with thermal emission (between $-0.1$ and $+2.0$) for 73\% of them. Based on the association with shock activity signposts, we have identified 28 radio jet candidates. Out these, we have identified 7 as the most probable radio jets. The radio luminosity of the radio jet candidates is correlated with the bolometric luminosity and the outflow momentum rate. About 7--36\% of the radio jet candidates are associated with non-thermal emission. The radio jet candidates associated with 6.7~GHz CH$_3$OH maser emission are preferentially thermal winds and jets, while a considerable fraction of radio jet candidates associated with H$_2$O masers show non-thermal emission, likely due to strong shocks.}
   {About 60\% of the radio continuum sources detected within the field of view of our VLA images are potential radio jets. The remaining sources could be compact \hii\ regions in their early stages of development, or radio jets for which we do not have yet further evidence of shock activity. Our sample of 18 regions is divided in 8 less evolved, infrared-dark regions and 10 more evolved, infrared-bright regions. We have found that $\approx$71\% of the identified radio jet candidates are located in the more evolved regions. Similarly, 25\% of the less evolved regions harbor one of the most probable radio jets, while up to 50\% of the more evolved regions contain one of these radio jet candidates. This suggests that the detection of radio jets in high-mass star forming regions is larger in slightly more evolved regions.}
   

   \titlerunning{A search for radio jets from massive young stellar objects}
   \authorrunning{Kavak et al.}
   \maketitle

%
\section{Introduction\label{s:introduction}}

High-mass stars (O and B-type stars with masses $\geq8\ \mathrm{M_\sun}$) play a crucial role in the chemical and physical composition of their host galaxies throughout their lifetimes, by injecting energy and material on different scales through energetic outflows, intense UV radiation, powerful stellar winds, and supernova explosions. Despite their importance, the formation process of massive stars is still not well understood due to observational and theoretical challenges (e.g., massive stars form in crowded environments and are located at far distances, see reviews by \citealt{Tan2014, Motte2018}). On the other hand, the formation of low-mass stars is better understood and is explained with a model based on accretion via a circumstellar disk and a collimated jet/outflow that removes angular momentum and enables accretion to proceed \citep[e.g.,][]{Larson1969, Andre2000}. Circumstellar disks have been indeed observed around low-mass protostars \citep[-e.g.,][]{Williams2011, Luhman2012}, while ejection of material has been mainly observed by large-scale, collimated jets and outflows \citep[e.g.,][]{Bachiller1996, Bally2016}. For high-mass stars, it remains to be understood the role that (accretion) disks and jets/outflows play in their formation, as well as how their properties vary with the mass of the forming star and the environment. As far as observations are concerned, some studies have concentrated on disks and jets/outflows in selected high-mass star-forming regions \citep [see e.g.,][]{Beuther2002a, Arce2007, LopezSepulcre2009, Bally2016}. The advent of facilities such as the Atacama Large Millimeter/Submillimeter Array (ALMA) or the upgraded Karl G. Jansky Very Large Array (VLA) provides the high spatial resolution and sensitivity necessary to fully resolve the structure of disks and jets/outflows in high-mass star-forming regions. While disks are bright at millimeter wavelengths and constitute perfect targets for ALMA observations \citep[e.g.,][]{SanchezMonge2013b, SanchezMonge2014, Beltran2014, Johnston2015, Cesaroni2017, Maud2019}, jets are found to be bright at centimeter wavelengths observable with the VLA \citep[e.g.,][]{CarrascoGonzalez2010, CarrascoGonzalez2015, Moscadelli2013, Moscadelli2016}.

Surveys of low-mass star-forming regions with the VLA \citep[e.g.,][]{Anglada1996, Anglada1998, Beltran2001} revealed radio-continuum sources elongated in the direction of the large-scale molecular outflows. These sources are called thermal radio jets, because their emission is interpreted as thermal (free-free) emission of ionised, collimated jets at the base of larger-scale optical jets and molecular outflows \citep[e.g.,][]{Curiel1987, Curiel1989, Rodriguez1995}. Due to the high spatial resolution that can be achieved at radio wavelengths with interferometers such as the VLA, thermal radio jets constitute strong evidence of collimated outflows on small scales ($\sim$100~au; \citealt{Torrelles1985, Anglada1996}) and permit to pinpoint the location of the star that is forming and powering the jet/outflow seen on larger scales. Although the emission of jets at radio-wavelengths is mainly thermal, some jets show a contribution from a non-thermal component \citep[e.g.,][]{Reid1995, CarrascoGonzalez2010, Moscadelli2013, Moscadelli2016}.

Following the strategy used in the study of low-mass star-forming regions, we aim to search for radio jets associated with high-mass star-forming regions in a large sample of sources. Until recently, only a limited number of regions harboring high-mass stellar objects were known to be associated with radio jets (e.g., HH80/81: \citealt{Marti1993, CarrascoGonzalez2010}, CepAHW2: \citealt{Rodriguez1994}, IRAS\,16547$-$4247: \citealt{Rodriguez2008}, IRAS\,16562$-$1732: \citealt{Guzman2010}, G35.20$-$0.74\,N: \citealt{Beltran2016}). In the last years, progress has been made in increasing the number of known jets associated with high-mass young stellar objects \citep[e.g.,][]{Moscadelli2016, Rosero2016, Sanna2018, Purser2018}. In this work, we used the VLA in two different frequency bands to search for radio jets in a sample of 18 high-mass star-forming regions associated with molecular outflow emission.

This paper is structured as follows. In Section~\ref{s:observations}, we present both the sample and the details of the observations. The results of the observations of the radio continuum (and maser) emission are presented in Section~\ref{s:results}. The analysis of the properties of the discovered sources is presented in Section~\ref{s:analysis}, while Appendix~\ref{s:individualsources} describes the properties of each region in more detail. In Section~\ref{s:discussion}, we discuss the implications of our results in the context of high-mass star formation, and in Section~\ref{s:summary} we summarize the most important conclusions.

%
\section{Observations\label{s:observations}}

%
\subsection{Selected sample\label{s:sample}}

We have selected 18 high-mass star-forming regions from the samples of \citet{LopezSepulcre2010, LopezSepulcre2011} and \citet{SanchezMonge2013d}, using the following criteria: (i) clump mass $> 100$~M$_{\sun}$, to exclude regions forming mainly low-mass stars, (ii) distance $< 4$~kpc, to resolve spatial scales $< 4000$~AU when observed with interferometers at a resolution of 1\arcsec, (iii) declination $>-15$\degr, to be observable from northern telescopes, (iv) association with an HCO$^+$ bipolar outflow and SiO emission with line widths broader than $> 20$~km~s$^{-1}$ \citep{LopezSepulcre2011, SanchezMonge2013d}, and (v) absence of bright centimeter continuum emission, to exclude developed \hii\ regions.

We used the NRAO VLA Sky Survey \citep [NVSS;][]{Condon1998}, the MAGPIS \citep{Helfand2006}, CORNISH\citep{Hoare2012, Purcell2013}, and RMS \citep{Urquhart2008, Lumsden2013} surveys to eliminate star-forming regions with developed \hii\ regions that would hinder the detection of faint radio jets. Our final sample of 18 high-mass star-forming regions is listed in Table~\ref{t:sample}.

\begin{table*}[ht!]
\centering
\caption{High-mass star-forming regions observed with the Very Large Array in this work}
\label{t:sample}
\begin{tabular}{l c c c c c c c c c}
\hline\hline

&R.A.~(J2000)
&Dec.~(J2000)
&$d$\tablefootmark{a}
&$M$\tablefootmark{a}
&\multicolumn{2}{c}{C~band\tablefootmark{b}}
&
&\multicolumn{2}{c}{K~band\tablefootmark{b}}
\\
\cline{6-7}\cline{9-10}
Region
&(h:m:s)
&(\degr:\arcmin:\arcsec)
&(kpc)
&(M$_\sun$)
&$\theta_\mathrm{beam}$, PA
&rms 
&
&$\theta_\mathrm{beam}$, PA
&rms
\\
\hline
IRAS\,05358$+$3543$^\dagger$  & 05:39:12.2 &$+$35:45:52.0 & 1.8 & \phn127 &$1.27\times1.23$, $+$61 & \phn8.1 && \ldots\tablefootmark{c}& \ldots \\ 
G189.78$+$0.34$^\dagger$      & 06:08:34.5 &$+$20:38:51.0 & 1.8 & \phn150 &$1.28\times1.09$, $+$23 & 16.0    && \ldots\tablefootmark{c}& \ldots \\ 
G192.58$-$0.04$^\dagger$      & 06:12:52.9 &$+$18:00:34.9 & 2.6 & \phn500 &$1.40\times1.19$, $+$21 & 23.6    && \ldots\tablefootmark{c}& \ldots \\ 
G192.60$-$0.05$^\dagger$      & 06:12:54.0 &$+$17:59:23.0 & 2.6 & \phn460 &$1.36\times1.14$, $+$20 & 26.5    && \ldots\tablefootmark{c}& \ldots \\ 
G18.18$-$0.30$^\dagger$       & 18:25:07.3 &$-$13:14:22.9 & 2.6 & \phn110 &$1.74\times1.05$, $-$16 & 10.0    &&$0.54\times0.31$, $+$25 & 16.7   \\ 
IRAS\,18223$-$1243$^\dagger$  & 18:25:10.9 &$-$12:42:27.0 & 3.7 & \phn980 &$1.89\times1.14$, $-$14 & 24.5    &&$0.58\times0.34$, $+$37 & 15.0   \\ 
IRAS\,18228$-$1312$^\dagger$  & 18:25:42.3 &$-$13:10:18.0 & 3.0 & \phn740 &$1.88\times1.16$, $-$14 & 35.0    &&$0.73\times0.67$, $-$07 & 59.1   \\ 
G19.27$+$0.1M2$^\ddag$      & 18:25:52.6 &$-$12:04:47.9 & 2.4 & \phn114 &$2.08\times1.14$, $-$14 & \phn9.8 &&$0.50\times0.32$, $-$26 & 20.7   \\ 
G19.27$+$0.1M1$^\ddag$      & 18:25:58.5 &$-$12:03:58.9 & 2.4 & \phn113 &$1.95\times1.19$, $-$16 & \phn9.6 &&$0.79\times0.42$, $+$49 & 20.0   \\ 
IRAS\,18236$-$1205$^\dagger$  & 18:26:25.4 &$-$12:03:50.9 & 2.7 & \phn780 &$1.99\times1.11$, $-$18 & 10.2    &&$0.51\times0.32$, $-$26 & 16.9   \\ 
G23.60$+$0.0M1$^\ddag$      & 18:34:11.6 &$-$08:19:05.9 & 2.5 & \phn365 &$1.85\times1.13$, $-$22 & \phn8.7 &&$0.54\times0.30$, $+$32 & 36.1   \\ 
IRAS\,18316$-$0602$^\dagger$  & 18:34:20.5 &$-$05:59:30.0 & 3.1 & 1000    &$1.71\times1.09$, $-$21 & \phn8.4    &&$0.53\times0.30$, $+$35 & 40.0   \\ 
G24.08$+$0.0M2$^\ddag$      & 18:34:51.1 &$-$07:45:32.0 & 2.5 & \phn201 &$1.80\times1.13$, $-$21 & 17.0    &&$0.53\times0.30$, $+$32 & 37.0   \\ 
G24.33$+$0.1M1$^\ddag$      & 18:35:07.8 &$-$07:35:04.0 & 3.8 & 1759    &$1.69\times1.25$, $-$13 & 21.0    &&$0.49\times0.31$, $-$31 & 26.1   \\ 
G24.60$+$0.1M2$^\ddag$      & 18:35:35.7 &$-$07:18:08.9 & 3.7 & \phn483 &$1.66\times1.03$, $-$20 & 18.2    &&$0.79\times0.41$, $+$51 & 21.8   \\ 
G24.60$+$0.1M1$^\ddag$      & 18:35:40.2 &$-$07:18:37.0 & 3.7 & \phn192 &$1.67\times1.22$, $-$10 & 12.7    &&$0.75\times0.38$, $+$57 & 22.5   \\ 
G34.43$+$0.2M3$^\ddag$      & 18:53:20.3 &$+$01:28:23.0 & 2.5 & \phn301 &$1.57\times1.46$, $-$59 & 17.9    &&$0.53\times0.29$, $+$40 & 19.0   \\ 
IRAS\,19095$+$0930$^\dagger$  & 19:11:54.0 &$+$09:35:52.0 & 3.0 & \phn500 &$1.61\times1.37$, $-$83 & 17.0    &&$0.53\times0.29$, $+$44 & 63.4   \\ 
\hline 
\end{tabular} 
\tablefoot{
\tablefoottext{a}{Distances ($d$) and clump masses ($M$) from \cite{LopezSepulcre2011} and \cite{SanchezMonge2013d}.}
\tablefoottext{b}{Synthesized beam ($\theta_\mathrm{beam}$) major and minor axis in arcsecond, and position angle (PA) in degrees. The rms noise level is given in units of $\mu$Jy~beam$^{-1}$. Regions marked with $^\dagger$ and $^\ddag$ in the first column indicate IR-loud and IR-dark sources, respectively, based on the classification of \citet{LopezSepulcre2010}.} 
\tablefoottext{c}{Region not observed in the K~band.}
}
\end{table*}

%
\subsection{VLA observations\label{s:vla}}

We used the VLA of the NRAO\footnote{The Very Large Array (VLA) is operated by the National Radio Astronomy Observatory (NRAO), a facility of the National Science Foundation operated under cooperative agreement by Associated Universities, Inc.} to observe the 18 selected regions (see Table~\ref{t:sample}). The observations were conducted between June and August 2012 (project number 12A-099), when the array was transitioning to its current upgraded phase and was known as `expanded VLA' (EVLA). During the observations, the array was in its B configuration, which provides a maximum baseline of 11~km. We observed the frequency bands C (4--8~GHz) and K (18--26.5~GHz) with 16 spectral windows of 128~MHz each, covering a total bandwidth of 2048~MHz in each band. Each spectral window has 128 channels, with a channel width of 1~MHz. The time spent per source is $\sim$20~minutes and $\sim$30~minutes at 6~cm (C~band) and 1.3~cm (K~band), respectively. Flux calibration was achieved by observing the quasars 3C286 ($F_\mathrm{1.3~cm}$=2.59~mJy, $F_\mathrm{6~cm}$=7.47~mJy) and 3C48 ($F_\mathrm{1.3~cm}$=1.13~mJy, $F_\mathrm{6~cm}$=5.48~mJy). The amplitude and phase were calibrated by monitoring the quasars J0555$+$3948, J0559$+$2353, J1832$-$1035, and J1851$+$0035. We used the standard guidelines for the calibration of VLA data. The data were processed using the Common Astronomy Software Applications \citep[CASA;][]{McMullin2007} software.

Continuum images of each source were obtained after excluding channels with line emission, corresponding to H$_2$O and CH$_3$OH maser lines. The images were done using the `clean' task with the Briggs weighting parameter set to 2, which results in a typical synthesized beam of 1\farcs5 and 0\farcs4 for the C and K~bands, respectively, and typical rms noise levels of $\sim$22~$\mu$Jy~beam$^{-1}$ at 6~cm and $\sim$30~$\mu$Jy~beam$^{-1}$ at 1.3~cm (see Table~\ref{t:sample}).

The spectral resolution of the observations is limited (about 50~km~s$^{-1}$ and 13~km~s$^{-1}$ for the C and K~bands, respectively) and insufficient to resolve spectral features. Despite this limitation, we have produced image cubes of spectral windows that cover the frequencies of the H$_2$O maser line at 22235.0798~MHz and the CH$_3$OH maser line at 6668.519~MHz. This allowed us to search for maser features that can be associated with the continuum emission. The rms noise levels of these cubes are $0.5$~mJy~beam$^{-1}$ and $0.3$~mJy~beam$^{-1}$ per channel of 13 and 50~km~s$^{-1}$ for the H$_2$O and CH$_3$OH images, respectively.

%
\section{Results\label{s:results}}

%
\subsection{Continuum emission\label{s:continuum}}

We have detected compact continuum emissions in all 18 observed high-mass star-forming regions. A total of 146 compact sources are identified with intensities above $3\sigma$ level, where $\sigma$ is the rms noise level of each map (see Table~\ref{t:sample}). 
In Table~\ref{t:catalogue}, we list the coordinates, fluxes and source sizes.

\begin{figure}[t!]
    \centering
    \includegraphics[width=1.0\columnwidth]{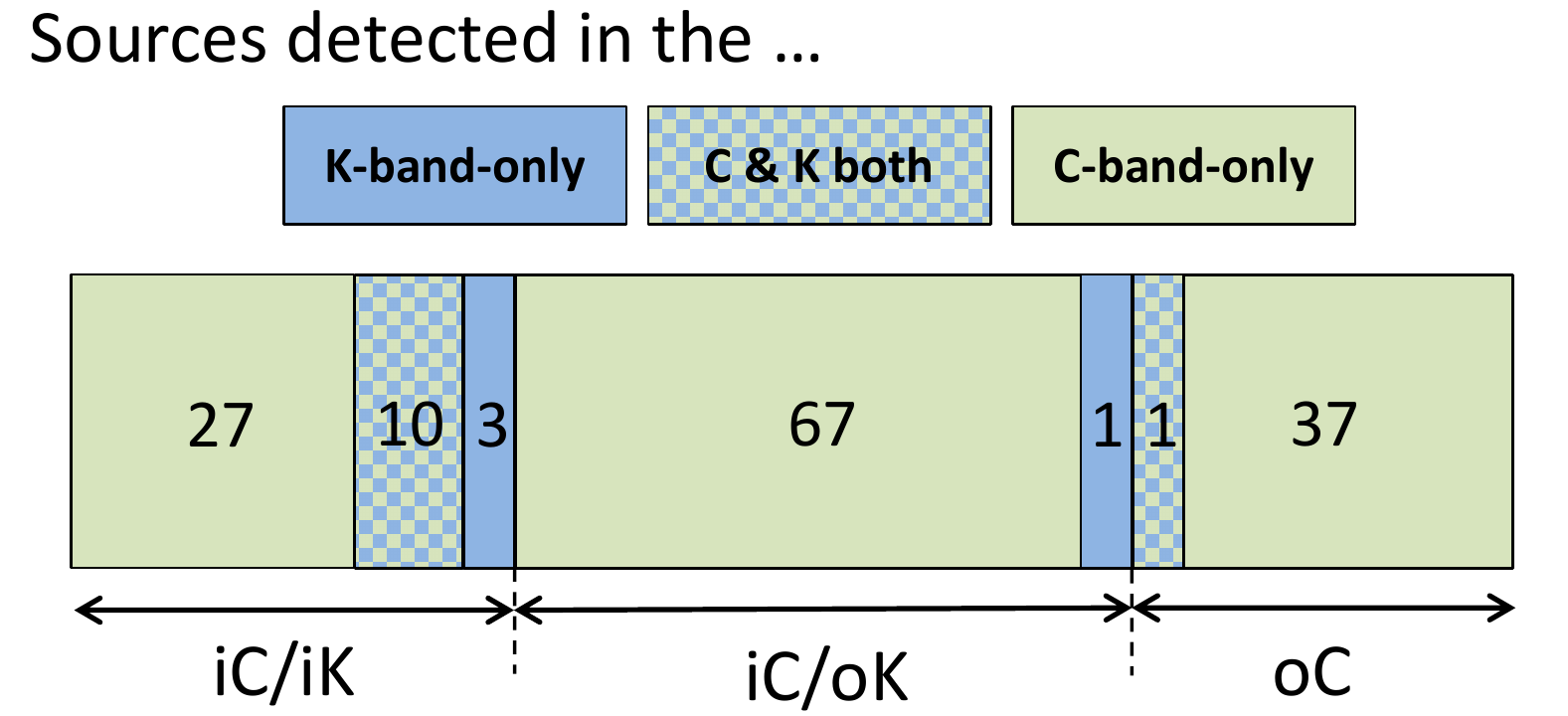}
    \caption{Number of radio continuum sources detected in the K~band images (marked in blue and corresponding to 15 sources) and in the C~band images (marked in green and corresponding to 142 sources). Out of the 146 detections, only four are detected only in the K~band images. The vast majority (131) are detected only in the C~band images (see Sect.~\ref{s:continuum} for more details). The bottom labels marked the sources located within the primary beams of the K~band and C~band images. We find 38 sources located outside the C~band primary beam (oC), 68 sources located inside the C~band primary beam but outside the K~band primary beam (iC/oK) and 40 sources located within the primary beam of both images (iC/iK). See Sects.~\ref{s:continuum} and \ref{s:spectralindex} for more details.}
    \label{f:source_detections}
\end{figure}

Most of the sources (a total of 131) are only detected in the C~band image, while four of them are only detected in the K~band (see Fig.~\ref{f:source_detections}). Only eleven sources are detected at both frequencies. The higher detection rate of sources in the C~band is due to several factors. First, four regions were only observed in the C~band (see Table~\ref{t:sample}). This results in 26 radio continuum sources for which we have no access to K~band images. Second, the field of view of the C~band images (primary beam $\approx$9\arcmin) is larger compared to the K~band primary beam ($\approx$2\arcmin). Only 40 sources are located within the K band primary beam (identified as iC/iK, see Fig.~\ref{f:source_detections}). This number reduces to only 24 when considering only the sources that have been observed in both frequency bands. A total of 68~sources are inside the C~band primary beam (identified as iC), but outside the K~band primary beam (identified as oK, see Fig.~\ref{f:source_detections}). The remaining 38~sources are outside the primary beam of the C~band observations (marked oC; see also Table~\ref{t:catalogue}). The sources that are outside the primary beam are bright enough to be detected even when the telescope's sensitivity is highly reduced. Thirdly, the spatial filtering of the interferometer is different at both frequencies. In the B-configuration, the VLA recovers scales up to 11\arcsec\ in the C~band, and only up to 4\arcsec\ in the K~band (see also Appendix~A of \citealt{Palau2010}). Finally, one cannot exclude the possibility that some of these sources are extragalactic objects that can only be detected at low frequencies. We have followed the approach of \citet{Anglada1998} to determine the possible contamination of background sources in our catalog. The expected number of background sources $N_\mathrm{bg}$ is given by

\begin{equation}
\begin{split}
N_\mathrm{bg} = & 1.4\left\{1-\exp\left[-0.0066\left(\frac{\theta_F}{\mathrm{arcmin}}\right)^2\left(\frac{\nu}{5~\mathrm{GHz}}\right)^2\right]\right\} \\
& \times \left(\frac{S_0}{\mathrm{mJy}}\right)^{-0.75}\,\left(\frac{\nu}{5~\mathrm{GHz}}\right)^{-2.52},
\end{split}
\end{equation}
where $\theta_F$ is the area of the sky that has been observed (18 fields in C~band, and 14 fields in K~band), $\nu$ is the frequency of the observations, and $S_{0}$ is the detectable flux density threshold ($3\times$rms, with an average rms of $22\ \mathrm{\mu Jy~beam^{-1}}$ in the C~band, and $30\ \mathrm{\mu Jy~beam^{-1}}$ in the K~band). This results in $N_\mathrm{bg}=11$ and $N_\mathrm{bg}=0.2$ for the C and K~band images, respectively. Less than 5\% of the sources detected in the C~band might be background objects not related to the star-forming regions, while we do not expect contamination in the K~band images.

%
\subsection{Spectral index analysis\label{s:spectralindex}}

The spectral index ($\alpha$) is defined as $S_{\nu} \propto \nu^{\alpha}$, where S$_{\nu}$ is the flux density and $\nu$ is the frequency. We have calculated the spectral index for the continuum sources using the measured flux densities at 1.3~cm (K~band) and 6~cm (C~band). For the sources without detection in one of the bands, we assume an upper limit of flux density equal to $5\sigma$. It should be noted here that the flux densities of the sources were corrected for the primary beam response of the antennas. The sources far away from the phase centers (listed in Table~\ref{t:sample}) have larger uncertainties in the correction factors of the primary beam and thus, in the final (corrected) flux. Therefore, the sources located within the primary beams in both frequency bands (i.e., sources listed as `iC/iK' in Table~\ref{t:catalogue}) have more accurate flux estimates. For the sources outside one of the primary beams (i.e., oK or oC), we do not determine the spectral index because of the high uncertainty involved in the fluxes. In the last column of Table~\ref{t:catalogue} we list the calculated spectral indices. For the sources detected at both frequencies, we have improved the determination of the spectral index by creating new images with the same $uv$ (visibility) coverage (see Table~\ref{t:uvsizes}). This ensures that the interferometer is sensitive to similar spatial scales at both frequencies.

\begin{figure}[t!]
\centering
\includegraphics[width=1.0\columnwidth]{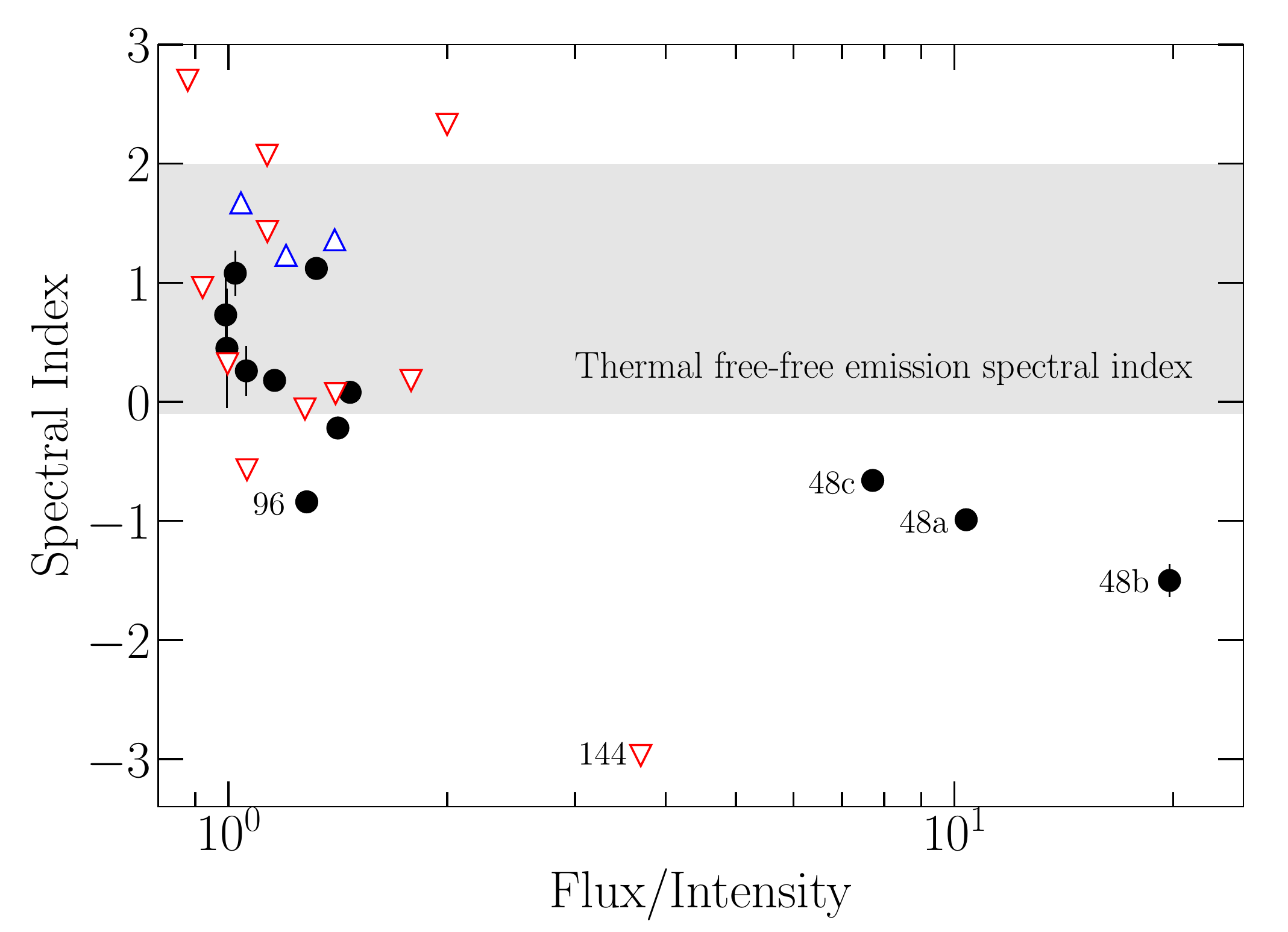}
\caption{Spectral index ($\alpha$, see Sect.~\ref{s:spectralindex}) against the flux to intensity ratio for the radio continuum sources detected in both frequency bands and inside the primary beam of both images (sources listed as `iC/iK' in Table~\ref{t:catalogue}). The grey-shaded region depicts the spectral index regime associated with thermal free-free emission (i.e., in the range from $-0.1$ to $+2$). Black dots correspond to sources detected in both bands (see spectral indices in Table~\ref{t:candidates}), blue up-pointing triangles correspond to sources detected only in the K~band (i.e., lower limits), and red down-pointing triangles correspond to sources detected only in the C~band (i.e., upper limits).}
\label{fig:spix_intensity}
\end{figure}

In Fig.~\ref{fig:spix_intensity}, we present the spectral index against the flux density to intensity peak ratio for the 24 continuum sources observed at both bands and located within the primary beams. For the sources detected at 6~cm only, we derive an upper limit\footnote{We note that the real spectral index may not be always an upper limit if the source emission is completely filtered out in our K~band images. Further observations at different wavelengths, with a similar $uv$-sampling and angular resolution are necessary to constrain the spectral index of the sources detected only in the C~band images.} to the spectral index (see red triangles pointing downwards), while we derive a lower limit for the spectral index for the sources detected only at 1.3~cm, (see blue triangles pointing upwards). The sources detected at both wavelengths (black dots) have a more precise determination of the spectral index. For most sources, we derive spectral indices consistent with thermal emission (i.e., in the range of $-0.1$ to $+2$), and in agreement with observations of other radio jets \cite[e.g.,][]{Anglada2018}. Only a few sources show very negative spectral indices (\#48, \#96 and \#144). These sources are likely to be partially filtered out in the K~band images, which may result in lower limits for the actual value of the spectral index. In particular, source \#48 appears as three distinguishable peaks, which we refer to as a, b, and c, surrounded by a more diffuse and extended structure that is mainly visible in the C~band image. It is worth noting that large flux-to-intensity ratios indicate that the source is likely extended and most likely partially filtered out in the K band images, which may result in negative spectral indices.

\begin{table*}[ht!]
\centering
\caption{H$_2$O and CH$_3$OH maser features}
\begin{tabular}{l c c c c c c c}
\hline\hline

& 
& R.A.~(J2000)
& Dec.~(J2000)
& $V_\mathrm{maser}^\mathrm{a}$
& $V_\mathrm{LSR}^\mathrm{H^{13}CO^{+}}$
& Intensity
& Continuum
\\
Region
& Maser
& (h:m:s)
& ($^{\degr}$:$^{\arcmin}$:$^{\arcsec}$)
& (km~s$^{-1}$)
& (km~s$^{-1}$)
& (Jy~beam$^{-1}$)
& source ID$^\mathrm{b}$
\\
\hline      
IRAS\,05358$+$3543 & CH$_3$OH & 05:39:13.071  & $+$35:45:50.938 & $-$304    & \phn$-$15.8 & \phn0.028     & 2\\
G189.78$+$0.34     & CH$_3$OH & 06:08:35.304  & $+$20:39:06.405 & \phn$-$13 & \phnn$+$9.2 & \phn0.014     & 14  \\ 
G192.58$-$0.04     & CH$_3$OH & 06:12:54.026  & $+$17:59:23.060 & \phn$-$14 & \phnn$+$9.1 & \phn0.72\phn  & 22  \\ 
G18.18$-$0.30      & H$_2$O   & 18:25:07.575  & $-$13:14:31.487 & \phnn$-$3 & \phn$+$50.0 & \phn0.57\phn  & --   \\
IRAS\,18223$-$1243 & H$_2$O   & 18:25:10.804  & $-$12:42:26.234 & \phn$+$24 & \phn$+$45.2 & \phn0.006     & --   \\
IRAS\,18228$-$1312 & H$_2$O   & 18:25:41.935  & $-$13:10:19.591 & \phn$+$24 & \phn$+$33.1 & \phn0.022     & 48   \\
IRAS\,18236$-$1205 & H$_2$O   & 18:26:25.677  & $-$12:03:48.402 & \phn$+$28 & \phn$+$26.5 & \phn0.010     & 63   \\
                   & H$_2$O   & 18:26:25.575  & $-$12:03:48.502 & \phn$+$28 & \phn$+$26.5 & \phn0.006     & 63   \\
                   & H$_2$O   & 18:26:25.782  & $-$12:03:53.263 & \phn$+$15 & \phn$+$26.5 & \phn0.010     & 64   \\
                   & H$_2$O   & 18:26:27.149  & $-$12:03:54.888 & \phn$+$15 & \phn$+$26.5 & \phn0.014     & --   \\
                   & CH$_3$OH & 18:26:25.788  & $-$12:03:53.456 & \phnn$+$5 & \phn$+$26.5 & \phn0.26\phn  & 64   \\
G19.27$+$0.1M1     & H$_2$O   & 18:25:58.546  & $-$12:03:58.516 & \phn$+$28 & \phn$+$26.5 & \phn0.022     & --   \\
G23.60$+$0.0M1     & H$_2$O   & 18:34:11.237  & $-$08:19:07.680 & $+$108    & $+$106.5    & \phn0.44\phn  & --   \\
                   & H$_2$O   & 18:34:11.452  & $-$08:19:07.138 & $+$108    & $+$106.5    & \phn0.10\phn  & --   \\
IRAS\,18316$-$0602 & H$_2$O   & 18:34:20.918  & $-$05:59:41.638 & \phn$+$41 & \phn$+$42.5 & 11.1\phnn     & 83   \\
                   & CH$_3$OH & 18:34:20.913  & $-$05:59:42.087 & $-$233    & \phn$+$42.5 & \phn0.014     & 83   \\
G24.33$+$0.1M1     & H$_2$O   & 18:35:08.123  & $-$07:35:04.216 & $+$108    & $+$113.6    & \phn4.03\phn  & 110  \\
                   & CH$_3$OH & 18:35:08.147  & $-$07:35:04.260 & $-$182    & $+$113.6    & \phn0.010     & 110  \\
G24.60$+$0.1M2     & H$_2$O   & 18:35:35.728  & $-$07:18:08.796 & $+$122    & $+$115.3    & \phn0.031     & --   \\
G24.60$+$0.1M2     & H$_2$O   & 18:35:40.120  & $-$07:18:37.417 & \phn$+$54 & \phn$+$53.2 & \phn1.02\phn  & 136  \\
IRAS\,19095$+$0930 & H$_2$O   & 19:11:53.975  & $+$09:35:50.559 & \phn$+$37 & \phn$+$43.9 & 26.8\phnn     & 143  \\
                   & H$_2$O   & 19:11:53.990  & $+$09:35:49.848 & \phn$+$37 & \phn$+$43.9 & 10.3\phnn     & 143  \\
                   & CH$_3$OH & 19:11:53.993  & $+$09:35:50.641 & \phn$+$39 & \phn$+$43.9 & \phn0.043     & 143  \\
\hline
\end{tabular}
\tablefoot{
\tablefoottext{a}{Uncertainties in the reported maser velocities ($V_\mathrm{maser}$) are expected to be $\sim$50~km~s$^{-1}$ for the CH$_3$OH masers and $\sim$13~km~s$^{-1}$ for the H$_2$O masers (see Sect.~\ref{s:observations})}. The systemic velocities ($V_\mathrm{LSR}^\mathrm{H^{13}CO^{+}}$) are reported in \citet{LopezSepulcre2011}.
\tablefoottext{b}{Radio continuum source spatially associated with the maser feature and listed as identified in Table~\ref{t:catalogue}.}}
\label{t:masers}
\end{table*}


%
\subsection{Maser Emission}\label{s:maserresults}

H$_2$O and CH$_3$OH masers are excellent indicators of star formation activity \citep[e.g.,][]{Beuther2002b, Moscadelli2005, deVilliers2015}. We created image cubes of the H$_{2}$O and CH$_{3}$OH spectral lines and searched for maser features by scanning the entire velocity range. Despite the limited spectral resolution of our observation setup (see Sect.~\ref{s:observations}), we found maser emission in 14 of the 18 regions. In Table~\ref{t:masers}, we list the coordinates of the maser features detected in each region, together with the velocity at which the feature is detected and its intensity. We also compare the velocities of the maser features with the systemic velocities determined from H$^{13}$CO$^{+}$\,(1--0) observations \citep{LopezSepulcre2011}. We find that the H$_2$O masers have velocities that match the H$^{13}$CO$^{+}$\,(1--0) velocities, while the CH$_3$OH masers have a larger discrepancy, probably due to the lower spectral resolution.

The low spectral resolution in our observations compared to the typical maser line widths (a few km~s$^{-1}$; \citealt{Elitzur1982, Kalenski2016}) leads to smearing of the maser intensities. The intensities given in Table~\ref{t:masers} should be considered as lower limits. Despite this limitation, the high-angular resolution of our observations can be used to spatially associate the H$_2$O and CH$_3$OH masers to the detected continuum sources. If the angular separation between the continuum source and the maser is smaller than the synthesized beam size (listed in Table~\ref{t:sample}), we assume that the maser is associated with the continuum source. In the last column of Table~\ref{t:masers}, we specify the identifier of the continuum source (see Table~\ref{t:catalogue}) with which the maser is associated. We find 10 continuum sources associated with maser features (see Section~\ref{s:masers} for more details).

%
\section{Analysis and discussion\label{s:analysis}}

In this section, we determine how many sources in our sample are potential radio jets. For this purpose we study the nature of the detected radio continuum emission, and investigate the association with molecular outflows, masers and EGOs\footnote{The so-called EGOs (extended green objects) are sources with bright emission in the Spitzer}/IRAC 4.5~$\mu$m band and are usually found associated with strong shocks and jets \citep[e.g.,][]{Cyganowski2008, Cyganowski2009}. By using these criteria, we identify the best radio jet candidates in our sample and characterize their properties.

%
\subsection{Nature of the radio continuum emission\label{s:nature}}

\begin{figure}[ht!]
\centering
\includegraphics[width=\columnwidth]{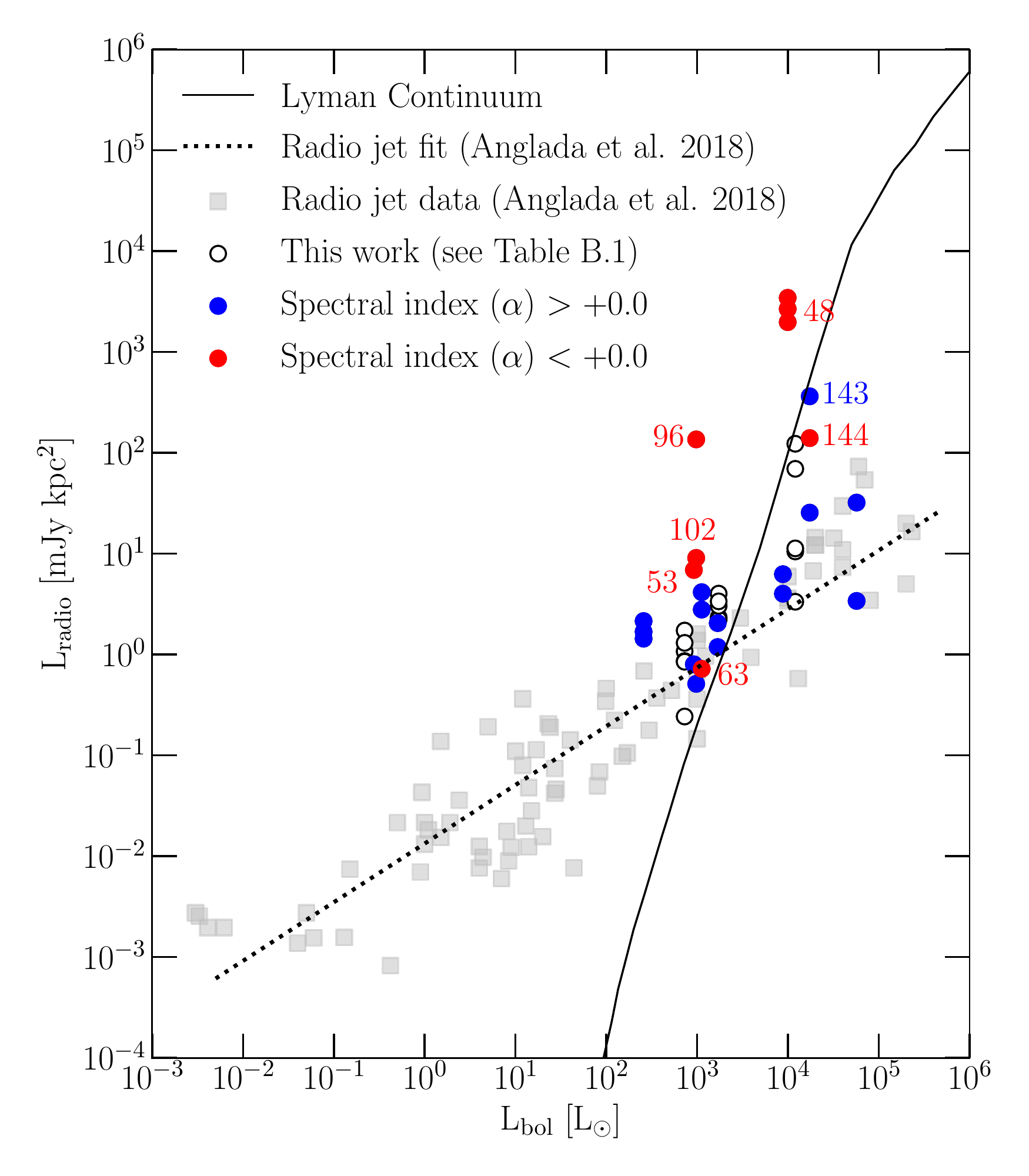}
\caption{Scatter plot of bolometric luminosity ($L_\mathrm{bol}$) and observed radio continuum luminosity ($L_\mathrm{radio}$) at 6~cm (C~band). Open black circles correspond to the continuum sources detected in our work and located within the primary beam of the K~band (1.3~cm) images (i.e., `iC/iK' in Table~\ref{t:catalogue}). Blue and red symbols mark the sources with positive and negative spectral indices, respectively, as listed in Table~\ref{t:catalogue} and shown in Fig.~\ref{fig:spix_intensity}). 
The solid line represents the values expected from Lyman continuum radiation for a zero-age main-sequence star of a given luminosity \citep{Thompson1984}.
The dashed line is a least-squares fit to the radio jets reported by \citet[][shown as grey squares]{Anglada2018}, corresponding to [$L_\mathrm{radio}$/mJy~kpc$^2$]~=~$10^{-1.90}$~[$L_\mathrm{bol}$/$L_\odot$]$^{+0.59}$ (see their Eq.~28).}
\label{fig:Lrad_Lbol}
\end{figure}

Usually two mechanisms are invoked to explain the origin of thermal free-free radiation from ionized gas in star-forming regions: Photoionization and ionization through shocks \citep[e.g.,][]{GordonSorochenko2002, Kurtz2005, SanchezMonge2008, SanchezMonge2013c, Anglada2018}. In the case of photoionization, ultraviolet (UV) photons with energies above 13.6~eV are emitted by massive stars and ionize the surrounding atomic hydrogen. In the second scenario, the ionization is produced when ejected material associated with outflows and jets interacts in a shock with neutral and dense material surrounding the forming star \citep[e.g.,][]{Curiel1987, Curiel1989, Anglada1992}.

\citet{Anglada1995, Anglada1996} show that the relation between the radio luminosity and the bolometric luminosity of young stellar objects (YSOs) depends on the origin of the ionization: stellar UV radiation or shocks \citep[see also][]{Anglada2018}. We use this relation to investigate the nature of our continuum sources. The solid line in Fig.~\ref{fig:Lrad_Lbol} shows the maximum radio luminosity that a high-mass object of a given luminosity may have according to its UV radiation, the so-called Lyman continuum limit usually associated with \hii\ regions. The radio luminosity decreases fast with decreasing bolometric luminosity. In contrast, the radio luminosity originated in shocks (i.e., radio jets) has a less steep curve. The dotted line in Fig.~\ref{fig:Lrad_Lbol} shows the least-squares fit to the sample of radio jets studied in \cite{Anglada2018} and shown as grey squares in the figure.

We have calculated the radio luminosity of our continuum sources as $L_\mathrm{radio}=S_\nu\,d^2$, where $S_\nu$ is the observed flux density in the C~band (listed in Table~\ref{t:catalogue}) and $d$ is the distance to the source (listed in Table~\ref{t:sample}). The bolometric luminosity ($L_\mathrm{bol}$) of each source is uncertain due to the lack of high-resolution data at far-infrared wavelengths. The bolometric luminosity of each region is given in Table~A.1 of \cite{LopezSepulcre2011} and provides an upper limit to the actual luminosity. As a simple approach, we divide the bolometric luminosity by the number of radio sources detected within the primary beam to have an estimate of the expected average luminosity for the continuum sources in the region. Circle symbols in Fig.~\ref{fig:Lrad_Lbol} show the continuum sources detected in our work and located inside the K~band primary beam (i.e., with reliable flux measurements and listed as iC/iK in Table~\ref{t:catalogue}). Colored symbols correspond to those sources for which we could derive the spectral index (see Fig.~\ref{fig:spix_intensity}), with blue symbols corresponding to positive spectral indices (i.e., $\alpha>+0.0$, mainly thermal emission) and red symbols corresponding to negative spectral indices. In general, our sources lie in between the two lines defining the radio jet and \hii\ region regimes\footnote{Some sources lie above the solid curve depicting the Lyman continuum limit. This is in agreement with other studies that report the existence of a population of \hii\ regions with radio fluxes larger than the Lyman continuum limit \citep[see e.g.,][]{SanchezMonge2013a, Cesaroni2016}.}. Interestingly, sources with positive spectral indices (blue symbols) seem to preferentially follow, although with some dispersion, the relation found for radio jets, while sources with negative spectral indices (red symbols) are located closer to the Lyman continuum regime. This favours our previous interpretation that sources with negative spectral indices may be slightly extended \hii\ regions that are partially filtered out in the K~band images.

%
\subsection{Association with molecular outflows\label{s:outflows}}

We investigate the association of radio continuum sources with molecular outflows by comparing the location of radio sources with respect to the molecular outflow emission reported mainly by \citet{LopezSepulcre2010} and \citet{SanchezMonge2013d}. It is expected that the most promising radio jet candidates will be in the center of the molecular outflow emission.

We find a total of twenty-four radio continuum sources that are spatially associated with molecular outflow emission (see Table~\ref{t:candidates} and Fig.~\ref{f:diagrams} for more details). Out of these sources, eighteen (\#2, \#4, \#13, \#14, \#15, \#16, \#22, \#23, \#25, \#48a, \#48b, \#48c, \#74, \#83, \#95, \#110, \#137 and \#143) are located at or near the geometric center of the molecular outflow emission, while the remaining six (\#12, \#63, \#64, \#65, \#73, and \#144) are located within the outflow lobes. Although we cannot confirm that these six sources are at the base of the outflows detected with single-dish telescopes (with angular resolutions of 11--29$^{\arcsec}$), we cannot exclude that they might drive molecular outflows. Further observations of outflow tracers at higher angular resolution are necessary to confirm and better associated the molecular outflows with the radio continuum sources. In Table~\ref{t:candidates}, we list the outflow momentum rates reported in the literature \citep[see][]{LopezSepulcre2010, SanchezMonge2013d}. For \#137, no outflow momentum rate has been reported \citep{Hatchell2001, Liu2013}.

%
\subsection{Association with EGOs\label{s:EGOs}}

In this section, we investigate the association of radio continuum sources with \textit{Spitzer}/IRAC~4.5~$\mu$m emission tracing EGOs, which are considered related to the shocked gas. For the association with EGOs we used the catalogues of \cite{Cyganowski2008, Cyganowski2009}. In total, we found six sources (\#42, \#63, \#64, \#119, \#137 and \#139) with an EGO counterpart (see Table~\ref{t:candidates}).

We have also inspected the \textit{Spitzer}/IRAC images of the different regions to search for other possible EGOs not included in previous catalogues. We have identified nine radio continuum sources in this category (see sources \#48, \#65, \#73, \#74, \#83, \#110, and \#143, marked with a '?' symbol in Table~\ref{t:candidates}). The association of these sources with bright 4.5~$\mu$m emission suggests their association with strong shocks and favors the hypothesis of a radio jet origin for the radio continuum emission of these objects. However, a more detailed characterization of the infrared properties of the nine additional sources is necessary to confirm whether these objects are EGOs or not.

\begin{table*}[ht!]
\caption{Properties of the radio jet candidates.}
\label{t:candidates}
\centering
\begin{tabular}{l c c c c c c c c c c c}
\hline

& \multicolumn{3}{c}{Flux properties\tablefootmark{a}}
& 
& \multicolumn{3}{c}{Source size properties\tablefootmark{b}}
& 
& \multicolumn{3}{c}{Outflow/shock activity\tablefootmark{c}}
\\
\cline{2-4}\cline{6-8}\cline{10-12}
\multicolumn{1}{c}{ID}
& $S_\mathrm{C~band}$
& $S_\mathrm{K~band}$
& $\alpha$
&
& $\theta_\mathrm{C~band}$
& $\theta_\mathrm{K~band}$
& $\beta$
&
& log($\dot{P}_\mathrm{out}$)
& EGOs
& Masers
\\
\hline\hline
 & \\
\multicolumn{10}{l}{Radio jet candidates with signposts of outflow activity} \\

\hline
\phnn2\tablefootmark{d}  & \phnn0.53$\pm$0.01 & ---                & ---               && \phs0.75 & ---          & ---         && $-3.9\pht$           & n  & CH$_{3}$OH \\
\phnn4\tablefootmark{d}  & \phnn0.33$\pm$0.01 & ---                & ---               && \phs1.23 & ---          & ---         && $-3.9\pht$           & n  & \ldots \\

\phn12\tablefootmark{d}  & \phnn0.72$\pm$0.03 & ---                & ---               && \phs0.97 & ---          & ---         && $-3.1^\dagger$       & n  & \ldots \\
\phn13\tablefootmark{d}  & \phnn1.24$\pm$0.05 & ---                & ---               && \phs0.75 & ---          & ---         && $-3.1\pht$       & n  & \ldots \\
\phn14\tablefootmark{d}  & \phnn0.69$\pm$0.05 & ---                & ---               && \phs1.08 & ---          & ---         && $-3.1\pht$       & n  & CH$_{3}$OH \\
\phn15\tablefootmark{d}  & \phnn0.94$\pm$0.04 & ---                & ---               && \phs1.60 & ---          & ---         && $-3.1\pht$       & n  & \ldots\\
\phn16\tablefootmark{d}  & \phnn1.04$\pm$0.04 & ---                & ---               && \phs1.16 & ---          & ---         && $-3.1\pht$       & n  & \ldots \\

\phn22\tablefootmark{d}  & \phn10.27$\pm$0.22 & ---                & ---               && \phs1.55 & ---          & ---         && $-3.3\pht$       & n  & CH$_{3}$OH \\
\phn23\tablefootmark{d}  & \phn1.56 $\pm$0.05 & ---                & ---               && \ldots   & ---          & ---         && $-3.3\pht$       & n  & \ldots \\
\phn25\tablefootmark{d}  & \phn0.49 $\pm$0.02 & ---                & ---               && \ldots   & ---          & ---         && $-3.3\pht$       & n  & \ldots \\

\phn42                   & \phnn4.16$\pm$0.08 & \ldots             & \ldots            && \phs0.81 & \ldots       & \ldots      && \ldots            & Y  & \ldots \\ 
\phn48a\tablefootmark{e} & \phn55.32$\pm$3.50 & \phn15.19$\pm$1.51 & $-0.99\pm0.09$    && \phs2.70 & \phn\phs2.07 & \phs$-0.20$ && $-2.9\pht$           & ?  & H$_{2}$O \\
\phn48b\tablefootmark{e} & \phn75.41$\pm$9.50 & \phn10.53$\pm$1.30 & $-1.50\pm0.14$    && \phs3.84 & \phn\phs1.64 & \phs$-0.65$ && $-2.9\pht$           & ?  & \ldots \\
\phn48c\tablefootmark{e} &    129.52$\pm$8.30 & \phn54.41$\pm$3.51 & $-0.66\pm0.07$    && \phs2.28 & \phn\phs1.89 & \phs$-0.14$ && $-2.9\pht$           & ?  & \ldots \\
\phn63\tablefootmark{e}  & \phnn0.99$\pm$0.06 & \phnn0.75$\pm$0.13 & $-0.22\pm0.07$    && \phs1.19 & \phn\phs1.11 & \phs$-0.05$ && $-2.6^\dagger$       & Y  & H$_{2}$O \\
\phn64\tablefootmark{e}  & \phnn0.28$\pm$0.02 & \phnn0.35$\pm$0.08 & $+0.18\pm0.04$    && $<$1.38  & \phn\phs0.66 & $>-0.56$    && $-2.6^\dagger$           & Y  & H$_{2}$O, CH$_{3}$OH \\
\phn65\tablefootmark{e}  & \phnn0.57$\pm$0.14 & \phnn2.31$\pm$0.09 & $+1.08\pm0.19$    && $<$1.46  & \phn$<$0.75  & \ldots      && $-2.6^\dagger$       & ?  & \ldots \\
\phn73\tablefootmark{e}  & \phnn0.24$\pm$0.15 & \phnn0.43$\pm$0.09 & $+0.45\pm0.50$    && \phs0.38 & \phn$<$0.74  & $<+0.51$    && $-2.4^\dagger$       & ?  & \ldots \\
\phn74\tablefootmark{e}  & \phnn0.35$\pm$0.03 & \phnn0.49$\pm$0.13 & $+0.26\pm0.21$    && \phs0.53 & \phn\phs0.42 & \phs$-0.18$ && $-2.4\pht$           & ?  & \ldots \\
\phn83\tablefootmark{e}  & \phnn3.35$\pm$0.21 & \phnn3.73$\pm$0.29 & $+0.08\pm0.08$    && \phs0.87 & \phn\phs0.85 & \phs$-0.02$ && $-1.8\pht$           & ?  & H$_{2}$O, CH$_{3}$OH \\
\phn95                   & $<$0.042            & \phnn0.25$\pm$0.07 & $>+1.36$          && \ldots   & \phn$<$0.42  & \ldots      && $-1.8\pht$           & n  & \ldots \\ 
110\tablefootmark{e}     & \phnn0.46$\pm$0.21 & \phnn1.20$\pm$0.06 & $+0.73\pm0.35$    && $<$1.39  & \phn\phs0.37 & $>-1.01$    && $-2.9\pht$           & ?  & H$_{2}$O, CH$_{3}$OH\\
119                      & \phnn1.11$\pm$0.06 & $<0.022$           & \ldots            && \phs2.28 & \ldots       & \ldots      && \ldots                & Y  & \ldots \\ 
136                      & $<$0.10            & \phnn0.85$\pm$0.12 & $>+1.67$          && \ldots   & \phn\phs1.50 & \ldots      && \ldots                & n  & H$_{2}$O \\ 
137\tablefootmark{e}     & \phn14.40$\pm$1.20 & \phnn2.21$\pm$0.20 & \ldots            && \phs2.53 & \phn\phs1.60 & \phs$-0.35$ && $\pht$\ldots$^\ddag$  & Y  & \ldots \\
139                      & \phnn0.73$\pm$0.03 & $<0.019$           & \ldots            && \phs0.87 & \ldots       & \ldots      && \ldots                & Y  & \ldots \\ 
143\tablefootmark{e}     & \phn39.50$\pm$1.60 &    130.87$\pm$2.60 & $+1.12\pm0.04$    && \phs0.55 & \phn\phs0.28 & \phs$-0.52$ && $-3.4\pht$           & ?  & H$_{2}$O, CH$_{3}$OH \\
144                      & \phn15.57$\pm$0.89 & $<$0.32            & $<-2.97$          && \phs2.44 & \ldots       &             && $-3.4^\dagger$           & n  & \ldots \\
\hline
 & \\
\multicolumn{10}{l}{Radio continuum sources consistent with positive spectral index, but with no signposts of outflow activity} \\
\hline
\phn61                   & \phnn0.14$\pm$0.03 & $<$2.10            & $<+2.07$          && \phs1.79 &\ldots        & \ldots      && \ldots   & n  & \ldots \\
\phn62                   & $<$0.05            & \phn0.24$\pm$0.09  & $>+1.23$          && \ldots   & \phn$<$0.44  & \ldots      && \ldots   & n  & \ldots \\
\phn86                   & \phnn0.35$\pm$0.01 & $<$2.32            & $<+1.43$          && $<$1.45  & \ldots       & \ldots      && \ldots   & n  & \ldots \\
109                      & \phnn0.08$\pm$0.03 & $<$1.75            & $<+2.33$          && $<$0.94  & \ldots       & \ldots      && \ldots   & n  & \ldots \\
113                      & \phnn0.28$\pm$0.01 & $<$0.31            & $<+0.07$          && \phs0.85 & \ldots       & \ldots      && \ldots   & n  & \ldots \\
126                      & \phnn0.11$\pm$0.01 & $<$0.37            & $<+0.96$          && $<$1.37  & \ldots       & \ldots      && \ldots   & n  & \ldots \\
129                      & \phnn0.16$\pm$0.01 & $<$0.20            & $<+0.18$          && \phs1.23 & \ldots       & \ldots      && \ldots   & n  & \ldots \\
145                      & \phnn2.84$\pm$0.02 & $<$4.34            & $<+0.32$          && $<$1.48  & \ldots       & \ldots      && \ldots   & n  & \ldots \\

\hline
\end{tabular}
\tablefoot{
\tablefoottext{a}{Primary beam corrected fluxes in mJy as listed in Table~\ref{t:catalogue}. For sources \#42, \#119, \#137 and \#139 it was not possible primary beam correct the fluxes at both bands (see Sect~\ref{s:results}), resulting in not usable spectral indices. The spectral index $\alpha$ is defined in Eq.~\ref{eq:sizeindex}.}
\tablefoottext{b}{Source sizes in arcsec determined as $\sqrt{\theta_\mathrm{major}\times\theta_\mathrm{minor}}$, with $\theta_\mathrm{major}$ and $\theta_\mathrm{minor}$ listed in Table~\ref{t:sizes}. Upper limits corresponds to sources for which we could not determine a deconvolved source size. The source size index $\beta$ is defined in Eq.~\ref{eq:sizeindex}.}
\tablefoottext{c}{Association of the radio continuum source with outflow and shock activity. The associations correspond to: (i) Molecular outflows, with the outflow momentum rate $\dot{P}_\mathrm{out}$ given in units of $M_\sun$~yr$^{-1}$~km~s$^{-1}$ \citep[from][]{LopezSepulcre2010, SanchezMonge2013d}, with the symbol $^\dagger$ indicating those radio continuum sources located within the outflow lobes and not at the center of the outflow, (ii) EGOs (or extended green objects), based on the catalogue of \citet[][question mark symbols indicate the presence of bright \textit{Spitzer}/IRAC 4.5~$\mu$m emission although without confirmation of the object being an EGO]{Cyganowski2008}, and (iii) H$_2$O and CH$_3$OH masers, as listed in Table~\ref{t:masers}. Source \#137, marked with a $^\ddag$ symbol, is associated with molecular outflow emission \citep{Hatchell2001, Liu2013}, but no outflow momentum rate has been reported.}
\tablefoottext{d}{Sources not observed in the K~band. For these sources we do not have information on the K~band flux and presence of H$_2$O masers.}
\tablefoottext{e}{Sources detected at both frequency bands and for which we have created new images using a common \textit{uv}-range that allows us to sample similar spatial scales. Fluxes and source sizes for these sources are taken from Table~\ref{t:uvsizes}. Fluxes for source \#137 can not be primary beam corrected and are not usable for spectral index determination.}}
\end{table*}

%
\subsection{Association with masers\label{s:masers}}

Our VLA observations (see Sect.~\ref{s:vla}) allow us to search for H$_2$O and CH$_3$OH maser spots associated with radio continuum sources. As shown in Table~\ref{t:masers}, we have found a total of sixteen H$_2$O and seven CH$_3$OH maser spots.

\begin{figure*}[ht!]
    \centering
    \includegraphics[width=0.85\textwidth]{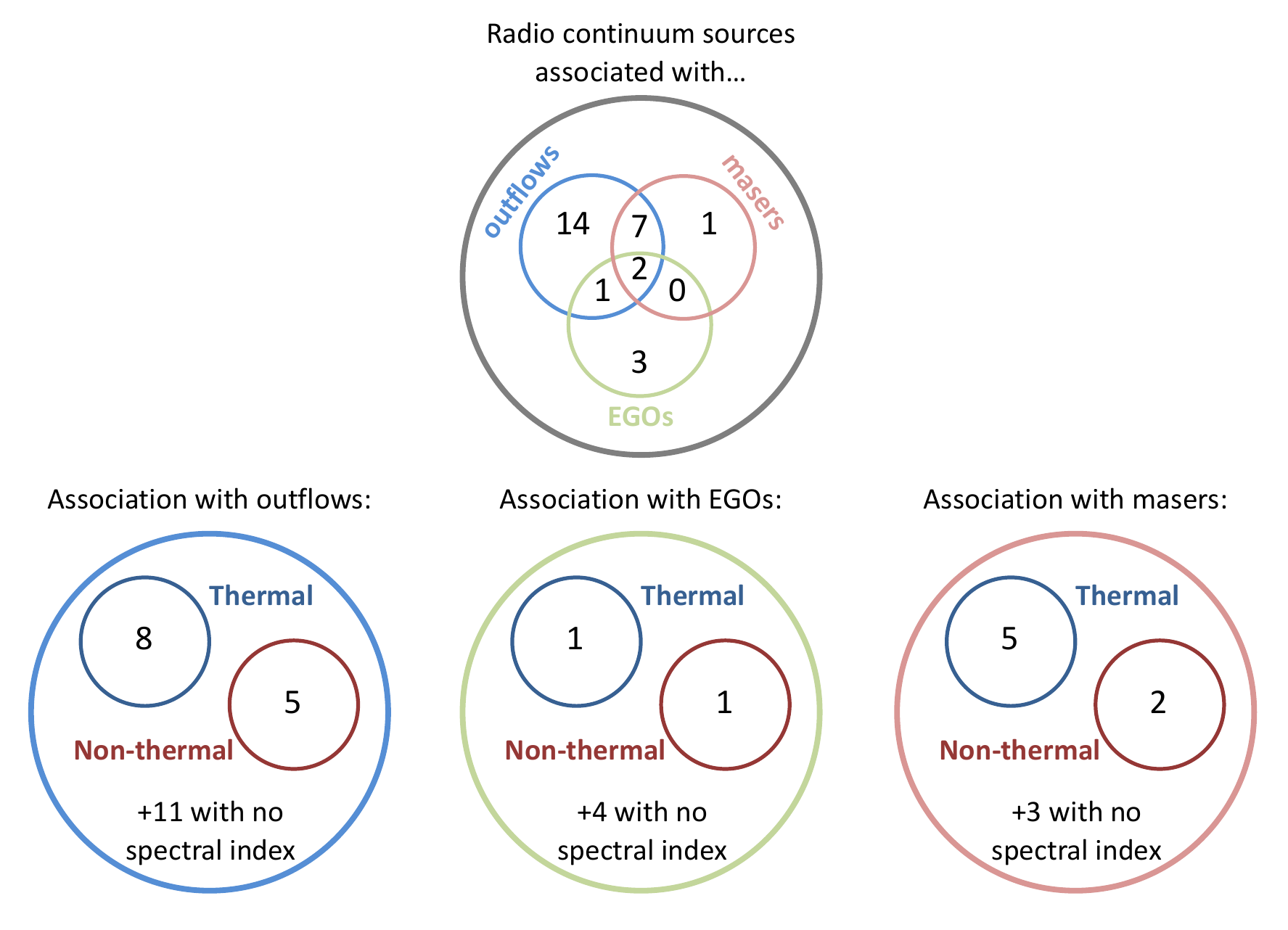}
    \caption{Diagrams summarizing the outflow-activity associations of the radio jet candidates studied in this work. The integer numbers indicate the number radio jet candidates in a specific group (see Table~\ref{t:candidates}). The top diagram summarizes the association of the radio jet candidates with molecular outflows, masers and EGOs (see Sect.~\ref{s:analysis}, for EGOs we only consider association if the source is labeled as `Y' in Table~\ref{t:candidates}). The bottom-row diagrams summarize the results regarding the thermal (spectral index $>-0.1$) and non-thermal (spectral index $<-0.1$) properties of the radio continuum emission. The number of sources for which we could not derive the spectral index is also indicated.}
    \label{f:diagrams}
\end{figure*}

We find ten radio continuum sources associated with maser features, of which three (\#2, \#14 and \#22) are associated with CH$_3$OH masers only, three sources (\#48, \#63, and \#136) are associated only with H$_2$O masers, and four sources (\#64, \#83, \#110 and \#143) are associated with both types of masers (see Table~\ref{t:candidates} and Fig.~\ref{f:diagrams}). It is worth noting that the three sources associated only with CH$_3$OH masers correspond to regions observed only in the C~band. Future observations of these sources in the K~band together with observations of the H$_2$O maser line could confirm that all sources associated with CH$_3$OH maser are also associated with H$_2$O maser features.

The observed Class~II 6.7~GHz CH$_3$OH masers are ideal signposts for embedded young stellar objects and mark the location of deeply embedded massive protostars \citep[e.g.,][]{Breen2013}. On the other hand, 22~GHz H$_2$O masers have been found associated with outflow activity \citep[e.g.,][]{Torrelles2011} as well as tracing disk-like structures around young stellar objects \cite[e.g.,][]{Moscadelli2019}. Our maser observations have therefore enabled us to identify at least seven potential candidates for a radio jet (i.e., sources associated with outflow activity).

\begin{table}[h]
    \caption{Number of the sources with thermal and non-thermal radio continuum emission associated with different outflow activity signatures.}
    \centering
    \begin{tabular}{l c c}
    \hline\hline
                    & Non-thermal Sources & Thermal Sources \\ 
    \hline
       Outflows          &  5/5 (100\%)        &  7/8 (88\%)    \\
       EGOs              &  1/5 (20\%)         &  1/8 (13\%)    \\
       Masers (all)      &  2/5 (40\%)         &  5/8 (63\%)    \\
    \hspace{4mm}H$_2$O   &  2/5 (40\%)         &  5/8 (63\%)    \\
    \hspace{4mm}CH$_3$OH &  0/5 (0\%)          &  4/8 (50\%)   \\
    \hline
    \end{tabular}
    \label{t:percentages}
\end{table}

%
\subsection{Radio jet candidates\label{s:candidates}}

Out of the 146 radio continuum sources detected in our study, we have identified twenty-eight sources (see list at the beginning of Table~\ref{t:candidates}) as possible radio jet candidates, based on their association with outflow and shock activity. We find twenty-four of these sources associated with molecular outflow emission, six of them with EGOs, and ten with masers. In the sketch presented in Fig.~\ref{f:diagrams}, we summarize these findings.

In addition to these twenty-eight sources, we have also identified eight radio-continuum sources with spectral indices consistent with thermal emission (see bottom list in Table~\ref{t:candidates}). Based on the results shown in Fig.~\ref{fig:Lrad_Lbol}, these sources could also be radio jet candidates, despite their lack of association with tracers of outflow and shock activity. In the following, we build on the properties of the identified radio jet candidates.

%
\subsubsection{Radio continuum properties\label{s:alphasize}}

\citet{Reynolds1986} describe radio jets with a model that assumes a jet of varying temperature, velocity, and ionization fraction. In case of constant temperature, the relations of the flux density ($S_\nu$) and source size ($\theta_\nu$) with frequency are given by
\begin{equation}
S_{\nu} \propto \nu^\alpha = \nu^{1.3-0.7/\epsilon} \\ \mathrm{and} \\ \theta_{\nu} \propto \nu^\beta = \nu^{-0.7/\epsilon},
\label{eq:sizeindex}
\end{equation}
where $\epsilon$ depends only on the geometry of the jet and is the power-law index that describes how the width of the jet varies with the distance from the central object. In this model, the spectral index $\alpha$ is always smaller than 1.3 and drops to values $<0.6$ for confined jets ($\epsilon<1$; \citealt{Anglada1998}).

In Table~\ref{t:candidates}, we list the spectral index ($\alpha$) and the source size index ($\beta$) for our radio jet candidates. The latter only for the sources detected at both frequencies. Nine of the radio jet candidates associated with outflow/shock activity have spectral indices consistent with thermal emission ($>-0.1$), with six showing clear positive ($>+0.4$) spectral indices. These values are consistent with the model of \citet{Reynolds1986} for values of $\epsilon>0.6$. For such geometries of the jet, the source size index ($\beta$) is expected to be about $-1$. The source size indices reported in Table~\ref{t:candidates} are mainly in the range $-0.1$ to $-1.0$, in agreement with the model of \citet{Reynolds1986} for radio jets.

Despite most of our radio jet candidates have spectral indices consistent with thermal emission (64\% of the sample, see Table~\ref{t:candidates}), we find some sources (accounting for 36\% of the sample, five sources\footnote{As discussed in Sect.~\ref{s:bestradiojets}, four of these five sources are most likely \hii\ regions. This would reduce the number of non-thermal radio jets to only one out of 14 (7\% of our sample).}) that show negative spectral indices. This finding is in agreement with some recent works. For example, \citet{Moscadelli2016} find about 20\% of their sample of 15 radio continuum sources to be associated with non-thermal emission. The presence of non-thermal emission is explained in terms of synchrotron emission from relativistic electrons accelerated in strong shocks within the jets, and a number of cases have been studied in more detail \citep[e.g.,][]{CarrascoGonzalez2010, Sanna2019}. Further detailed observations towards these new four non-thermal radio jet candidates, can provide further constraints to understand the characteristics of this kind of objects.

We have searched for possible relations between the presence of thermal and non-thermal radio jets and different outflow/shock activity signposts (i.e., outflows, masers and EGOs). We summarize our findings in the bottom panels of Fig.~\ref{f:diagrams}). We do not find a preferred relation between thermal and non-thermal radio jets with the outflow activity signposts, since we find similar percentages (see Table~\ref{t:percentages}) for the association with outflows (88\% and 100\%, respectively), EGOs (13\% and 20\%), and masers (55\% and 40\%). The low number of objects included in our analysis prevents us from deriving further conclusions, and we indicate that a larger sample of radio jets needs to be studied to better understand the properties and differences between thermal and non-thermal radio jets. It is also worth noting that all the four objects associated with both CH$_3$OH and H$_2$O masers are thermal radio jet candidates (see Table~\ref{t:candidates}), while only one of the three objects associated with only H$_2$O masers shows thermal emission. This might suggest that radio jets associated with CH$_3$OH masers tend to have positive spectral indices (i.e., thermal emission), while radio jets associated with only H$_2$O masers might preferentially have negative spectral indices (i.e., non-thermal emission). However, the low number of sources studied in our sample prevents from deriving further conclusions. One should also note that the different levels of association of the radio continuum sources with maser emission may be affected by the variability of masers \citep{Felli2007, Sugiyama2017, Ashimbaeva2017}. Moreover, we can not discard that the poor spectral resolution of our observations, which may smear out the intensity of the maser lines making some of them undetectable with our sensitivity limits, may also affect our detectability limits. Despite these limitations, our results are in agreement with the 6.7~GHz CH$_3$OH masers tracing the actual location of the newly-born YSOs usually associated with thermal winds/jets, while H$_2$O masers may be originated in strong shocks where non-thermal synchrotron emission can be relevant \cite[see e.g.,][]{Moscadelli2013, Moscadelli2016}.

\begin{figure}[h]
\centering
\includegraphics[width=1.0\columnwidth]{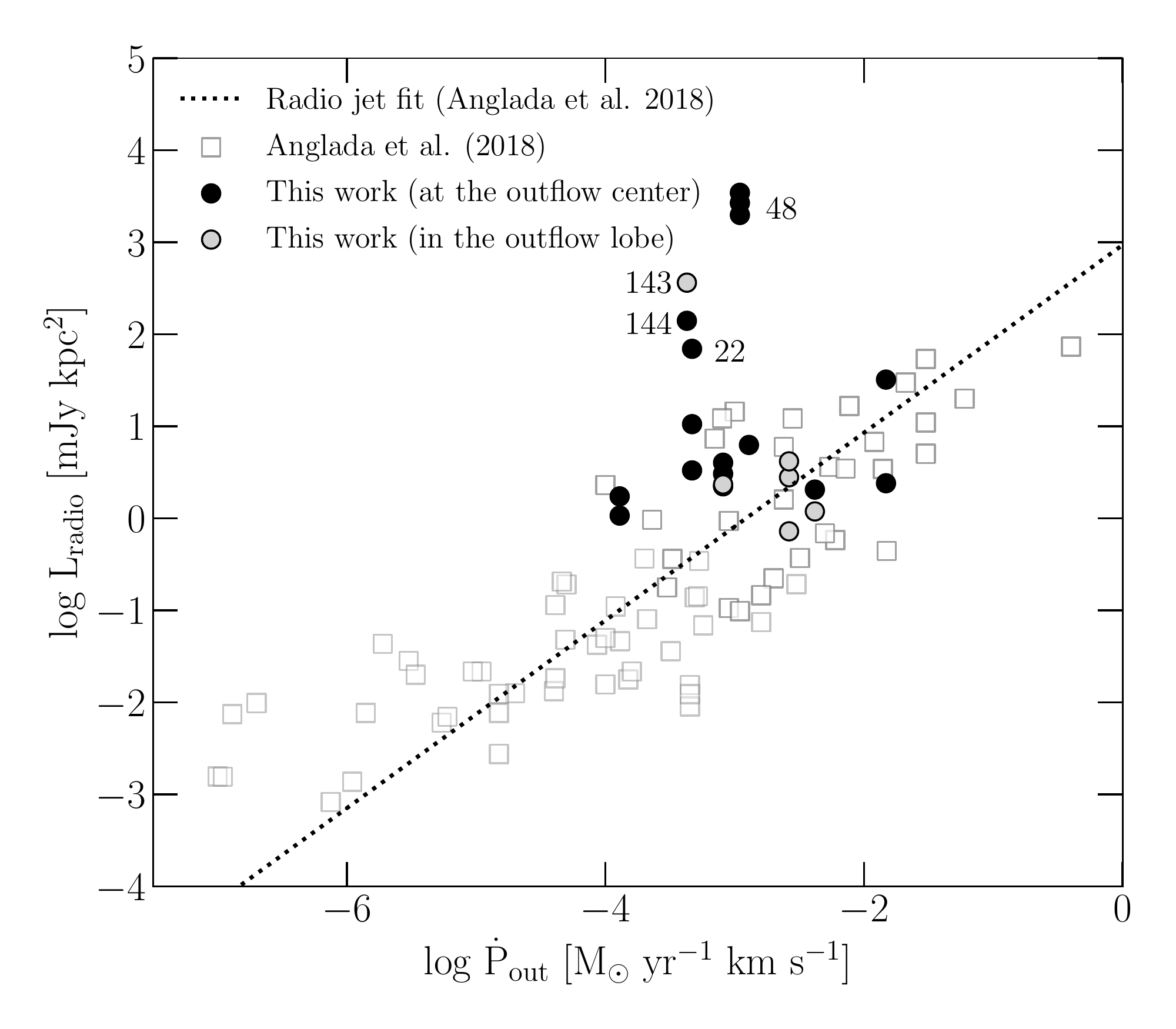}
\caption{Relation between radio luminosity ($L_\mathrm{radio}$) and outflow momentum rate ($\dot{P}_\mathrm{out}$). Open boxes show data from \cite[][see also \citealt{Anglada1992}]{Anglada2018}. The dashed line is a least-squares fit to the radio jets reported by \citet{Anglada2018}, corresponding to [$L_\mathrm{radio}$/mJy~kpc$^2$]~=~10$^{+2.97}$~[$\dot{P}$/$M_\odot$~yr$^{-1}$~km~s$^{-1}$]$^{+1.02}$ (see their Eq.~31).}
\label{f:LradPout}
\end{figure}

%
\subsubsection{Jet-outflow connection\label{s:jetoutflow}}

It has been found that the radio luminosity ($L_\mathrm{rad} = S_{\nu}d^2$) of thermal radio jets is correlated with the energetics of the associated molecular outflows. The relationship between radio luminosity and momentum rate in the outflow ($\dot{P}_\mathrm{out}$) is empirically derived by \citet[][see also \citealt{Anglada2018}]{Anglada1992}. In Fig.~\ref{f:LradPout}, we compare our radio jet candidates (see Table~\ref{t:candidates}) with the radio jets studied by \citet{Anglada2018}. As reported by \citet{Anglada2018}, there is a tight correlation between the radio luminosity of the jet and the outflow momentum rate. This relationship is interpreted as proof that shocks are the ionization mechanism of radio jets \citep[see e.g.,][]{Rodriguez2008, Anglada2018}. Most of the radio jet candidates investigated in this work, with the exception of only a few sources, follow this relationship, suggesting a radio jet origin for the detected radio continuum emission. The exceptions are mainly the sources \#48a, \#48b, \#48c, \#143 and \#144, which have a much larger radio luminosity compared to the associated outflow momentum rate. This excess suggests that another mechanism could be responsible for a large fraction of the observed radio continuum emission. Based on the location of these sources in the diagram shown in Fig.~\ref{fig:Lrad_Lbol}, these sources may correspond to more evolved and extended \hii\ regions, instead of radio jets, thus explaining the discrepancy between the observed radio luminosity and outflow momentum rate. In this case, we could be facing two possible scenarios. The first is that the sources are indeed radio jets transitioning into a more evolved \hii\ region phase (similar to what has been proposed for G35.20$-$0.74N, \citealt{Beltran2016}). The second scenario is that the radio continuum sources that we are detecting are only associated with an \hii\ region, and the spatial coincidence with the molecular outflow emission is due to the presence of another (lower-mass) object powering the outflow but with non-detectable radio continuum emission in our maps. Higher angular resolution observations of the molecular outflow can better establish the location of the powering source and its association with the detected radio continuum sources.

\begin{figure}[t!]
\centering
\includegraphics[width=1.0\columnwidth]{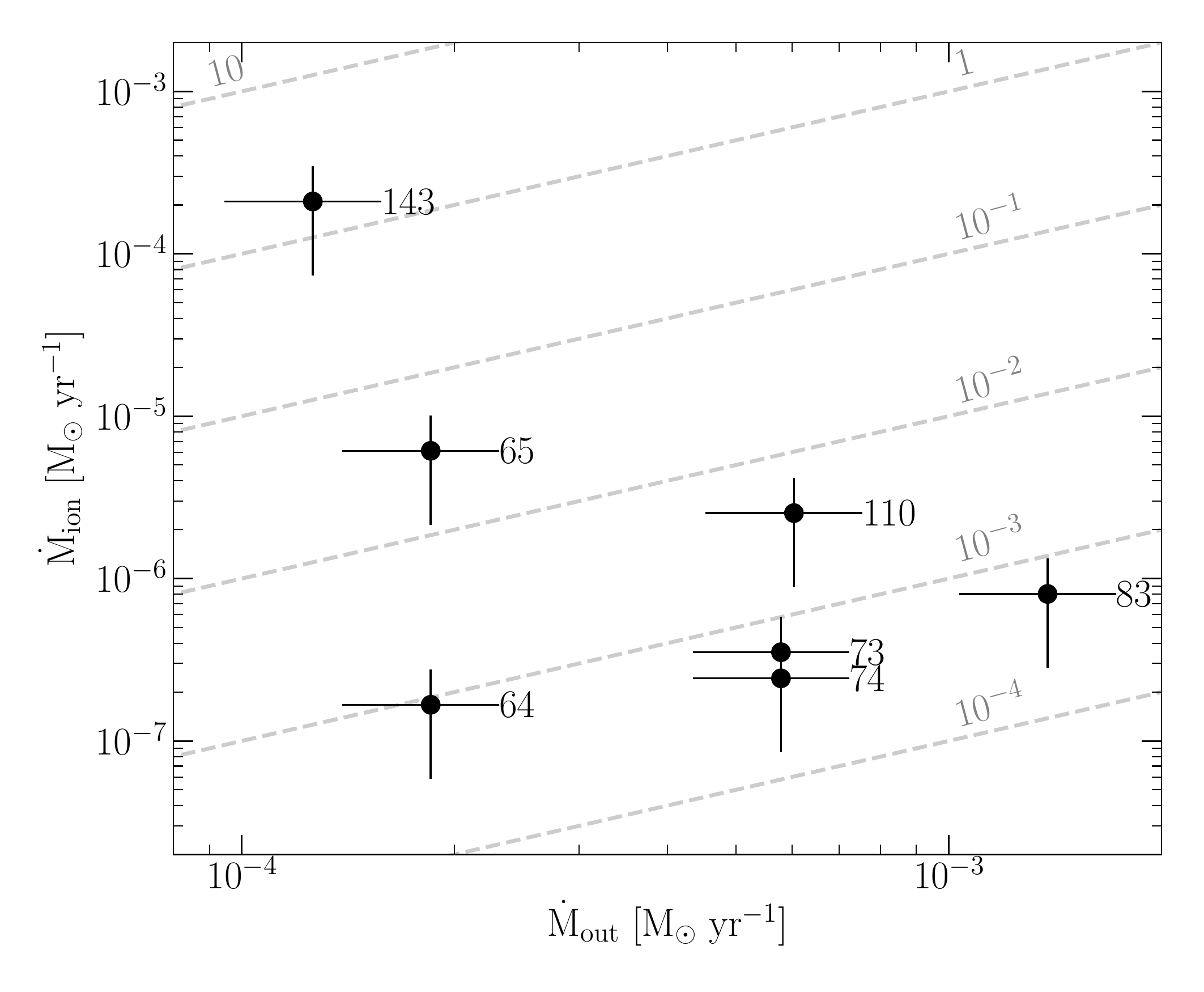}
\caption{
Relation between ionized ($\dot{M}_\mathrm{ion}$) and molecular outflow ($\dot{M}_\mathrm{out}$) mass loss rates for the radio jet candidates listed in Table~\ref{t:candidates}. The ionized mass loss rate is derived for the thermal radio jets using Eq.~\ref{eq:ionmasslossrate}, while the molecular outflow mass loss rate is provided in \citet{LopezSepulcre2010} and \citet{SanchezMonge2013d}. The dashed lines indicate different ionization levels given by the ratio $\dot{M}_\mathrm{ion}$ to $\dot{M}_\mathrm{out}$.}
\label{f:MionMout}
\end{figure}

Following Eq.~8 of \citet[][see also \citealt{Reynolds1986}]{Anglada2018}, we estimate the ionized mass-loss rate ($\dot{M}_\mathrm{ion}$) of our radio jet candidates as
\begin{equation}\label{eq:ionmasslossrate}
\begin{split}
    \left(\frac{\dot{M}_\mathrm{ion}}{10^{-6}~M_{\sun}~\mathrm{yr}^{-1}}\right) = &
    0.108 \left(\frac{d}{\mathrm{kpc}}\right)^{1.5} \left[\frac{\left(2-\alpha\right)\left(0.1+\alpha\right)}{1.3-\alpha}\right]^{0.75} \\
    & \times\left(\frac{T}{10^4~\mathrm{K}}\right)^{-0.075} \left[\left(\frac{S_\nu}{\mathrm{mJy}}\right)\left(\frac{\nu}{10~\mathrm{GHz}}\right)^{-\alpha}\right]^{0.75} \\
    & \times\left(\frac{V_\mathrm{jet}}{200~\mathrm{km}~\mathrm{s}^{-1}}\right) \left(\frac{\nu_m}{10~\mathrm{GHz}}\right)^{0.75\alpha-0.45} \\
    & \times\left(\frac{\theta_0}{\mathrm{rad}}\right)^{0.75} (\sin~i)^{-0.25}, 
\end{split}
\end{equation}
where $\alpha$ is the spectral index and $S_\nu$ is the radio continuum flux, both listed in Table~\ref{t:candidates}, and $d$ is the distance to the source. The opening angle of the jet $\theta_0$ can be approximated as $2\arctan(\theta_\mathrm{min}/\theta_\mathrm{maj})$ \citep{Beltran2001, Anglada2018}. We assume a value of 0.5 for the ratio of the minor and major axis of the jet. We also assume that the velocity of the jet ($V_\mathrm{jet}$) is 500~km~s$^{-1}$ and that it lies in the plane of the sky (i.e., $\sin~i=1$). For a random orientation of the jet on the celestial plane, the value of $\sin~i$ is on average $\pi/4$ \citep[e.g.,][]{Beltran2001}. Usually, a value of $T=10^4$~K is adopted for ionized gas. For the turnover frequency $\nu_m$, we assume a value of 40~GHz \citep[see discussion in][]{Anglada2018}. In Fig.~\ref{f:MionMout}, we show the relationship between the mass loss rates of the ionized and the molecular outflow for the thermal radio jet candidates listed in Table~\ref{t:candidates} and associated with the molecular outflows. Major uncertainties in the determination of $\dot{M}_\mathrm{ion}$ may arise from parameters such as the jet velocity, the turnover frequency or the aspect ratio of the jet, as they cannot be determined from our observational data. However, their effects are almost negligible and variations within reasonable ranges result in variations of the ionized mass loss rate of less than a factor of 10. Our derived $\dot{M}_\mathrm{ion}$ are mainly in the range of $10^{-7}$ to $10^{-5}$~$M_\odot$~yr$^{-1}$, consistent with the values reported for low-mass radio jets ($\approx10^{-10}$~$M_\odot$~yr$^{-1}$) and high-mass radio jets ($\approx10^{-5}$~$M_\odot$~yr$^{-1}$, see \citealt{Rodriguez1994, Beltran2001, Guzman2010, Guzman2012}). The dashed lines in Fig.~\ref{f:MionMout} indicate different degrees of ionization for the mass loss rate. Most of our radio jet candidates, with the exception of source \#143, which is probably associated with an already developed \hii\ region (see Fig.~\ref{f:LradPout} and discussion above), have ionization levels of $\dot{M}_\mathrm{ion}$ = 10$^{-3}$~$\times$~$\dot{M}_\mathrm{out}$. These values are about one to two orders of magnitude smaller than those reported in previous studies \citep[see e.g.,][]{Rodriguez1990, Hartigan1994, Bacciotti1995, Anglada2018}. This difference may be due to uncertainties in the assumed parameters of Eq.~\ref{eq:ionmasslossrate}, as well as to the fact that our molecular outflow emission is studied with a single-dish (sensitive to all scale structures), while the radio jet observations were carried out with a large interferometric configuration likely resolving out part of the radio jet emission. 

%
\subsubsection{Best radio jet candidates\label{s:bestradiojets}}

In previous sections, we have analyzed the properties of the 146 detected radio continuum sources and built a sample of possible radio jet candidates based on their association with outflow activity: molecular outflows, EGOs, and masers (see Table~\ref{t:candidates}). In Sect.~\ref{s:alphasize} and \ref{s:jetoutflow}, we have investigated in more detail the possible nature of the radio continuum emission and its relation to the outflow activity tracers, in particular, the outflow momentum rate. The results presented in Figs.~\ref{fig:Lrad_Lbol} and \ref{f:LradPout} allow us to identify sources with properties that differ from those expected from radio jets, therefore suggesting that these sources may actually not be radio jets. From this analysis and the individual description of selected sources (see Appendix~\ref{s:individualsources}), we discuss in this section which objects are most likely to be radio jets.

Out of the 28 sources listed in Table~\ref{t:candidates}, five of them have radio luminosities similar to those expected for \hii\ regions: \#48a, \#48b, \#48c, \#143 and \#144 (see Fig.~\ref{fig:Lrad_Lbol}). In addition, all of these sources, with the exception of \#143, have negative spectral indices. These negative values could be due to the sources being slightly extended (as expected for \hii\ regions) and partially filtered out in the K band images. Moreover, these five sources also exhibit large radio luminosities compared to their associated outflow momentum rates (see Fig.~\ref{f:LradPout}), which supports the interpretation that there may be an excess of radio continuum emission not necessarily associated with a radio jet, but with an \hii\ region. In the absence of further evidence, we are not in a position to interpret further and we can not consider these sources to be among the best radio jet candidates. Further observations, sensitive to extended emission, can provide the necessary information to better characterize these sources in terms of their spatial extend and the nature of the emission. It is also worth noting that sources with negative spectral indices could correspond to background sources with synchrotron radiation since we expect about 11 objects in our sample to have this possible origin (see Sect.~\ref{s:continuum}). Source \#22 also shows an excess of radio-continuum emission compared to its outflow momentum rate, which suggests that this is also a dubious radio jet candidate. From the individual source descriptions presented in Appendix~\ref{s:individualsources}, sources \#42 and \#137 seem to be radio continuum sources with most of their emission dominated by cometary/ultracompact \hii\ regions, which makes it difficult to identify a radio jet in our data.

Out of the remaining sources listed in Table~\ref{t:candidates}, we can identify seven of them having a high probability to be radio jets. These are sources \#2, \#14, \#22, \#64, \#74, \#83 and \#110, which are associated with additional outflow/shock activity such as masers and EGOs. One should note that source \#74 is adjacent to two H$_2$O maser features, which are only 2\arcsec\ away and coincident with extended 4.5~$\mu$m emission (see Fig.~\ref{f:source7374}). The remaining sources do still classify as radio jet candidates, since we do not have clear evidence against that. Some of them are located at the center of molecular outflows (e.g., \#95) but are not associated with additional outflow/shock signposts. This could be related to the variability of H$_2$O masers (see Sect.~\ref{s:alphasize}). Others are located within molecular outflow lobes (e.g., \#12, \#63, \#64, \#65, \#73, and \#144), and for which higher angular resolution observations of outflow tracers are necessary to confirm if they are powering some of the molecular outflows. Other sources, despite not being associated with molecular outflows, show other shock activity signposts such as the presence of EGOs (e.g., \#119, \#136, \#139). Further observational constrains are therefore needed to fully confirm or discard these objects as radio jets. The results acquired so far, allow us to classify them as radio jet candidates.

%
\section{Implications for high-mass star formation\label{s:discussion}}

Recently, \citet{Rosero2019} studied the properties of 70 radio continuum sources associated with the earliest stages of high-mass star formation. They find that $\approx$30--50\% of their sample are ionized jets. This fraction is in agreement with our findings. Out of the 146 radio continuum sources detected in our study, we identify 28 possible radio jets (i.e., about 19\% of our sample). However, if we focus on the sources for which we have more accurate information (i.e., sources classified as iC/iK in Table~\ref{t:catalogue}, see also Sect.~\ref{s:continuum}), we have 24 out of 40 sources being potential radio jets. Therefore, the percentage of radio continuum sources being radio jets increases up 60\%. This suggests that about half of the radio continuum sources found in star-forming regions at early evolutionary stages may indeed be radio jets powered by young stars. The remaining $\approx$50\% of objects could still be radio jets for which we have not yet identified shock activity signposts, or they could represent extremely compact \hii\ regions in their early stages of development. These objects could be powered by early B-type stars and not necessarily by the most massive stars, and could be an intermediate stage between radio jets and already developed \hii\ regions \citep[see e.g.,][]{Beltran2016, RiveraSoto2020}.

\citet{LopezSepulcre2010} classified the regions studied in our work as infrared dark (IRDC, infrared dark cloud) and infrared bright (HMSFR, high-mass star-forming region) based on their detectable infrared emission. Our sample, therefore, consists of two sub-classes: IRDC (8~regions) and HMSFR (10~regions; see Table~\ref{t:sample}). We assume that these two types belong to different evolutionary phases of massive star formation, with the IRDC regions being less evolved than the HMSRF regions. Considering the 40 sources located within the primary beam of our images (i.e., sources classified as iC/iK), we find $\approx$2.8 radio continuum sources per HMSFR region, and $\approx$1.8 radio continuum sources per IRDC region. This suggests that it is more probable to detect compact radio continuum emission in more evolved regions. Regarding the presence of radio jets in these two evolutionary stages, we find 21 out of the 28 radio jet candidates listed in Table~\ref{t:candidates} in HMSFR regions (corresponding to 75\%), while we only find 7 (corresponding to 25\%) in the less evolved IRDC regions. If we consider only the best radio jet candidates (see Sect.~\ref{s:bestradiojets}), we find five radio jets (\#2, \#14, \#22, \#64 and \#83; corresponding to 71\%) in HMSRF regions and two radio jets (\#74 and \#110; corresponding to 29\%) in IRDC regions. This shows a preference of radio jets to be found in more evolved clouds. Complementary to this, we can determine the fraction of IRDC and HMSFR regions that harbor radio jets. Out of the 8 IRDC regions studied in this work, only 2 (corresponding to 25\%) harbor one of the best radio jet candidates. This increases up to 50\% (5 out of 10) for the HMSFR regions. Therefore, the frequency of radio jets in IRDC regions is lower than in HMSFR regions. One possible explanation is that the jets may become larger and brighter with time. Our limited data do not show that IRDC jets are smaller or fainter than HMSFR jets, but future work on larger source samples may provide further insight.

%
\section{Summary\label{s:summary}}

In this work, we have used of the VLA in two different bands (C and K band, corresponding to 6 and 1.3~cm wavelengths) to search for radio jets powered by high-mass YSOs. We have studied a sample of 18 high-mass star-forming regions with signposts of SiO and HCO$^+$ outflow activity. In the following we summarize our main results.

\begin{itemize}

    \item We have identified 146 radio continuum sources in the 18 high-mass star forming regions, with 40 of the radio continuum sources located within the primary beams of our images (i.e., labeled as iC/iK and with reliable flux measurements). Out of these sources, 131 (27 iC/iKs) are only detected in the C band, 4 (3 iC/iKs) are only detected in the K band, and 11 (10 iC/iKs) are detected in both bands. This different detection level is likely due to different factors: (i) four regions were not observed in the K~band, (ii) the C~band images have a larger field of view, and (iii) the K~band images are affected by a larger interferometric spatial filtering. In addition to the continuum emission, we have detected 23 maser features in the 6.7~GHz CH$_3$OH and 22~GHz H$_2$O lines.
    
    \item Out of the 146 continuum sources, only 40 sources are located within the field of view of both images allowing for an accurate characterization of their radio properties. For these sources we have derived the spectral index, which we find to be consistent with thermal emission (i.e., in the range $-$0.1 to $+$2.0) for most of the objects (73\%).
    
    \item We have investigated the nature of the radio continuum emission by comparing the radio luminosity to the bolometric luminosity. We find that most sources with positive spectral indices (i.e., thermal emission) follow the trend expected for radio jets, while sources with large negative spectral indices seem to follow the relation expected for \hii\ regions. These large negative spectral indices are likely due to the emission in the K~band images being partially filtered out.
    
    \item Based on the association of the radio continuum sources with shock activity signposts (i.e., association with molecular outflows, EGOs or masers), we have compiled a list of 28 radio jet candidates. This corresponds to $\approx$60\% of the radio continuum sources located within the field of view of both VLA images. Out of these sample of radio jet candidates, we have identified 7 objects (\#2, \#14, \#22, \#64, \#74, \#83, and \#110) as the most probable radio jets. The remaining 21 require additional observations, either at different radio frequency bands or of molecular outflow tracers at higher resolution, to confirm or discard them as radio jets.
    
    \item We find about 7--36\% of the possible radio jet candidates to show non-thermal radio continuum emission. This is consistent with previous studies reporting $\approx$20\% of non-thermal radio jets. We do not find a clear association of the non-thermal emission with the presence of outflows, EGOs or masers. However, and despite the low statistics, we find that radio jet candidates associated with CH$_3$OH masers have thermal emission, while those radio jet candidates associated with only H$_2$O masers tend to have non-thermal emission. This is in agreement with the 6.7~GHz CH$_3$OH masers tracing the actual location of newly-born YSOs powering thermal winds and jets, while the H$_2$O masers may be originated in strong shocks where non-thermal emission becomes relevant.
    
    \item As previously found in other works, we find a correlation between the radio luminosity of our radio jet candidates and their associated outflow momentum rates. We derive an ionized mass loss rate in the range $10^{-7}$ to $10^{-5}$~$M_\odot$~yr$^{-1}$, which results in ionization levels of $\dot{M}_\mathrm{ion}=10^{-3}~\dot{M}_\mathrm{out}$ (i.e., $\approx$0.1\% of the outflow mass being ionized).
    
    \item The 18 high-mass star-forming regions studied in this work are classified in two different evolutionary stages: 8 less evolved IRDC and 10 more evolved HMSFR. We find more radio continuum sources ($\approx$2.8 sources per region) in the more evolved HMSFR compared to the IRDC ($\approx$1.8). Regarding radio jets, we find about 71\% of the radio jet candidates to be located in HMSFR regions, and only 29\% in IRDC regions. Complemenary to this, 25\% of the IRDC regions harbor one of the most probable radio jet candidates, while this percentage increases up to 50\% for the HMSFR regions. This suggests that the frequency of radio jets in the less evolved IRDC regions is lower compared to the more evolved HMSFR regions.

\end{itemize}

%
\begin{acknowledgements}
The authors thank the referee for his/her review and greatly appreciate the comments and suggestions that have contributed significantly to improve the quality of the publication. \"UK also thanks Jonathan Tan for useful discussions. This work has been partially supported by the Scientific and Technological Research Council of Turkey (T\"UBİTAK), project number: 116F003. Part of this work was supported by the Research Fund of Istanbul University, project number: 44659. ASM research is carried out within the Collaborative Research Centre 956, sub-project A6, funded by the Deutsche Forschungsgemeinschaft (DFG; project ID 184018867). \"UK would like to thank William Pearson for checking the language of the paper and Kyle Oman for helping with Python issues.
\end{acknowledgements}

%
\bibliographystyle{aa}
\bibliography{radiojets.bib}

%
\begin{appendix}

%
\section{Comments on individual sources}\label{s:individualsources}

In the following, we comment on different aspects of selected sources for which additional literature information is available.

%
\subsection*{IRAS\,05358$+$3543 (\#2 and \#4)\label{s:iras05358+3543}}

\begin{figure}[t]
\centering
\begin{tabular}{c}
    \includegraphics[width=0.9\columnwidth]{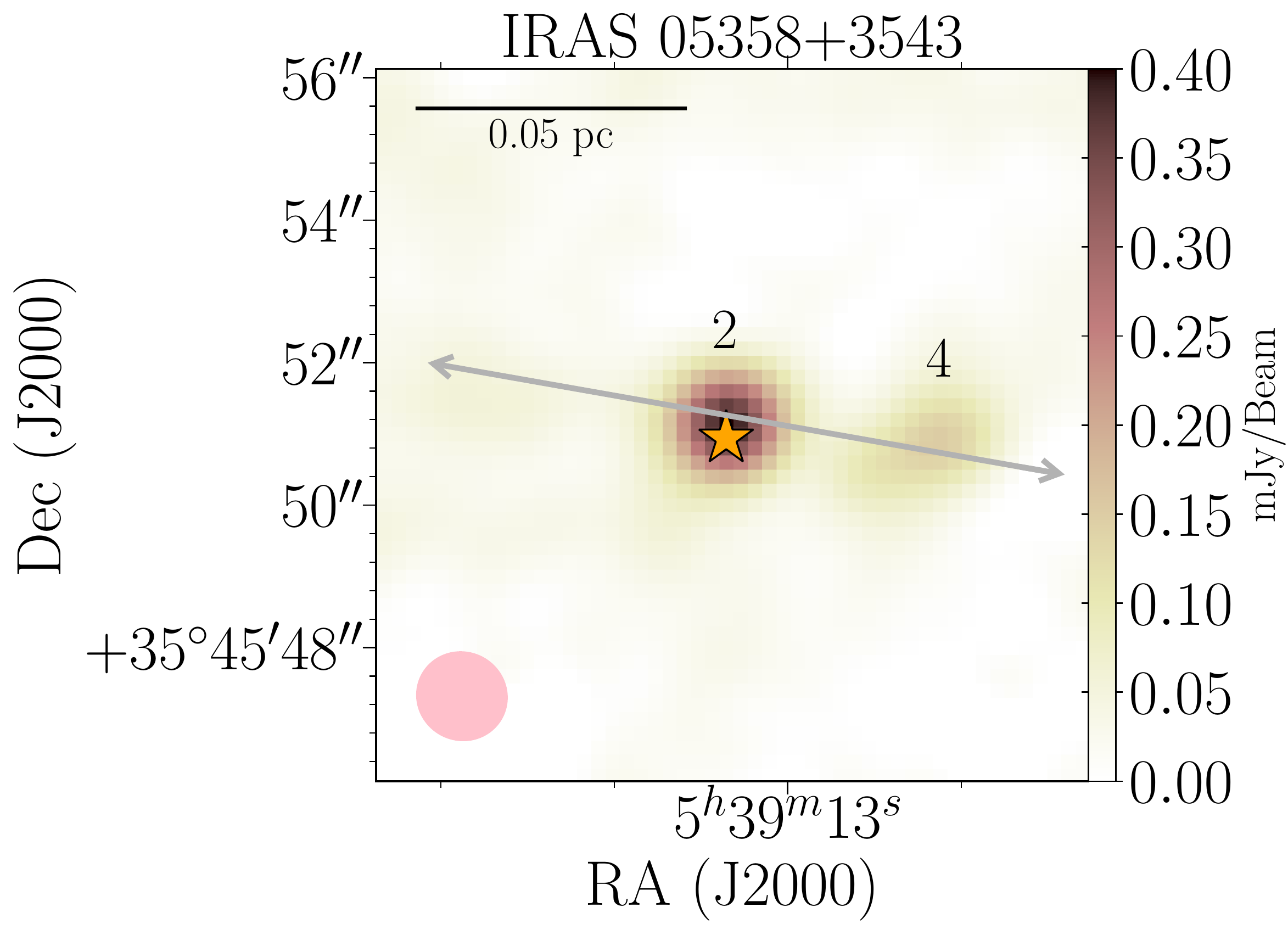} \\
\end{tabular}
\caption{VLA C~band (6~cm) continuum emission map of the radio jet candidates \#2 and \#4 located in the region IRAS\,05358$+$3543. The pink ellipse is the beam size of the C~band. The orange star marks the location of the CH$_3$OH maser (see Table~\ref{t:masers}). The grey double-headed arrow indicates the direction of the outflows.}
\label{f:source2}
\end{figure}

In region IRAS\,05358$+$3543, we have identified two radio continuum sources that can be potential radio jets (see Fig.~\ref{f:source2}). Sources \#2 and \#4 were observed only in the C~band, and therefore we cannot determine a spectral index for these sources. Despite this limitation, both sources are located at the center of the molecular outflow reported by \citet{LopezSepulcre2010}. Furthermore, source \#2 is associated with a 6.7~GHz CH$_3$OH maser emission, suggesting that this marks the position of a massive YSO. Of the two radio continuum sources, source \#2 is likely the main object powering the molecular outflow for which its centimeter emission traces a radio jet. Further observations at different frequency bands are necessary to better constrain its properties.

%
\subsection*{G189.78$+$0.34 (\#12, \#13, \#14, \#15 and \#16)\label{s:g189.78+0.34}}

\begin{figure}[t]
\centering
\begin{tabular}{c}
    \includegraphics[width=0.9\columnwidth]{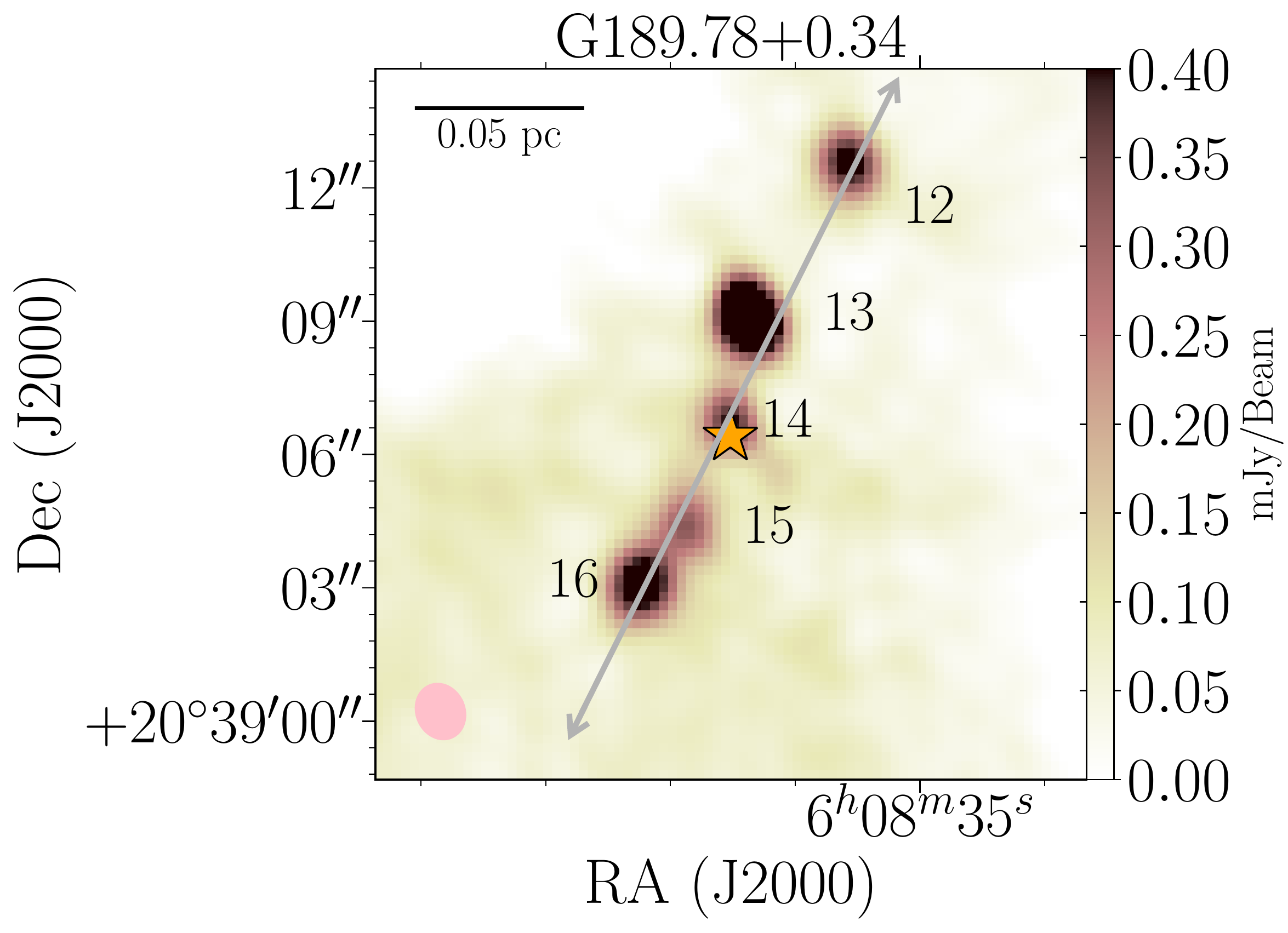} \\
\end{tabular}
\caption{VLA C~band (6~cm) continuum emission map of the radio jet candidates \#12, \#13, \#14, \#15 and \#16 located in the region G189.78$+$0.34. The pink ellipse is the beam size of the C~band. The orange star marks the location of the CH$_3$OH maser (see Table~\ref{t:masers}). The grey double-headed arrow indicates the direction of the outflows.}
\label{f:source14}
\end{figure}

In region G189.78+0.34, we have identified five radio continuum sources (sources \#12 to \#16) associated with the molecular outflow reported by \citet{LopezSepulcre2010}. All of them are located at the center of the outflow, with source \#12 slightly offset from the rest (see Fig.~\ref{f:source14}). All these sources were observed only in the C~band and no spectral index can be derived. Out of the five sources, source \#14 is associated with CH$_3$OH maser emission \citep[see also][]{Caswell2010} suggesting that this marks the location of a massive YSO. The radio continuum sources are found in an elongated chain extending from the south-east to the north-west. This direction is consistent with the orientation of the molecular outflow \citep{LopezSepulcre2010}. Overall, we consider that source \#14 is the powering source and most likely the main component of the radio jet. The other sources could correspond to different radio continuum knots located along the jet, as seen in other sources (e.g., HH\,80-81: \citealt{CarrascoGonzalez2010}, and G35.20$-$0.74\,N: \citealt{Beltran2016}), where the radio continuum knots usually show non-thermal spectral indices. Observations at different frequency bands are necessary to gather information on the spectral index and thermal/non-thermal nature of the different radio continuum sources in the region. 

%
\subsection*{G192.58$-$0.04 (\#22, \#23 and \#25)\label{s:g192.58-0.04}}

\begin{figure}[t]
\centering
\begin{tabular}{c}
    \includegraphics[width=0.9\columnwidth]{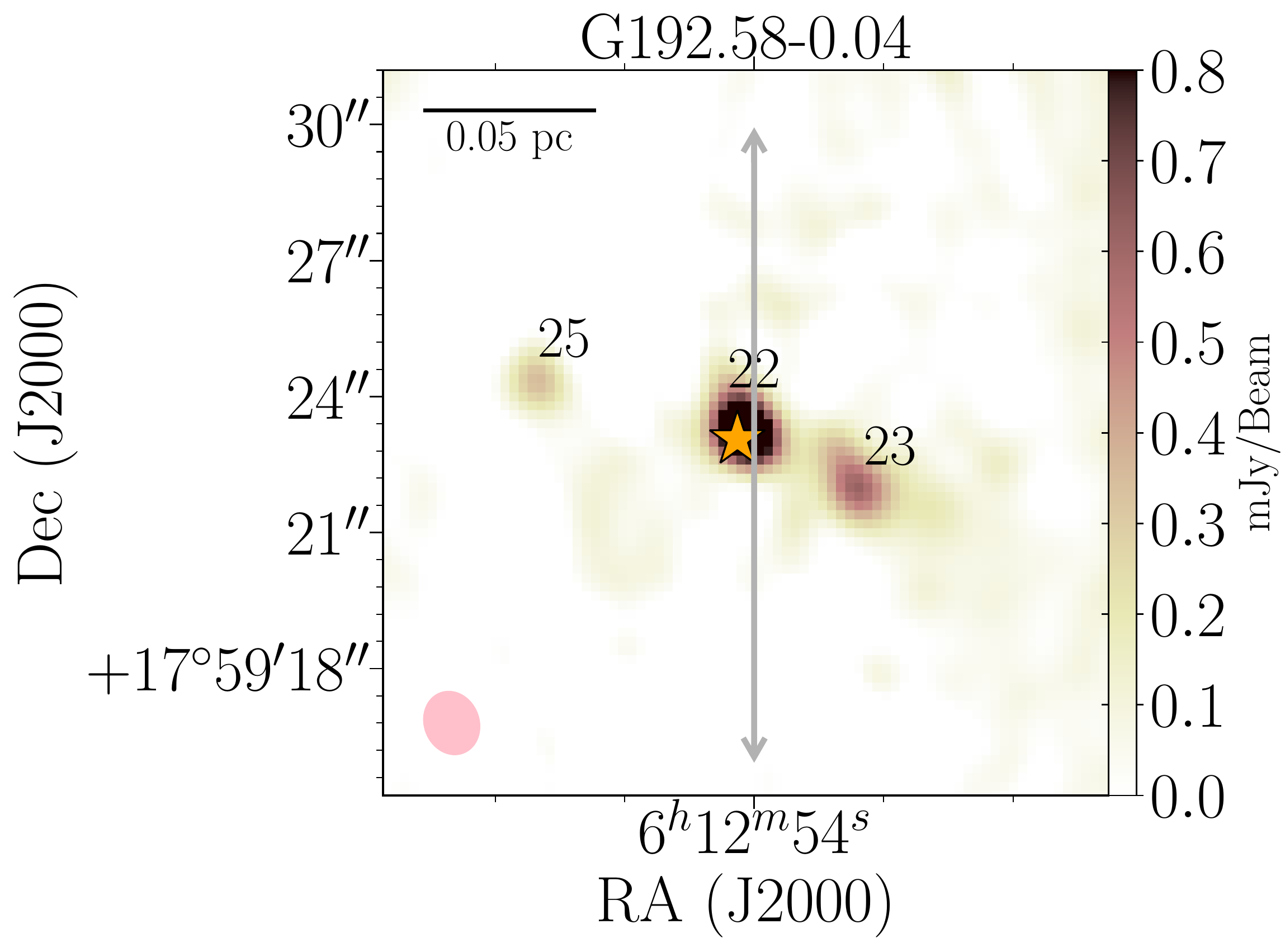} \\
\end{tabular}
\caption{VLA C~band (6~cm) continuum emission map of the radio jet candidates \#22, \#23 and \#25 located in the region G192.58$-$0.04. The pink ellipse is the beam size of the C~band. The orange star marks the location of the CH$_3$OH maser (see Table~\ref{t:masers}). The grey double-headed arrow indicates the direction of the outflows.}
\label{f:source22}
\end{figure}

In region G192.58$-$0.04, we have identified sources \#22, \#23, and \#25 as potential radio jets (see Fig.~\ref{f:source22}). These sources were observed only in the C~band and no spectral index can be derived. Out of the three sources, source \#22 is associated with CH$_3$OH maser emission suggesting that this source may mark the location of a massive YSO. The sources are located at the center of the molecular outflow reported by \citet{LopezSepulcre2010}. Out of the three sources, we consider that source \#22 is the most probable radio jet. The comparison between the radio continuum luminosity with the outflow momentum rate (see Fig.~\ref{f:LradPout}), also confirms this possibility, although there seems to be a slightly excess of radio continuum emission, suggesting that there can be an additional contribution to the radio continuum source (e.g., from an early-stage \hii\ region).

%
\subsection*{IRAS\,18223$-$1243 (\#42)\label{s:iras18223-1243}}

\begin{figure}[t]
\centering
\begin{tabular}{c}
    \includegraphics[width=0.9\columnwidth]{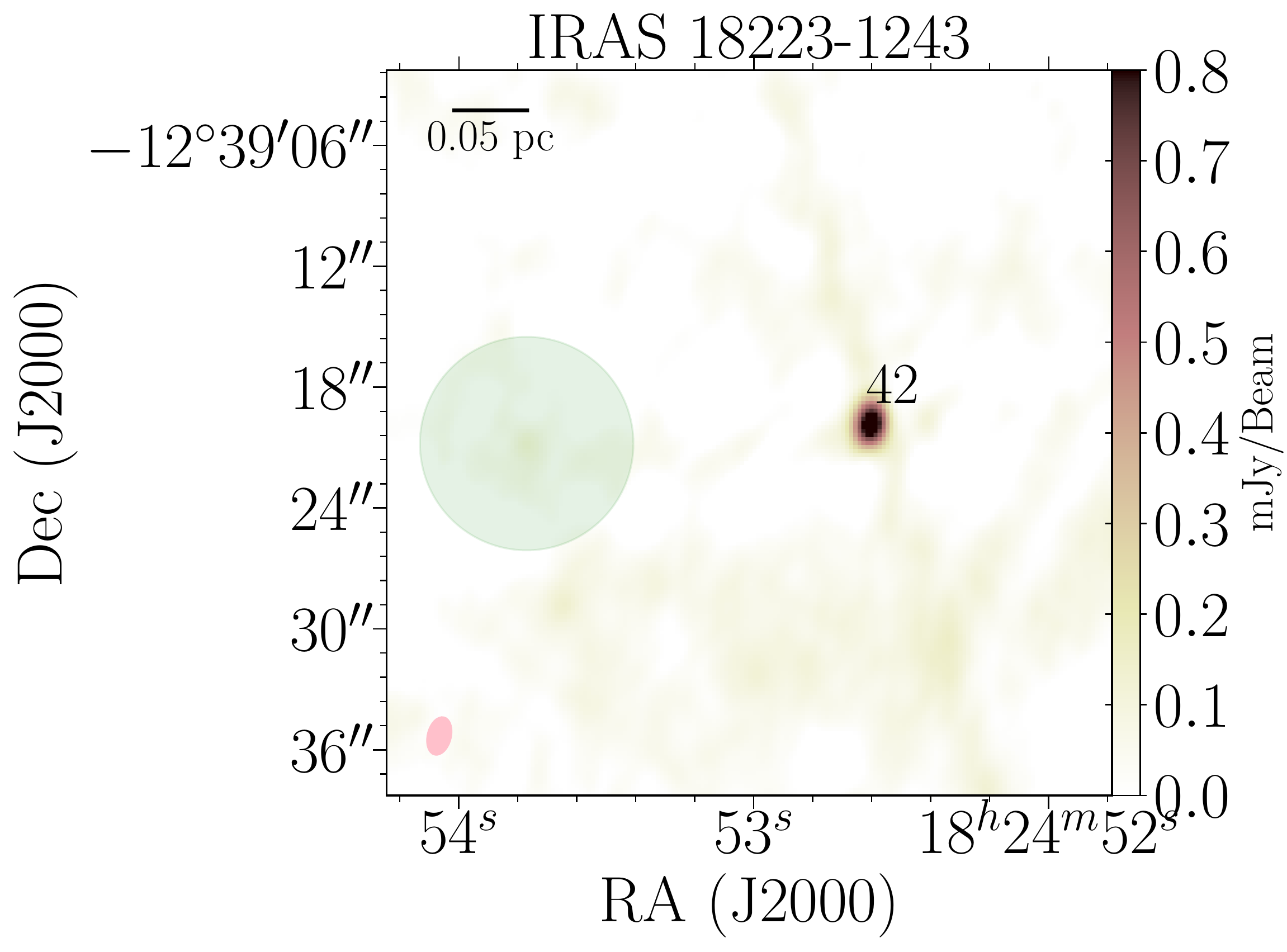} \\
\end{tabular}
\caption{VLA C~band (6~cm) continuum emission map of the radio jet candidate \#42 located in the region IRAS\,18223$-$1243. The pink ellipse is the beam size of the C~band. The green circle with a radius of $\sim$5$^{\arcsec}$ marks the EGO F G18.67$+$0.03-CM1 reported by \citet{Cyganowski2011}.}
\label{f:source42}
\end{figure}

In the region IRAS\,18223$-$1243, we identified the radio continuum source \#42 (see Fig.~\ref{f:source42}) as adjacent to the one reported in \citet{Cyganowski2011} EGO F~G18.67$+$0.03$-$CM1. This is the only signposts of shock activity, since no molecular outflow or maser emission are found for this object. In addition, \citet{Cyganowski2012} report the existence of a massive protocluster consisting of a hot molecular core and an ultracompact \hii\ region. Our source seems to be located at the same position of the ultracompact \hii\ region, which makes us to doubt if this can be a radio jet candidate.

%
\subsection*{IRAS\,18228$-$1312 (\#48)\label{s:iras18228-1312}}

\begin{figure}[t]
\centering
\begin{tabular}{c}
    \includegraphics[width=0.9\columnwidth]{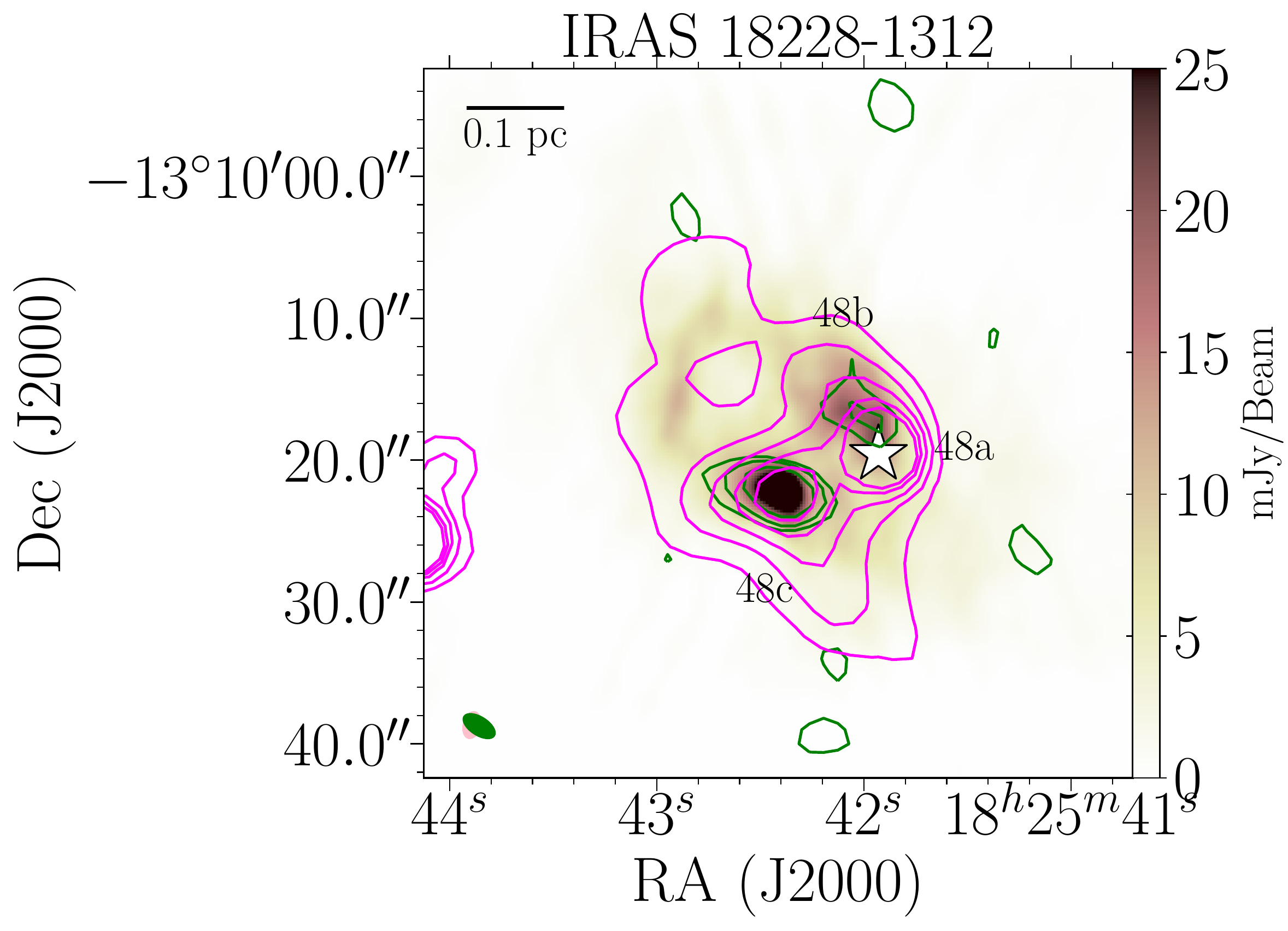} \\
\end{tabular}
\caption{VLA C~band (6~cm) continuum emission map of source \#48 located in the region IRAS\,18228$-$1312. Three bright peaks are visible and labelled as \#48a, \#48b and \#48c. The green contour levels of the K~band (1.3~cm) continuum emission are 3, 5 and 9 times 0.7~mJy~beam$^\mathrm{-1}$. The magenta contours show the \textit{Spitzer}/GLIMPSE 4.5~$\mu$m emission. The pink and green ellipses are the beam sizes of the C and K~bands, respectively. The white star marks the location of the H$_2$O maser (see Table~\ref{t:masers}).}
\label{f:source48}
\end{figure}

Radio continuum source \#48 is observed as a group of three compact sources (\#48a, \#48b and \#48c) surrounded by an extended and more diffuse structure. One of these compact sources (\#48a) is clearly associated with H$_2$O maser emission (see Fig.~\ref{f:source48}). The \textit{Spitzer}/IRAC 4.5~$\mu$m emission is extended and spatially coincident with the radio continuum extended emission. The spectral indices for these three compact sources are in the range $-0.6$ to $-1.5$, most likely due to additional filtering of the emission in the K~band image. Previous studies have classified this extended source as a region containing hypercompact (HC) and ultracompact (UC) \hii\ regions \citep[e.g.,][]{Chini1987, Lockman1989, Kurtz1994, Kuchar1997, Leto2009}, which is consistent with our derived radio continuum luminosity (see Figs.~\ref{fig:Lrad_Lbol} and \ref{f:LradPout}). It is worth noting that the the three sources are spatially located at the center of a molecular outflow \citep[e.g.,][]{LopezSepulcre2010}. This may suggest that one of the compact sources may be powering the molecular outflow. In this case, this object would be in a evolutionary stage where the radio jet still exists but a young \hii\ region has already developed, similar to the high-mass young stellar object G35.20$-$0.74\,N \citep[e.g,][]{SanchezMonge2013b, SanchezMonge2014, Beltran2016}.

%
\subsection*{IRAS\,18236$-$1205 (\#63, \#64 and \#65)\label{s:iras18236-1205}}

\begin{figure}[t!]
\centering
    \includegraphics[width=0.95\columnwidth]{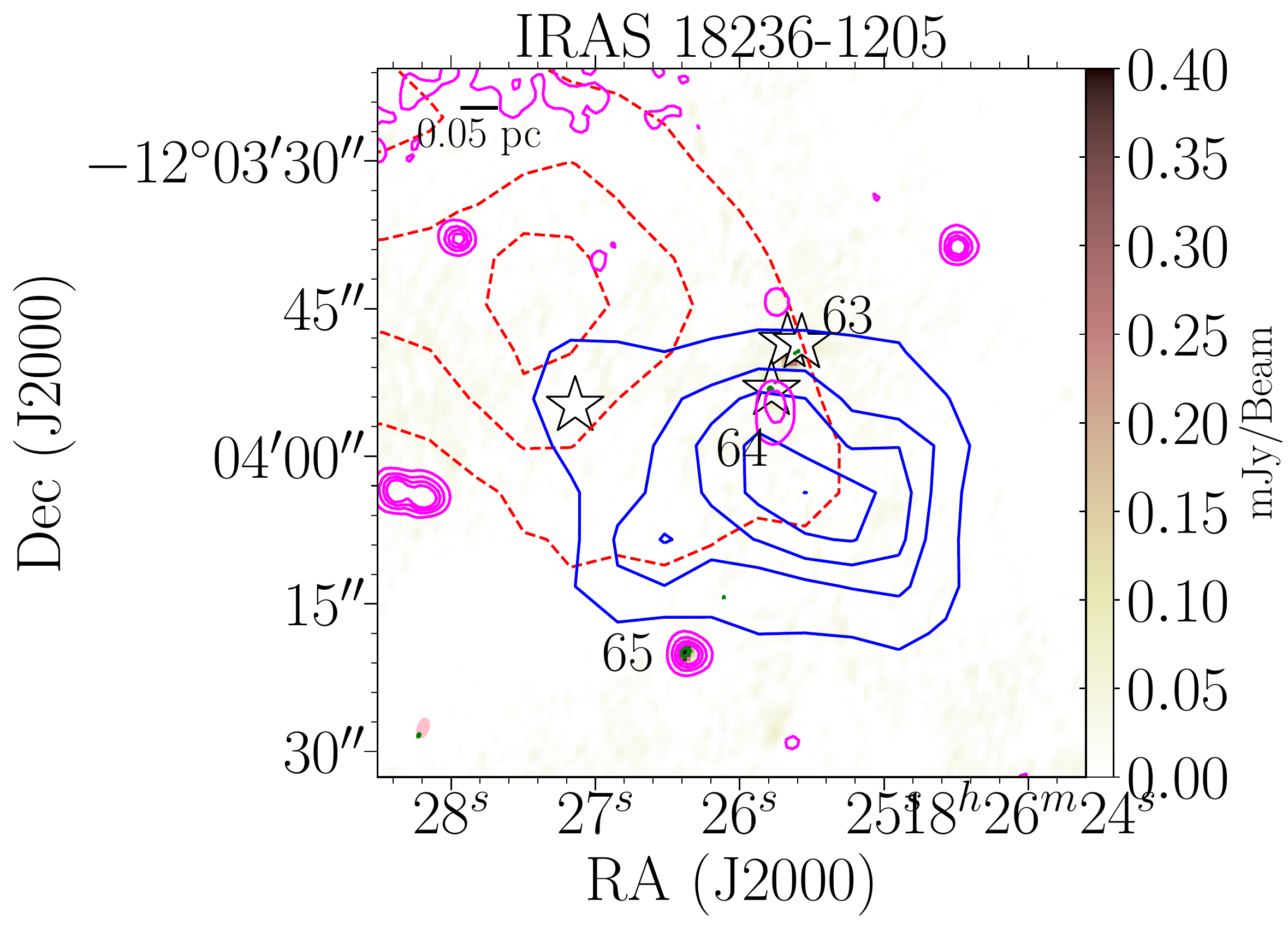} 
    \includegraphics[width=0.95\columnwidth]{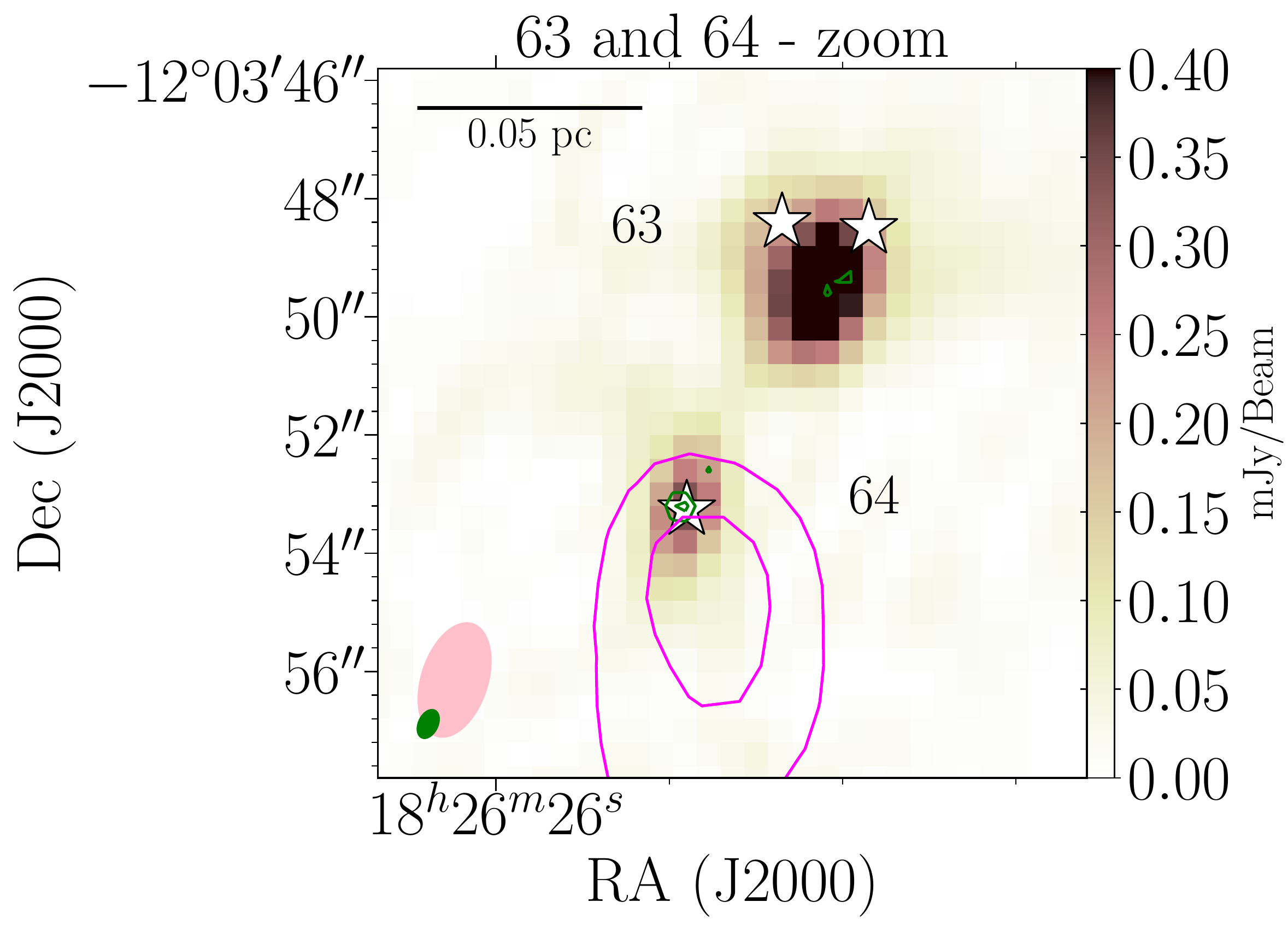} 
    \includegraphics[width=0.95\columnwidth]{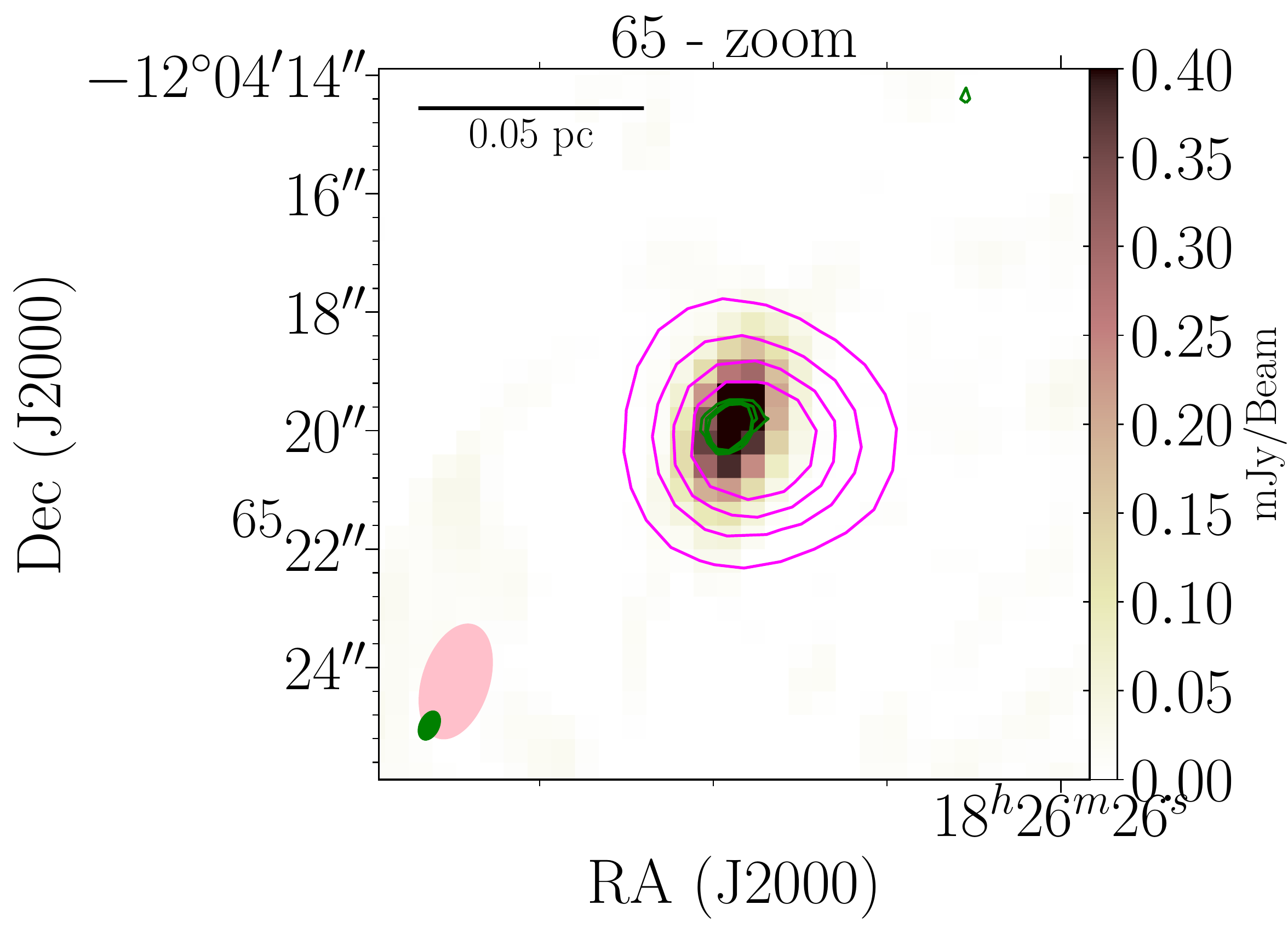} 
\caption{VLA C~band (6~cm) continuum emission map of the radio jet candidates \#63, \#64 and \#65 located in the region IRAS\,18236$-$1205. A close-up view of the three radio sources is shown in the bottom and right panels. The green contour levels of the K~band (1.3~cm) continuum emission are 3, 5 and 7 times 30~$\mu$Jy~beam$^\mathrm{-1}$. The magenta contours show the \textit{Spitzer}/GLIMPSE 4.5~$\mu$m emission. The blue- and red-shifted outflow lobed of SiO (2$-$1) are shown as blue-solid and red-dashed contours, respectively \citep[see][]{SanchezMonge2013d}. The pink and green ellipses are the beam sizes of the C and K~bands, respectively. The white stars mark the location of the H$_2$O masers (see Table~\ref{t:masers}).}
\label{f:source636465a}
\end{figure}

We have identified nine radio continuum sources in the IRAS\,18236$-$1205 region (also referred to in the literature as G19.36$-$0.03), three of which have been classified as radio jet candidates: Sources \#63, \#64 and \#65 with spectral indices of $-0.22\pm0.07$, $+0.18\pm0.04$ and $+1.08\pm0.19$. We have identified four H$_2$O maser features near these sources (see Fig.~\ref{f:source636465a}). Two maser features are associated with source \#63, one maser feature is associated with source \#64 (which also spatially coincides with a CH$_3$OH maser), and the last maser feature is located in the center of the red-shifted outflow lobe where no radio continuum emission is detected. The molecular outflow in this region has been mapped in the lines SiO\,(2--1) and HCO$^+$\,(1--0) by \citep{SanchezMonge2013d}.

Sources \#63 and \#64 are associated with EGOs \citep [see][]{Cyganowski2009}, indicating the presence of strong shock activity in these two sources. Their location near the center of the molecular outflow together with their association with H$_2$O maser emission and EGOs, suggests that these two sources could be candidates for radio jets. The source \#64 is spatially more coincident with the geometric center of the outflow, and its association with 6.7~GHz CH$_3$OH maser emission, suggests that a massive YSO exists at this position. This massive YSO could be the driving source of the molecular outflow seen on larger scales. The third candidate (source \#65) is located $\sim$28$\arcsec$ away from the source \#64 and the center of the outflow, and has been studied by \cite[G19.36-0.03-CM2]{Cyganowski2011}. This source is associated with an emission of 4.5~$\mu$m, although it is unclear whether it can be convincingly classified as an EGO \citep{Cyganowski2009}. The positive spectral index indicates the presence of thermal emission, which could come from a radio jet. However, there is no clear evidence of outflow or shock activity. The source \#65 is also located in the vicinity of a dense core (18236$-$1205\,P8) identified by \cite{Lu2014} in the VLA NH$_3$ maps, which supports the interpretation of this source as an embedded young stellar object. Overall, the source \#65 could be a YSO-powered radio jet in the vicinity of the more massive object (sources \#63 and \#64) in IRAS\,18236$-$1205.

%
\subsection*{G23.60$+$0.0M1 (\#73 and \#74)\label{s:g23.60+0.0m1}}

\begin{figure}[t]
\centering
\begin{tabular}{c}
    \includegraphics[width=0.95\columnwidth]{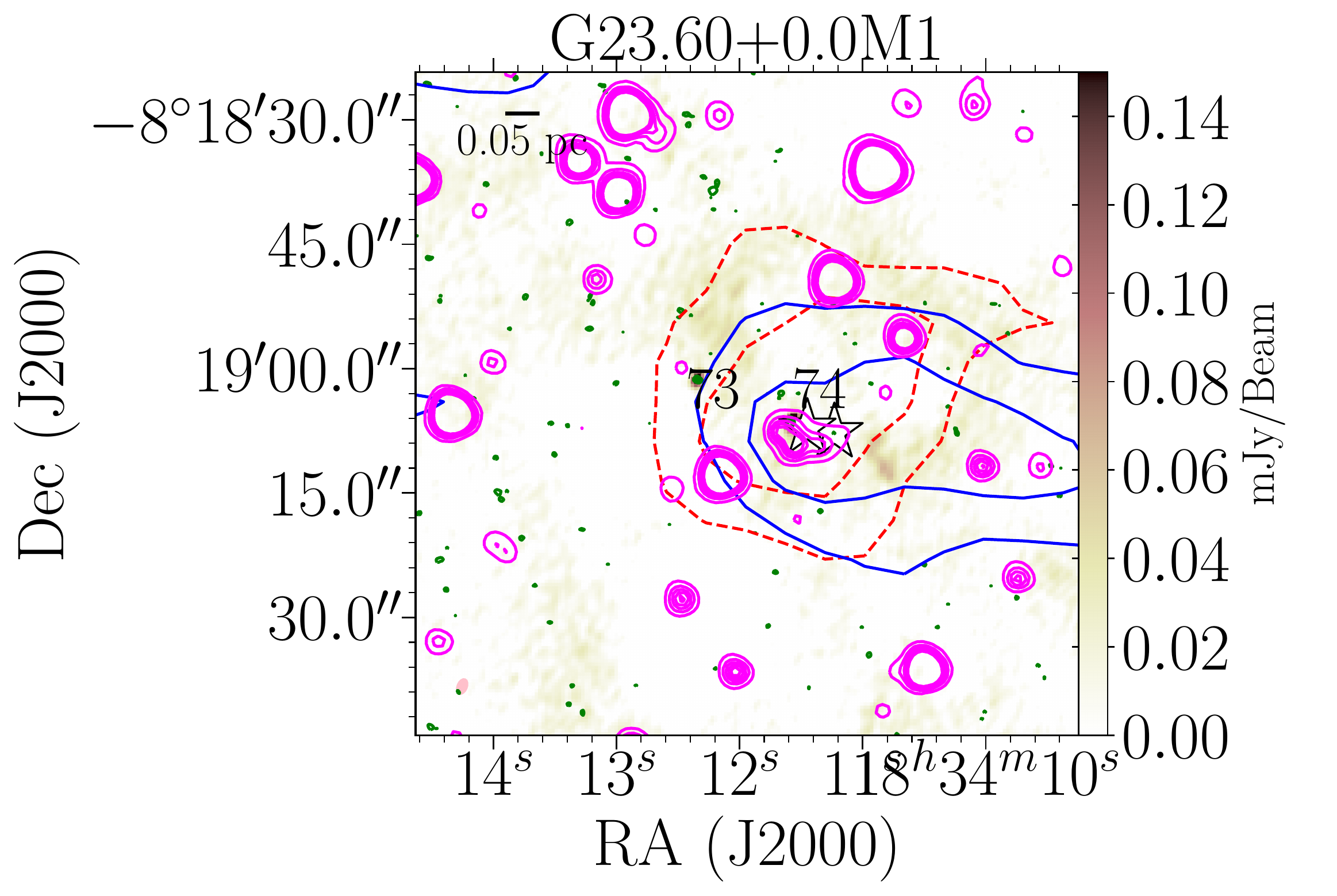} \\
    \includegraphics[width=0.95\columnwidth]{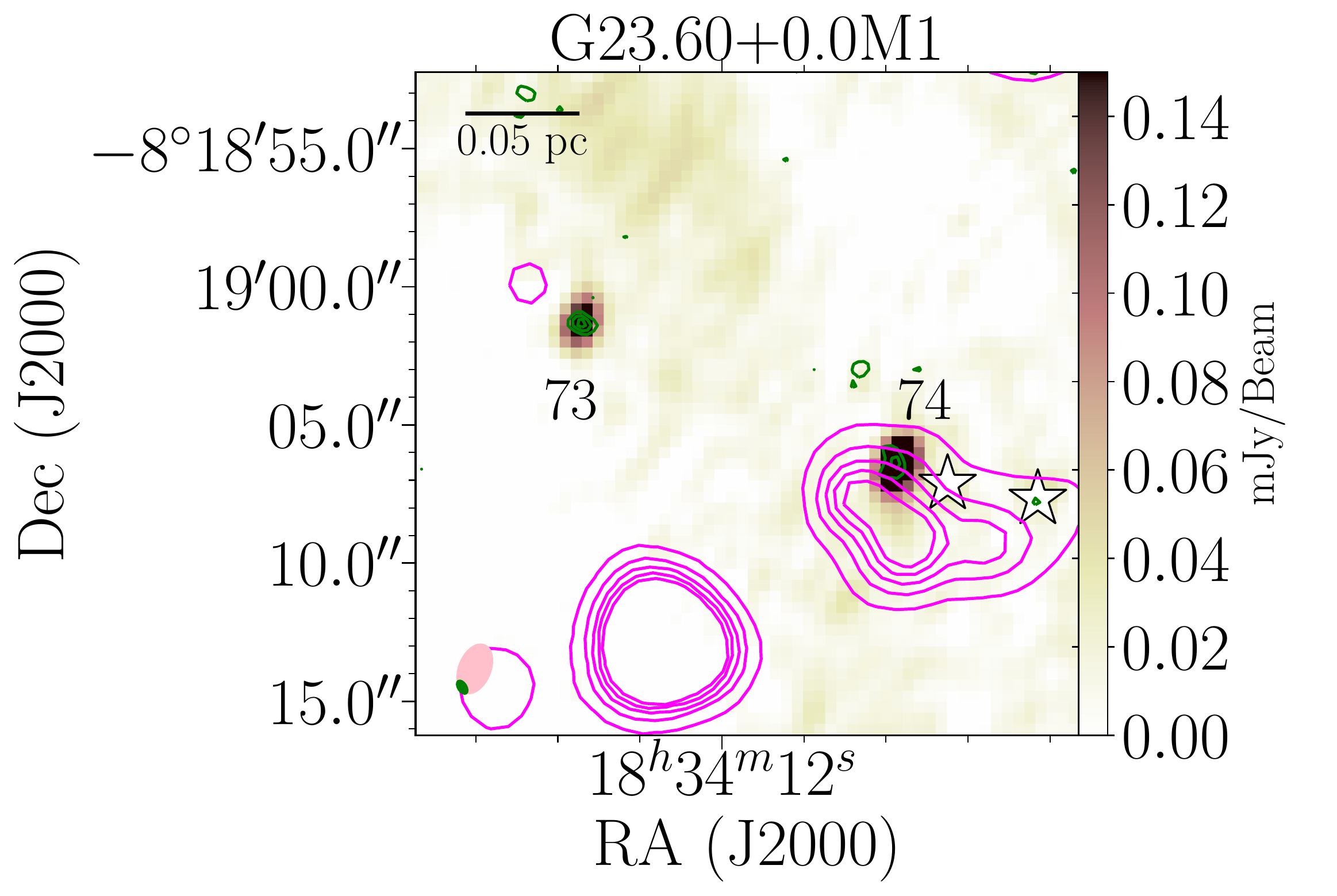} \\
\end{tabular}
\caption{VLA C~band (6~cm) continuum emission map of the radio jet candidates \#73 and \#74 located in the region G23.60$+$0.0M1. A close-up view of the two radio sources is shown in the bottom panel. The green contour levels of the K~band (1.3~cm) continuum emission are 3, 5, 7, 9 and 11 times 20~$\mu$Jy~beam$^\mathrm{-1}$. The magenta contours show the \textit{Spitzer}/GLIMPSE 4.5~$\mu$m emission. The blue- and red-shifted outflow lobed of SiO (2$-$1) are shown as blue-solid and red-dashed contours, respectively \citep[see][]{SanchezMonge2013d}. The pink and green ellipses are the beam sizes of the C and K~bands, respectively. The white stars mark the location of the H$_2$O masers (see Table~\ref{t:masers}).}
\label{f:source7374}
\end{figure}

The G23.60$+$0.01M1 star-forming region has been studied in the literature by various authors \citep[e.g.,][]{Rathborne2006, Battersby2010, Ginsburg2013}, who reported the presence of two massive dense clumps with masses of 100~$M_\odot$ and 120~$M_\odot$. The two candidate radio jets (\#73 and \#74) have positive spectral indices consistent with thermal emission of radio jets. In particular, the source \#74 is located at the center of the molecular outflow reported by \cite{SanchezMonge2013d} and is associated with a strong 4.5~$\mu$m emission (see Fig.~\ref{f:source7374}). The two H$_2$O maser detected in the region are slightly displaced from the radio continuum source but coincide with the extended 4.5~$\mu$m emission (see Fig.~\ref{f:source7374}-bottom). The association of molecular outflow emission, bright and extended 4.5~$\mu$m emission, and close H$_2$O maser features favor the interpretation of this source is a good radio jet candidate.

The second radio continuum source (\#73) is located relatively close to the center of the outflow. However, no maser or EGOs are found in connection with the source. Although we cannot reject this source as a radio jet, we prefer the source \#74 as the main object driving the outflow observed in the region.

%
\subsection*{IRAS\,18316$-$0602 (\#83 and \#95)\label{s:iras18316-0602}}

\begin{figure}[t]
\centering
\begin{tabular}{c}
\includegraphics[width=0.9\columnwidth]{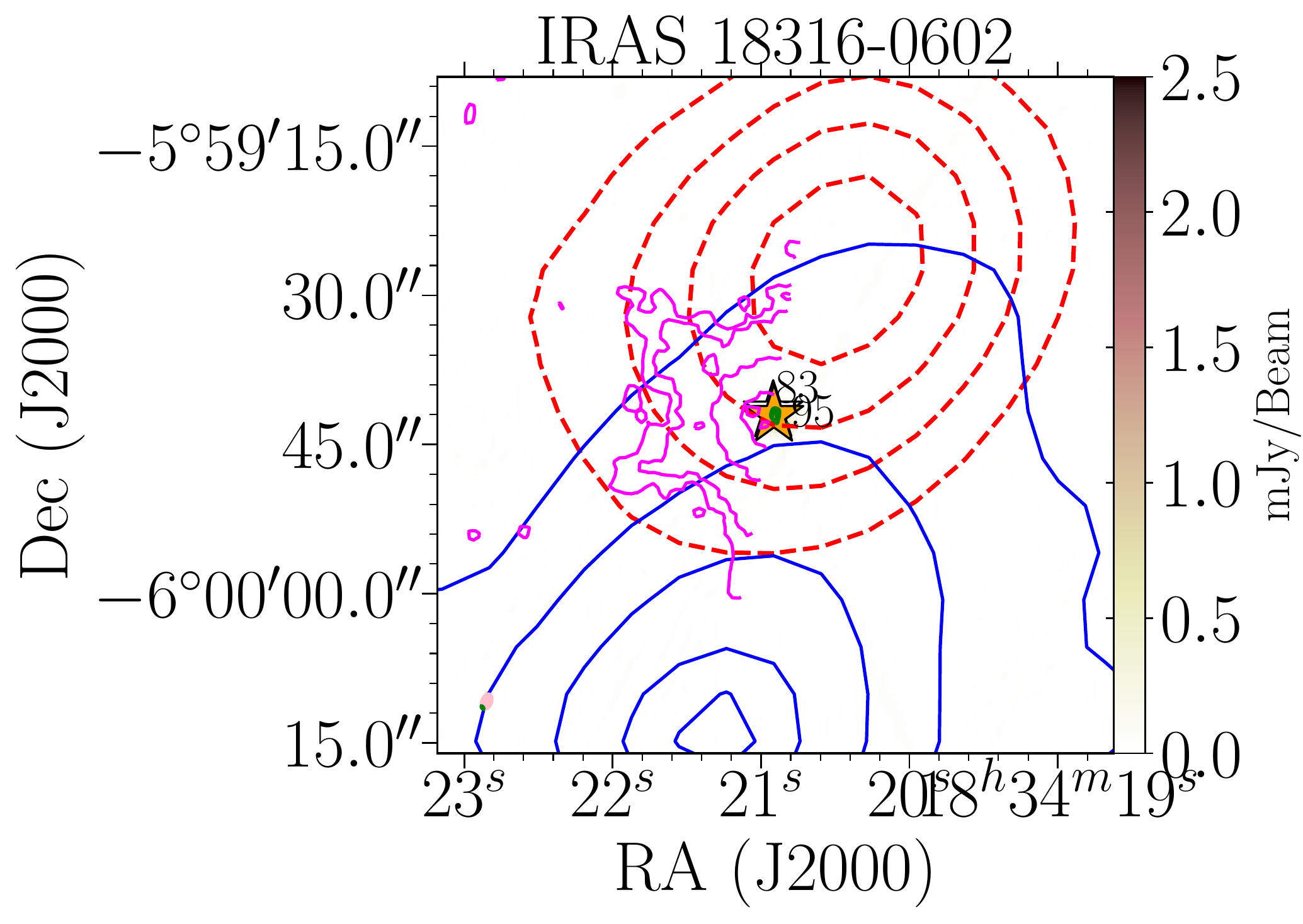} \\
\includegraphics[width=0.9\columnwidth]{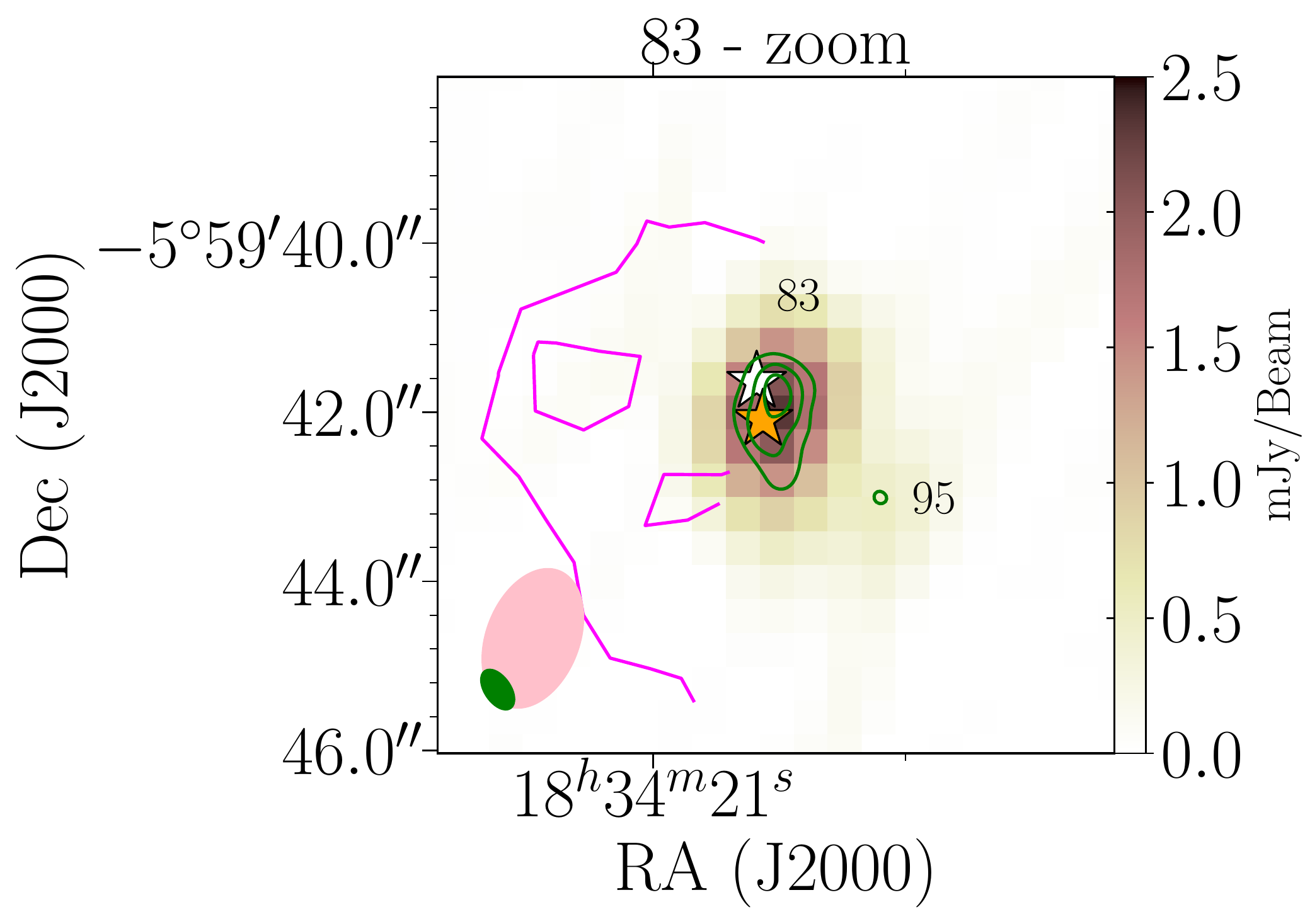} \\
\end{tabular}
\caption{VLA C~band (6~cm) continuum emission map of the radio jet candidates \#83 and \#95 located in the region IRAS\,18316$-$1602. A close-up view of the two radio sources is shown in the bottom panel. The green contour levels of the K~band (1.3~cm) continuum emission are 3, 5, 9 and 11 times 27~$\mu$Jy~beam$^\mathrm{-1}$. The magenta contours show the \textit{Spitzer}/GLIMPSE 4.5~$\mu$m emission (note that half of the region was not covered in the mapped area). The blue- and red-shifted outflow lobed of SiO (2$-$1) are shown as blue-solid and red-dashed contours, respectively \citep[see][]{SanchezMonge2013d}. The pink and green ellipses are the beam sizes of the C and K~bands, respectively. The white and orange stars mark the location of the H$_2$O and CH$_3$OH masers, respectively (see Table~\ref{t:masers}).}
\label{f:source8395}
\end{figure}

We have identified thirteen radio continuum sources in the IRAS\,18316$-$0602 region, two of which have been classified as radio jet candidates: Sources \#83 and \#95. The source \#83 has an almost flat spectral index ($0.08\pm0.08$) and shows a weak extension to the south, which is better resolved in the K~band image. Source \#95 is fainter, appears located about 3\arcsec to the south-east of source \#83, and is visible only in the K~band (some faint emission slightly above the noise level is visible in the C~band image, see Fig.~\ref{f:source8395}). As for the maser emission, we have identified both a H$_2$O and a CH$_3$OH maser feature associated with the brightest source \#83. This source has been studied in previous works \citep[e.g.,][]{Roueff2006, Cutri2012, Azatyan2016, Stecklum2017}, in some of them named RAFGL\,7009S. The source is detected in the near-infrared together with two other objects separated by about 10\arcsec, and surrounded by a diffuse and extended structure \citep[see e.g.,][]{Stecklum2017}.

\citet{LopezSepulcre2010} and \citet{SanchezMonge2013d} report on molecular outflow emission in the region. The blue and red contours in Fig.~\ref{f:source8395} show the SiO\,(2--1) blue-shifted and red-shifted outflow emission. Both radio continuum sources (\#83 and \#95) are located close to the center of the outflow. Although this region was not included in the \citet{Cyganowski2008, Cyganowski2009, Cyganowski2011} surveys, we have identified a bright 4.5~$\mu$m source in the \textit{Spitzer}/GLIMPSE data. However, the source is located at the edge of the area surveyed by \textit{Spitzer}, which prevents a detailed characterization of its infrared emission. The association of the source \#83 with a 6.7~GHz CH$_3$OH maser emission suggests that this source harbors a massive YSO together with its central location within the outflow and its association with a H$_2$O maser feature hints towards this source is the radio jet that drives the outflow. The almost flat spectral index may indicate that this is a fully ionized radio jet. However, further observations in different frequency bands may help to better narrow down the spectral index and the status of radio continuum emission.

%
\subsection*{G24.08$+$0.0\,M2 (\#96)\label{s:g24.08+0.0m2}}

\begin{figure}[t]
\centering
\begin{tabular}{c}
\includegraphics[width=0.9\columnwidth]{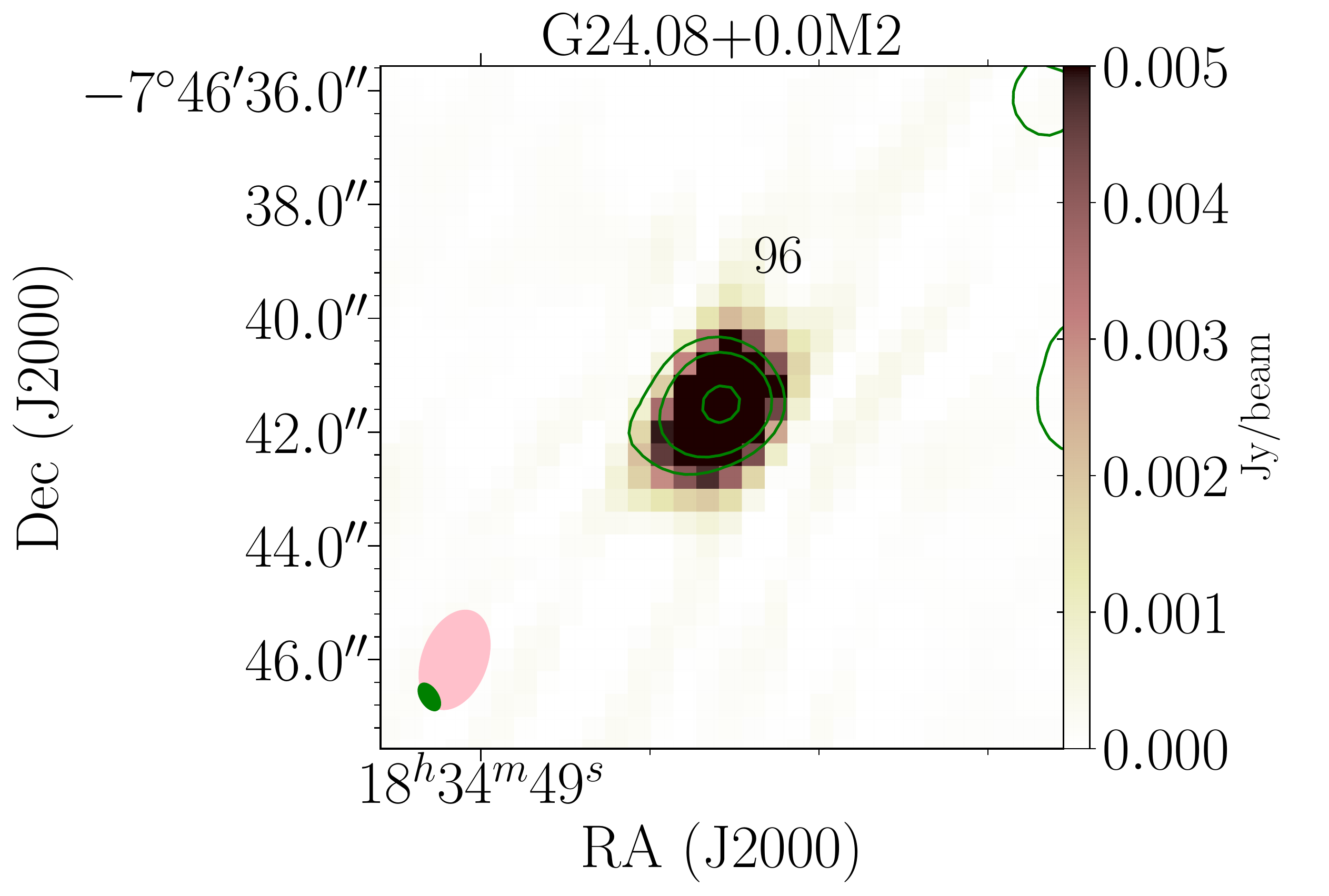} \\
\end{tabular}
\caption{VLA C~band (6~cm) continuum emission map of the radio source \#96 located in the region G24.08$+$0.0M2. The green contour levels of the K~band (1.3~cm) continuum emission are 3, 5 and 9 times 20~$\mu$Jy~beam$^\mathrm{-1}$. The pink and green ellipses are the beam sizes of the C and K~bands, respectively.}
\label{f:source96}
\end{figure}

We have identified fourteen radio continuum sources in the region G24.08$+$0.0\,M2, one of which, source \#96, is detected in both frequency bands and has a negative spectral index ($-0.84\pm+0.08$, see Fig.~\ref{f:source96}). The outflow activity in the region has been studied by \citet{LopezSepulcre2011} and \citet{SanchezMonge2013d} who found molecular outflow in different tracers. However, this outflow is not spatially related to any of the radio continuum sources identified in this work. Also, this source could be a background source.

%
\subsection*{G24.33$+$0.1\,M1 (\#110)\label{s:g24.33+0.1m1}}

\begin{figure}[t]
\centering
\begin{tabular}{c}
\includegraphics[width=0.9\columnwidth]{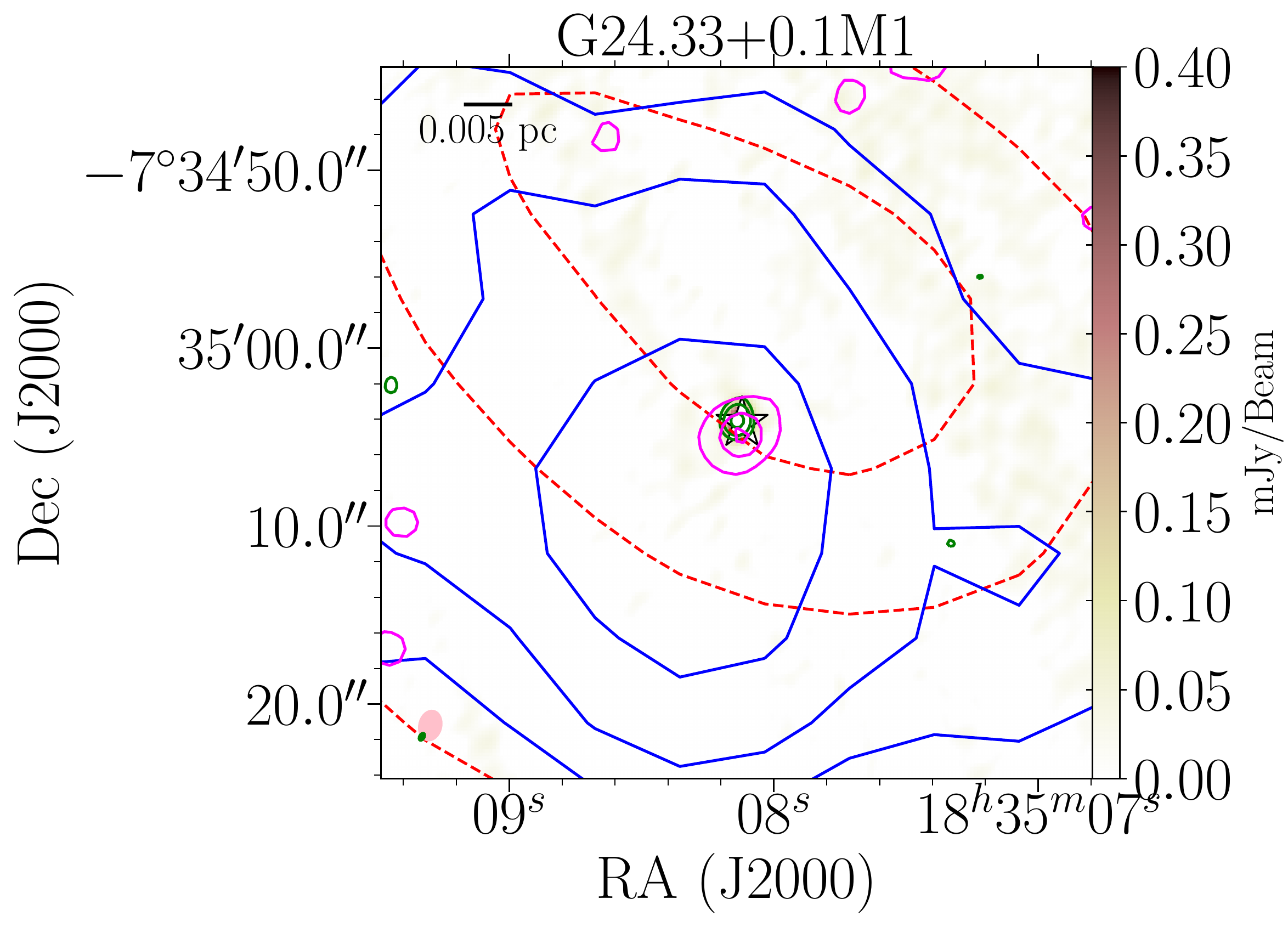} \\
\includegraphics[width=0.9\columnwidth]{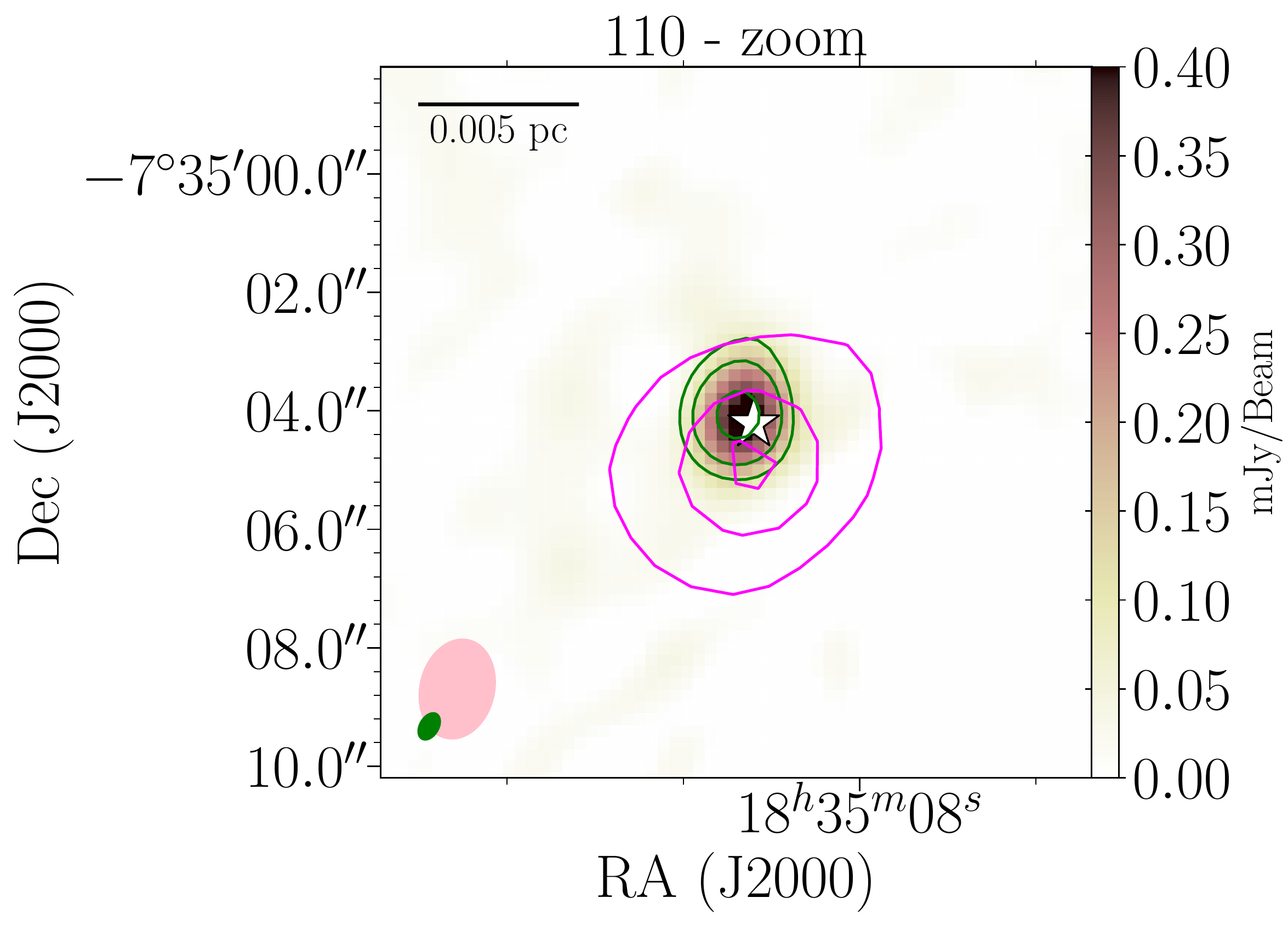} \\
\end{tabular}
\caption{VLA C~band (6~cm) continuum emission map of the radio jet candidate \#110 located in the region G24.33+0.1\,M1. A close-up view of the radio source is shown in the bottom panel. The green contour levels of the K~band (1.3~cm) continuum emission are 3, 5 and 9 times 7~$\mu$Jy~beam$^\mathrm{-1}$. The blue- and red-shifted outflow lobed of SiO (2$-$1) are shown as blue-solid and red-dashed contours, respectively \citep[see][]{SanchezMonge2013d}. The pink and green ellipses are the beam sizes of the C and K~bands, respectively. The white star marks the location of the H$_2$O (see Table~\ref{t:masers}).}
\label{f:source110}
\end{figure}

In the region G24.33$+$0.1\,M1 we identified a radio continuum source (\#110) located in the center of our field of view and detected in both frequency bands (see Fig.~\ref{f:source110}). This source has a positive spectral index of $+0.73\pm0.35$, which is consistent with thermal emission. We have also detected maser features of both H$_2$O and CH$_3$OH associated with the continuum source. In addition, other authors have reported the presence of OH masers towards this object \citep[see e.g.,][]{CaswellGreen2011}. \citet{Rathborne2007} studied this source in the millimeter regime, and identified a singular compact source with a rich chemistry characteristic of hot molecular cores. \citet{SanchezMonge2013d} reported molecular outflow activity, with the source \#110 at its geometric center. Overall, this source is one of the best candidates for a thermal radio jet.

%
\subsection*{G24.60$+$0.1M1 (\#119)\label{s:g24.60+0.1m1}}

As for \#119, the only information we have is that it is associated with an EGO. The nearest studied object is a extended H$_2$ emission that is 19$^{\arcsec}$ away from \#119 \citep{Froebrich2015}. We can only say that we cannot determine the spectral index because the source is outside the primary beam of the K~band images, but we propose that \#119 is a radio jet.

%
\subsection*{G24.60$+$0.1M2 (\#136)\label{s:g24.60+0.1m2}}

\begin{figure}[t]
\centering
\begin{tabular}{c}
    \includegraphics[width=0.9\columnwidth]{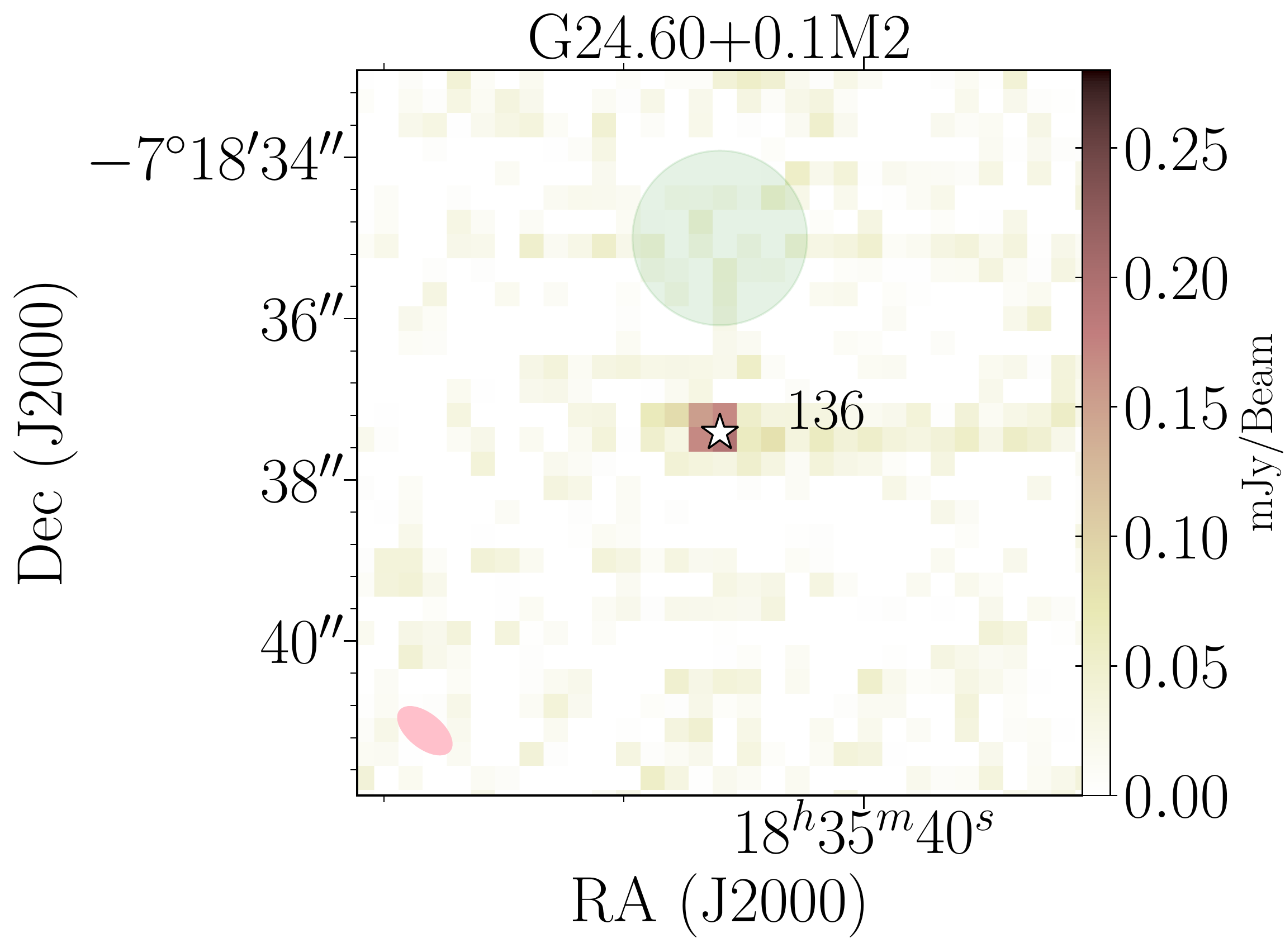} \\
\end{tabular}
\caption{VLA K~band (1.3~cm) continuum emission map of the radio jet candidate \#136 located in the region G24.60$+$0.1\,M2. The green circle with a radius of $\sim$4$^{\arcsec}$ marks EGO G24.63$+$0.15, reported by \citet{Cyganowski2008}. The pink ellipse is the beam size of the K~band. The white star marks the location of the H$_2$O (see Table~\ref{t:masers}).}
\label{f:source136}
\end{figure}

Source \#136 is detected only in the K~band, resulting in a spectral index limit ($>+1.67$) that is consistent with thermal emission. The source is associated with H$_2$O maser emission. This source, although not directly associated with one EGO, is located in the vicinity of G24.63$+$0.15 reported by \citet[see green circle in Fig.~\ref{f:source136}]{Cyganowski2008}. \citet{Rathborne2007} suggest that the main dense condensation, hosting source \#136, may contain several condensations, referred to as G024.60$+$00.08\,MM1 (A, B, and C). Our radio continuum source appears to be related to component C which is an IRDC condensation. Further observations of this object are necessary to confirm its possible nature as a radio jet.

%
\subsection*{G34.43$+$0.2M3 (\#137 and \#139)\label{s:g34.43+0.2m3}}

\begin{figure}[t]
\centering
\begin{tabular}{c}
\includegraphics[width=0.9\columnwidth]{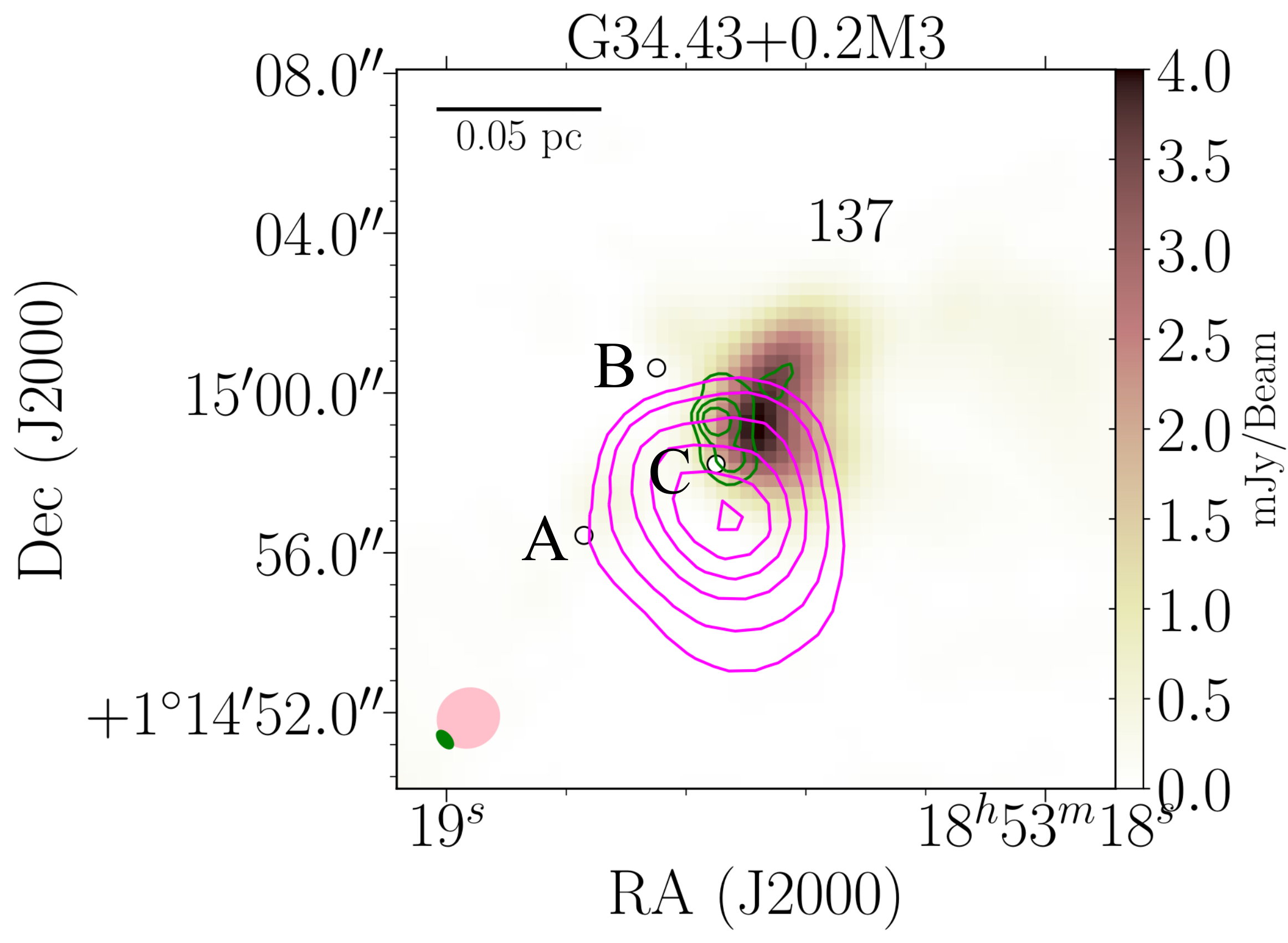} \\
\end{tabular}
\caption{VLA C~band (6~cm) continuum emission map of the radio jet candidate \#137 located in the region G34.43$+$0.2\,M3. The green contour levels of the K~band (1.3~cm) continuum emission are 3, 5 and 7 times 20~$\mu$Jy~beam$^\mathrm{-1}$. The magenta contours show the \textit{Spitzer}/GLIMPSE 4.5~$\mu$m emission. The pink and green ellipses are the beam sizes of the C and K~bands, respectively. The white circles (A, B, and C) mark the position of 2~mm continuum sources reported by \citet{Gasiprong2002}.}
\label{f:source137}
\end{figure}

Our radio continuum observations towards the region G34.43$+$0.2\,M3 have led to the discovery of six radio continuum sources, although most of them are located far from the central region studied in \citet{SanchezMonge2013d}. The brightest source is \#137, which is about 13\arcmin\ from the phase center of our observations, i.e., the source is outside the primary beam responses of the VLA antennas on both bands. This object is so bright that it is detected in both the C and K~bands. At 6~cm, the source appears as a comet-like structure (see Fig.~\ref{f:source137}), which resembles cometary \hii\ regions. The 1.3~cm continuum emission also shows an arc-shaped structure shifted to the east, probably tracing the head of the cometary object. This source, referred to in the literature as G34.26$+$0.15, has been studied by other authors who report the presence of two hyper-compact \hii\ regions (A and B) and one ultra-compact \hii\ region (C), all marked in the Figure \citep [see also][]{ReidHo1985, Gaume1994, Sewilo2011}. Various studies \citep[e.g.,][]{Hatchell2001, Liu2013} have reported the presence of outflow activity in this region, but no information on the outflow energetics such as the outflow momentum rate is reported. Despite this source is associated with an EGO \citep{Cyganowski2008} and with a molecular outflow, the bright emission together with previous studies suggests that a major fraction of the radio continuum emission we have detected is originated in a \hii\ region rather than in a radio jet.
Source \#139 in the region, is also classified as a radio jet candidate in Table~\ref{t:candidates}. However, the only information we have this object is its association with an EGO \citep [i.e., G34.41$+$0.24,][]{Cyganowski2008}. \citet{Shepherd2004} suggest that the embedded object (G34.4\,MM) appears to be a massive B2 protostar at an early stage of evolution. This region is also associated with the H$_2$O maser activity \citep{Cyganowski2013}, which may favor a radio jet origin for the detected radio continuum emission. Further observations are needed to better constrain its properties.

%
\subsection*{IRAS\,19095$+$0930 (\#143 and \#144)\label{s:iras19095+0930}}

\begin{figure}[t]
\centering
\begin{tabular}{c}
\includegraphics[width=0.9\columnwidth]{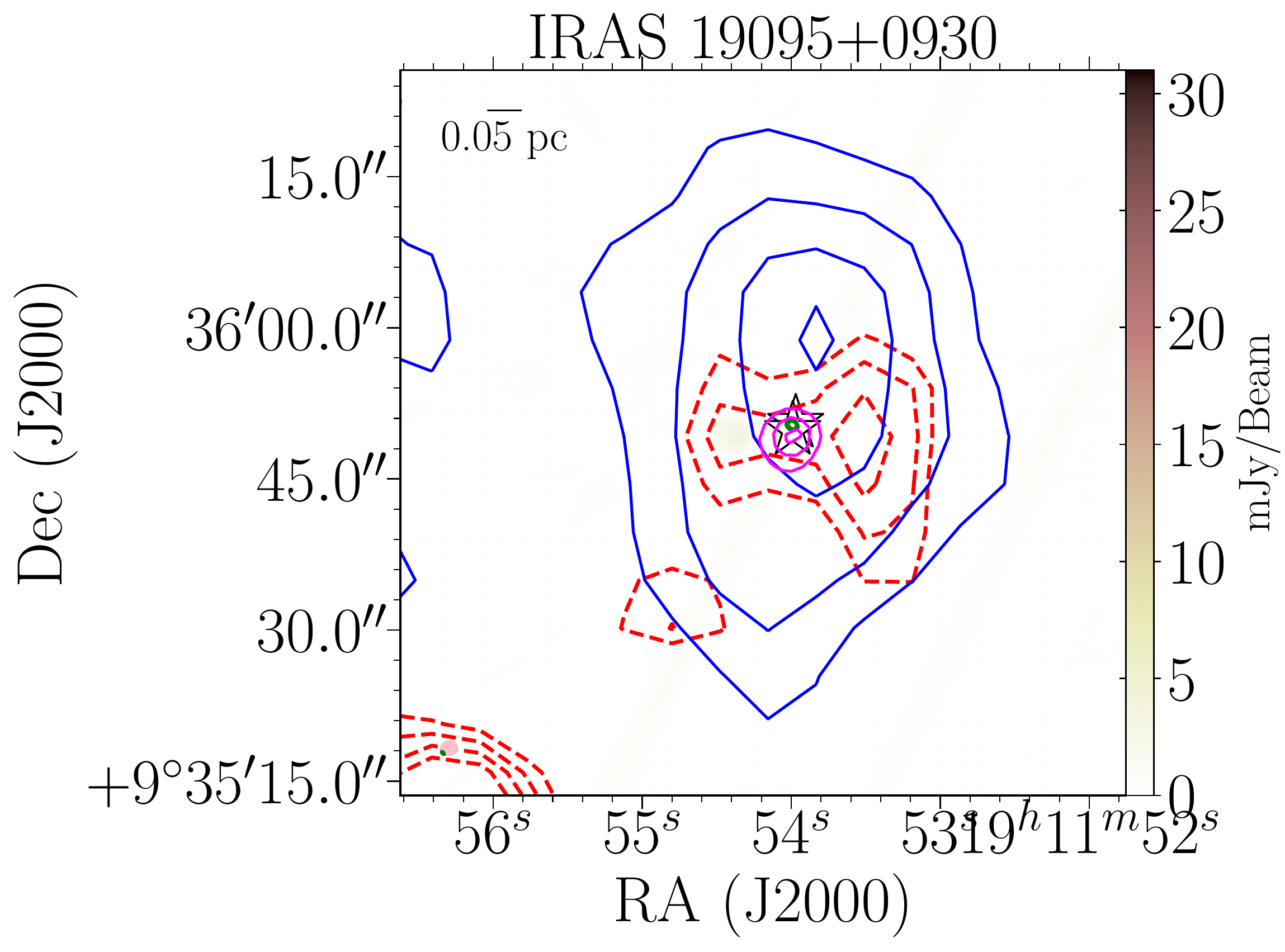} \\
\includegraphics[width=0.9\columnwidth]{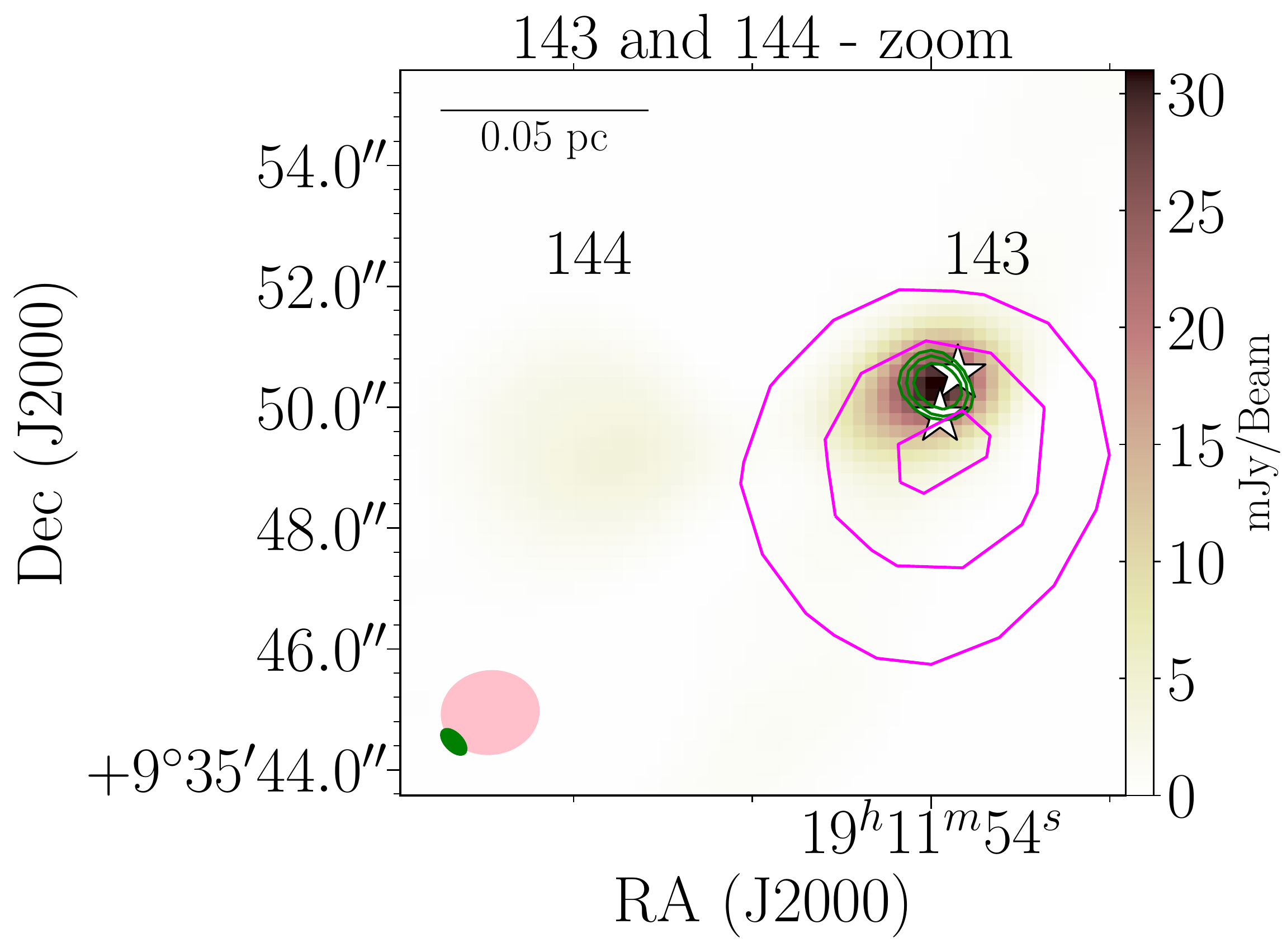} \\
\end{tabular}
\caption{VLA C~band (6~cm) continuum emission map of the radio sources \#143 and \#144 located in the region IRAS\,19095$+$0930. The green contour levels of the K~band (1.3~cm) continuum emission are 3, 5 and 9 times 2~mJy~beam$^\mathrm{-1}$. The blue- and red-shifted outflow lobed of SiO (2$-$1) are shown as blue-solid and red-dashed contours, respectively \citep[see][]{SanchezMonge2013d}. The magenta contours show the \textit{Spitzer}/GLIMPSE 4.5~$\mu$m emission. The pink and green ellipses are the beam sizes of the C and K~bands, respectively. The white and orange stars mark the of the H$_2$O and CH$_3$OH masers (see Table~\ref{t:masers}).}
\label{f:source143}
\end{figure}

We have identified four radio continuum sources in the region IRAS\,19095$+$0930, also known in the literature as G43.80$-$0.13). Two of these radio sources, \#143 and \#144, are located close to each other and in the center of a molecular outflow \citep[][see also Figure~\ref{f:source143}]{SanchezMonge2013d}. The source \#143 has a brighter flux, is also clearly visible in the K~band image, and is associated with H$_2$O maser emission features.

This region has been studied in the past by different authors at different wavelengths \citep{Kurtz1994, Lekht2000, DeBuizer2005}. \citet{DeBuizer2005} report a kidney-bean shape structure at mid-infrared wavelengths that match the radio continuum sources reported by \citet{Kurtz1994} refers to as a \hii\ region. The object is also associated with OH masers. We have not found this source in the EGO catalogues \citep{Cyganowski2008, Cyganowski2009}, but we have identified a 4.5~$\mu$m source associated with \#143. No 4.5~$\mu$m infrared source appears to be associated with the eastern source \#144.

We derive a spectral index of $+1.12\pm0.04$ for \#143, which is consistent with thermal emission. This, together with its location at the geometrical center of the outflow, and its association with masers may suggest that \#143 is a good radio jet candidate. However, the large radio continuum flux of this source seem to not be consistent with the typical properties of other radio jets (see Figs.~\ref{fig:Lrad_Lbol} and \ref{f:LradPout}). This might mean that this source is in a transition phase from a radio jet to an \hii\ region. However, this requires further investigation.

%
\section{Catalog of the continuum sources}\label{s:catalogue}

\begin{figure*}[!ht]
	\centering
	\includegraphics[scale=0.33]{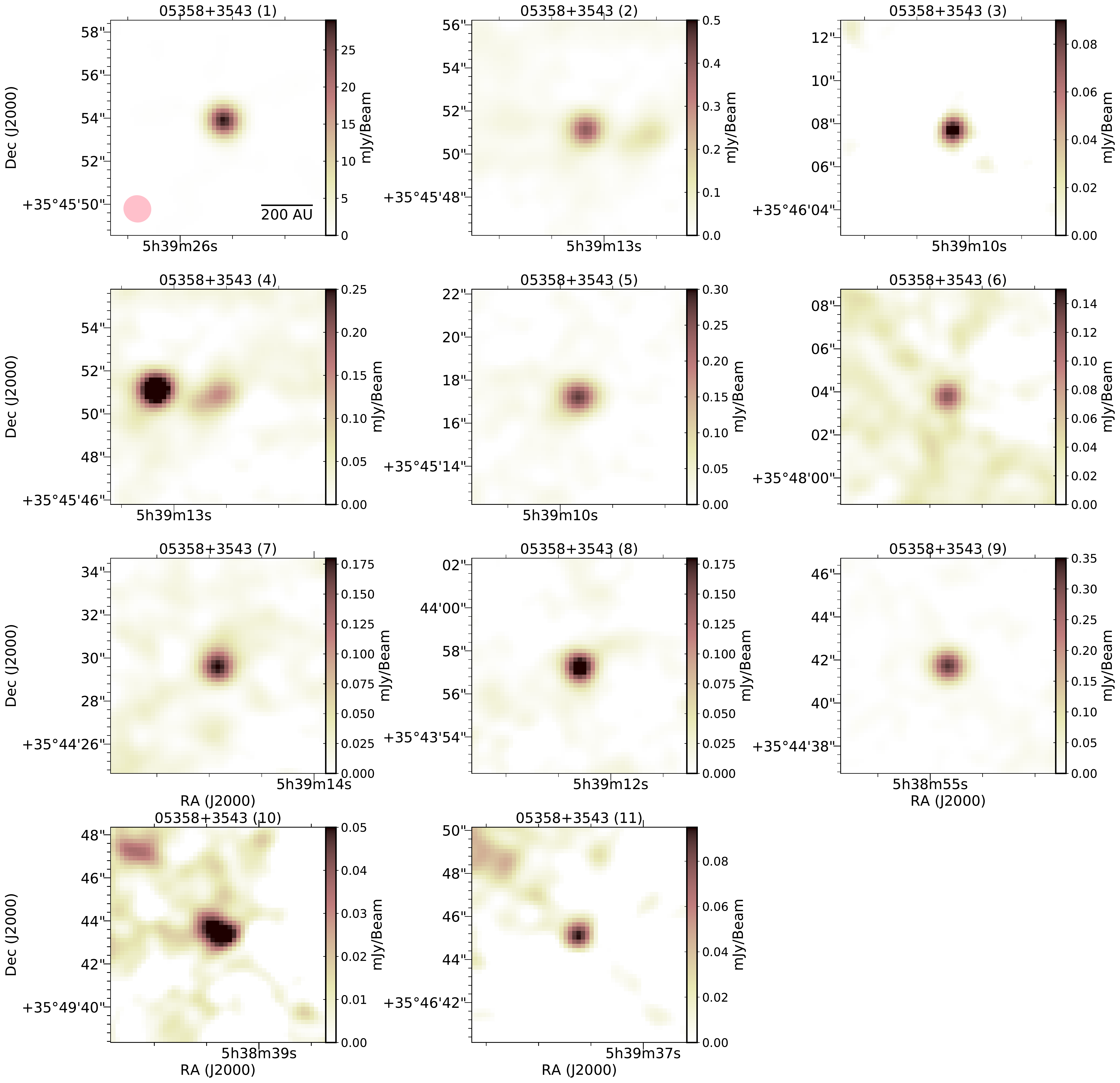}
	\caption{\label{f:stamps01} Close-up views of the C~band (color scale image) and K~band (contours) continuum images for the sources listed in Table~\ref{t:catalogue}. Maps for the sources detected in region IRAS\,05358$+$3543.}
\end{figure*}

\begin{figure*}[!ht]
	\centering
	\includegraphics[scale=0.33]{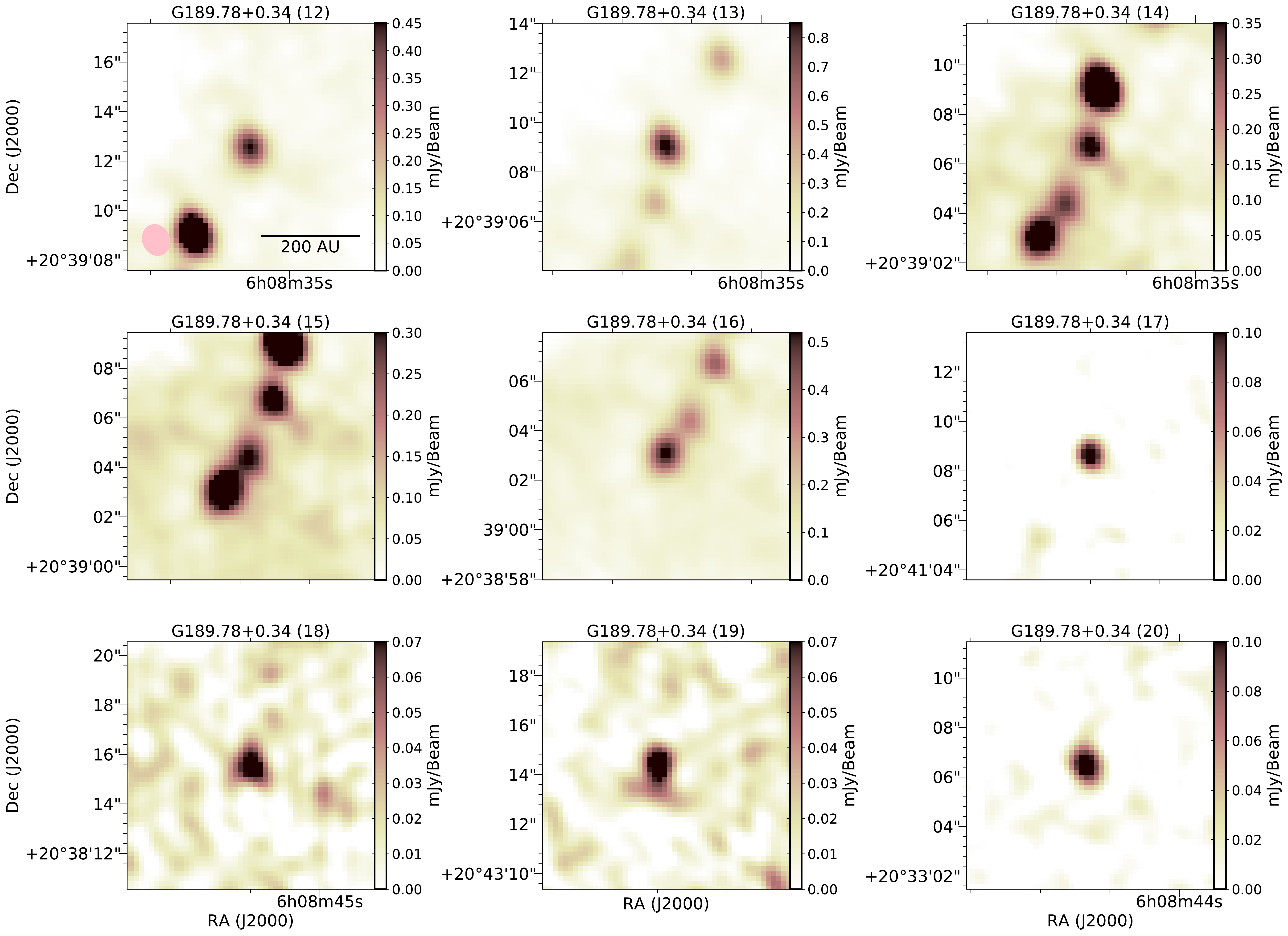}
	\caption{\label{f:stamps02} Close-up views of the C~band (color scale image) and K~band (contours) continuum images for the sources listed in Table~\ref{t:catalogue}. Maps for the sources detected in region G189.78$+$0.34.}
\end{figure*}

\begin{figure*}[!ht]
	\centering
	\includegraphics[scale=0.33]{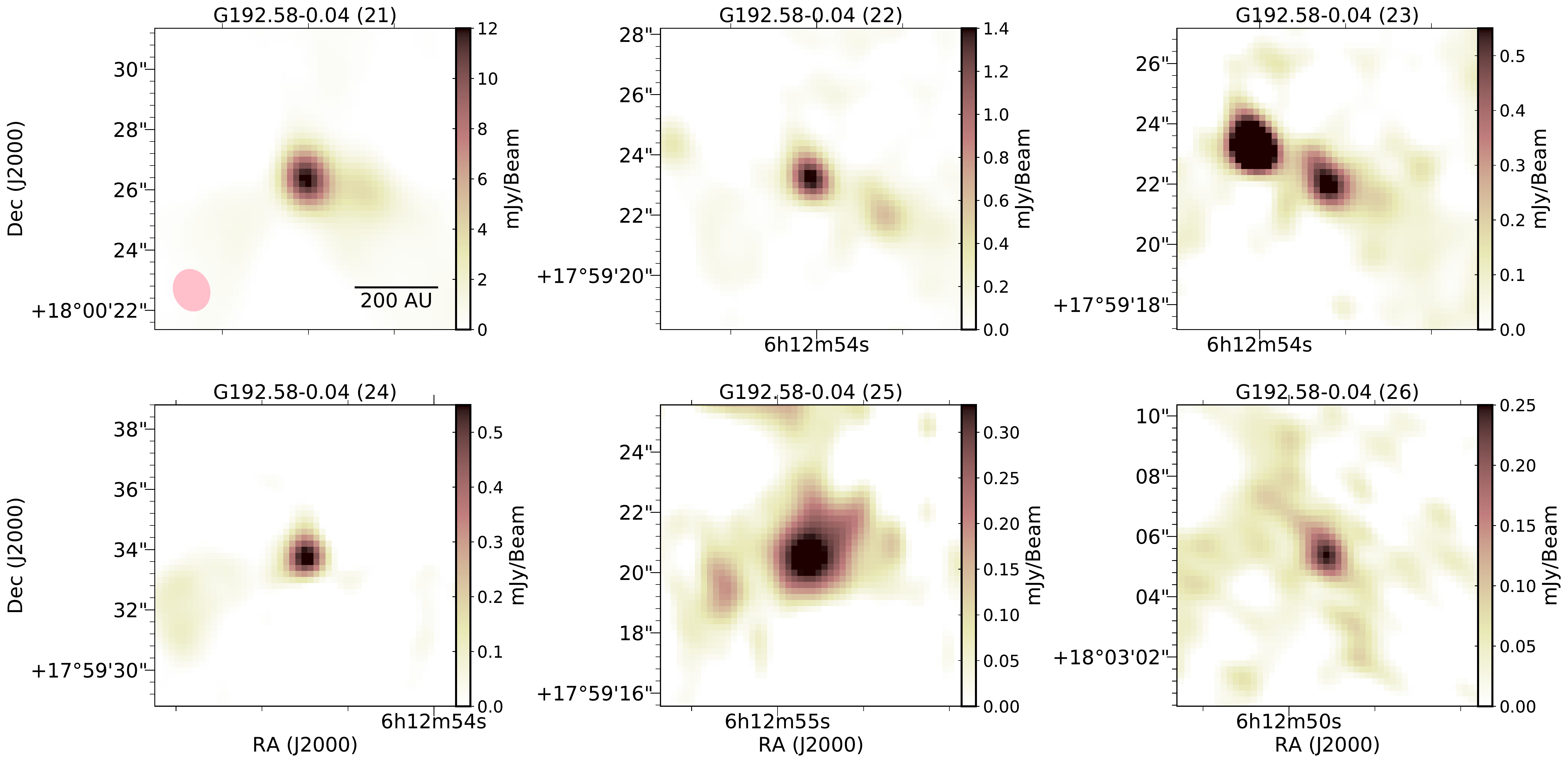}
	\caption{\label{f:stamps03} Close-up views of the C~band (color scale image) and K~band (contours) continuum images for the sources listed in Table~\ref{t:catalogue}. Maps for the sources detected in region G192.58$-$0.04.}
\end{figure*}

\begin{figure*}[!ht]
	\centering
	\includegraphics[scale=0.33]{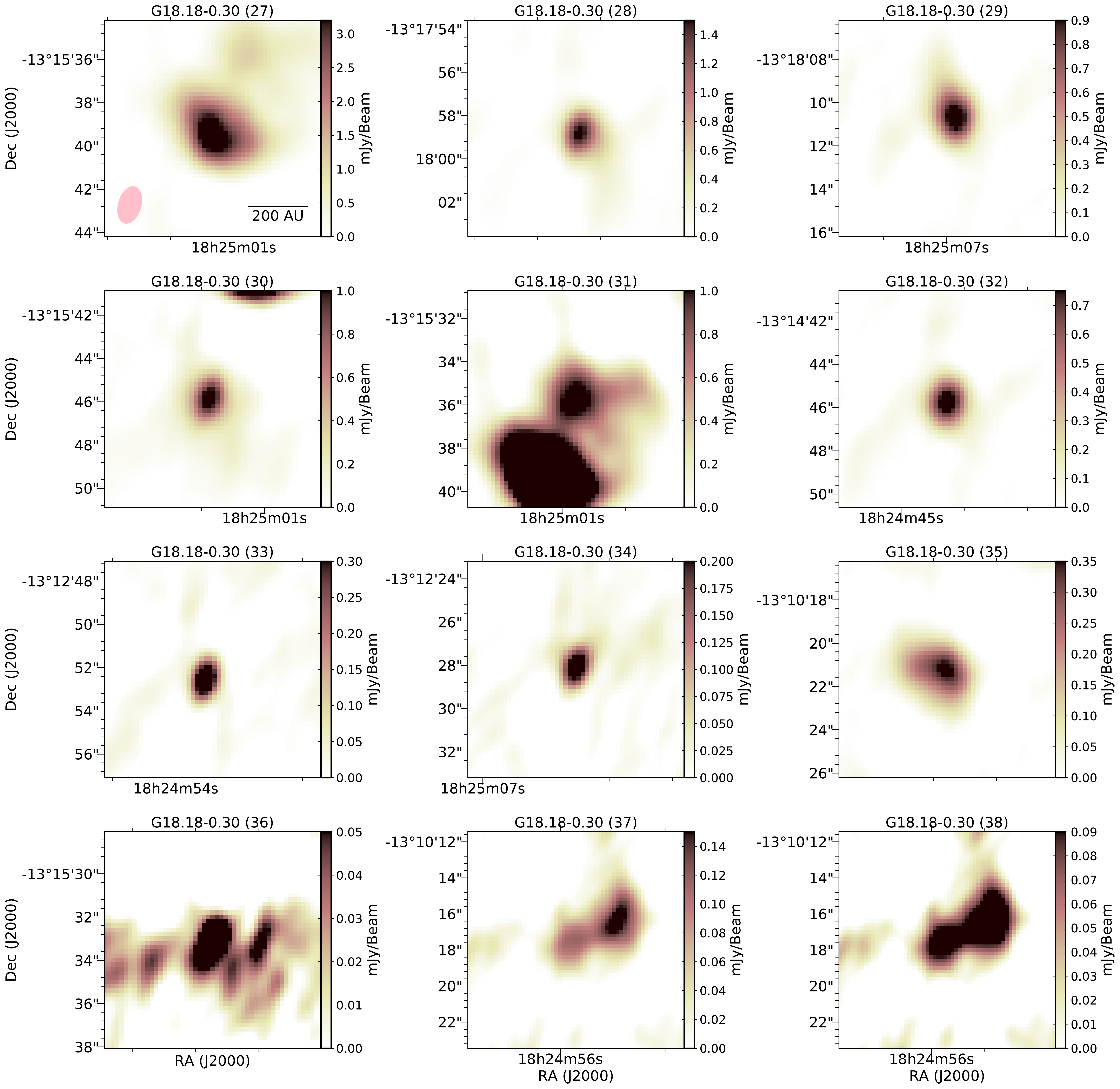}
	\caption{\label{f:stamps04} Close-up views of the C~band (color scale image) and K~band (contours) continuum images for the sources listed in Table~\ref{t:catalogue}. Maps for the sources detected in region G18.18$-$0.30.}
\end{figure*}

\begin{figure*}[!ht]
	\centering
	\includegraphics[scale=0.33]{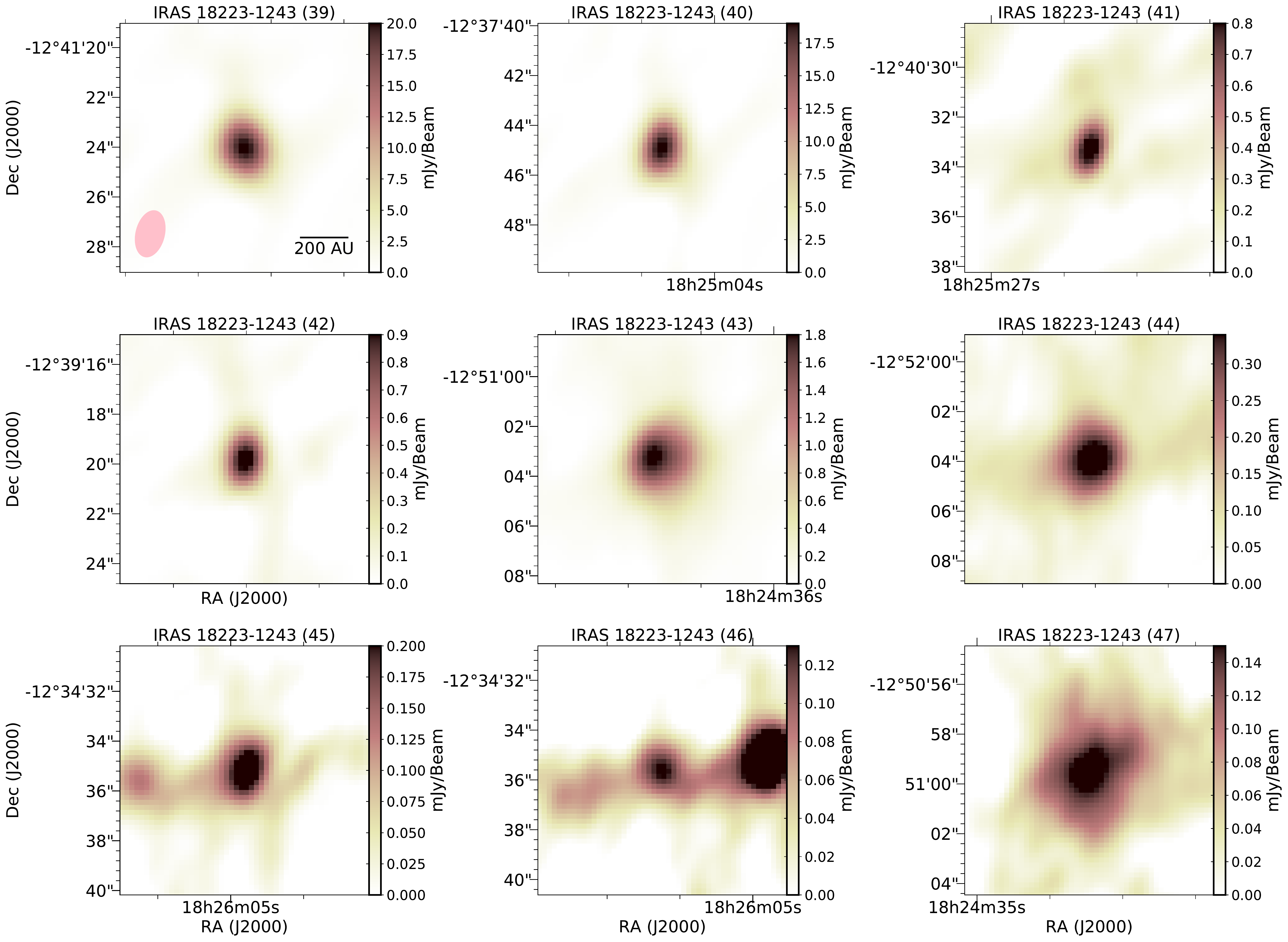}
	\caption{\label{f:stamps05} Close-up views of the C~band (color scale image) and K~band (contours) continuum images for the sources listed in Table~\ref{t:catalogue}. Maps for the sources detected in region IRAS~18223$-$1243.}
\end{figure*}

\begin{figure*}[!ht]
	\centering
	\includegraphics[scale=0.33]{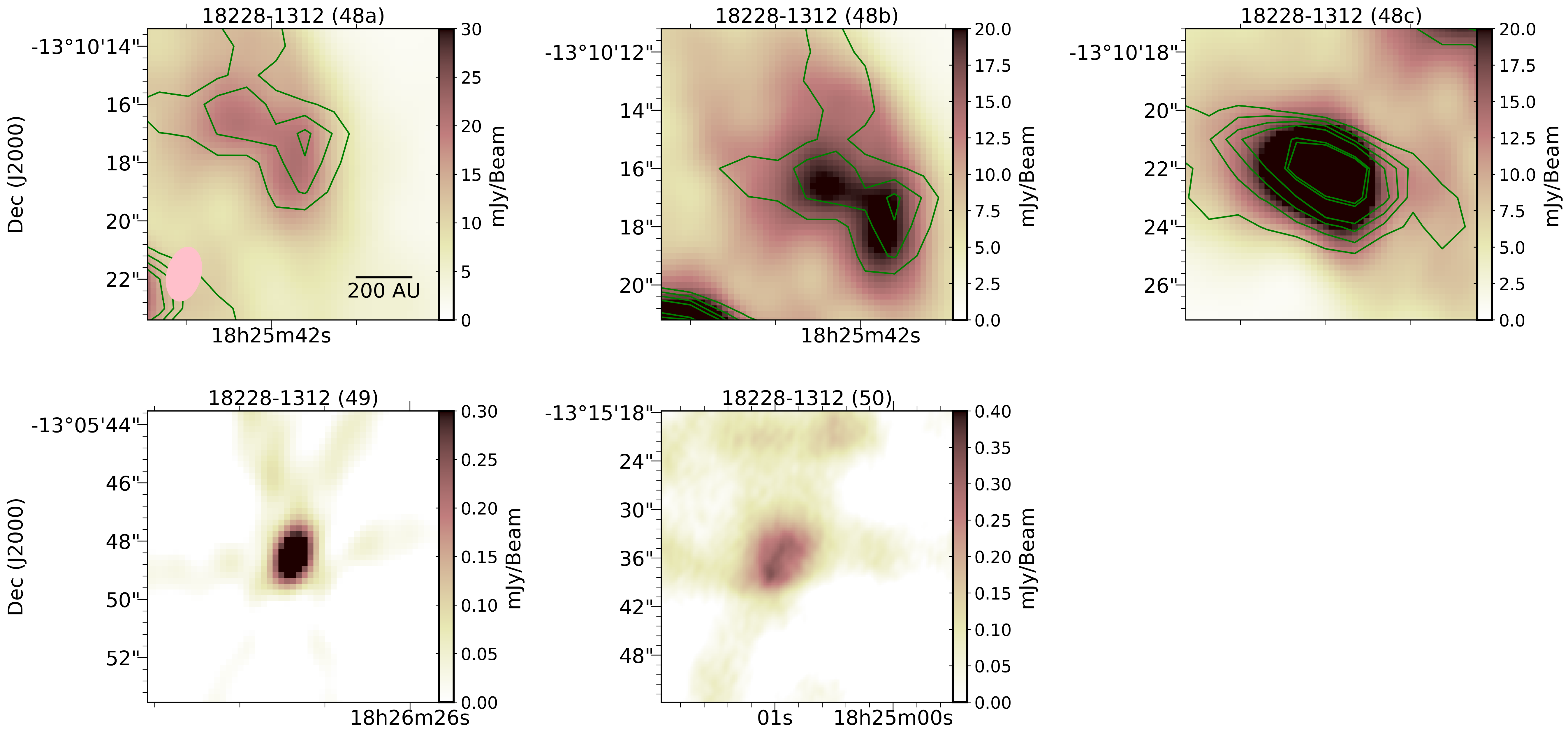}
	\caption{\label{f:stamps06} Close-up views of the C~band (color scale image) and K~band (contours) continuum images for the sources listed in Table~\ref{t:catalogue}. Maps for the sources detected in region IRAS~18228$-$1312.}
\end{figure*}

\begin{figure*}[!ht]
	\centering
	\includegraphics[scale=0.33]{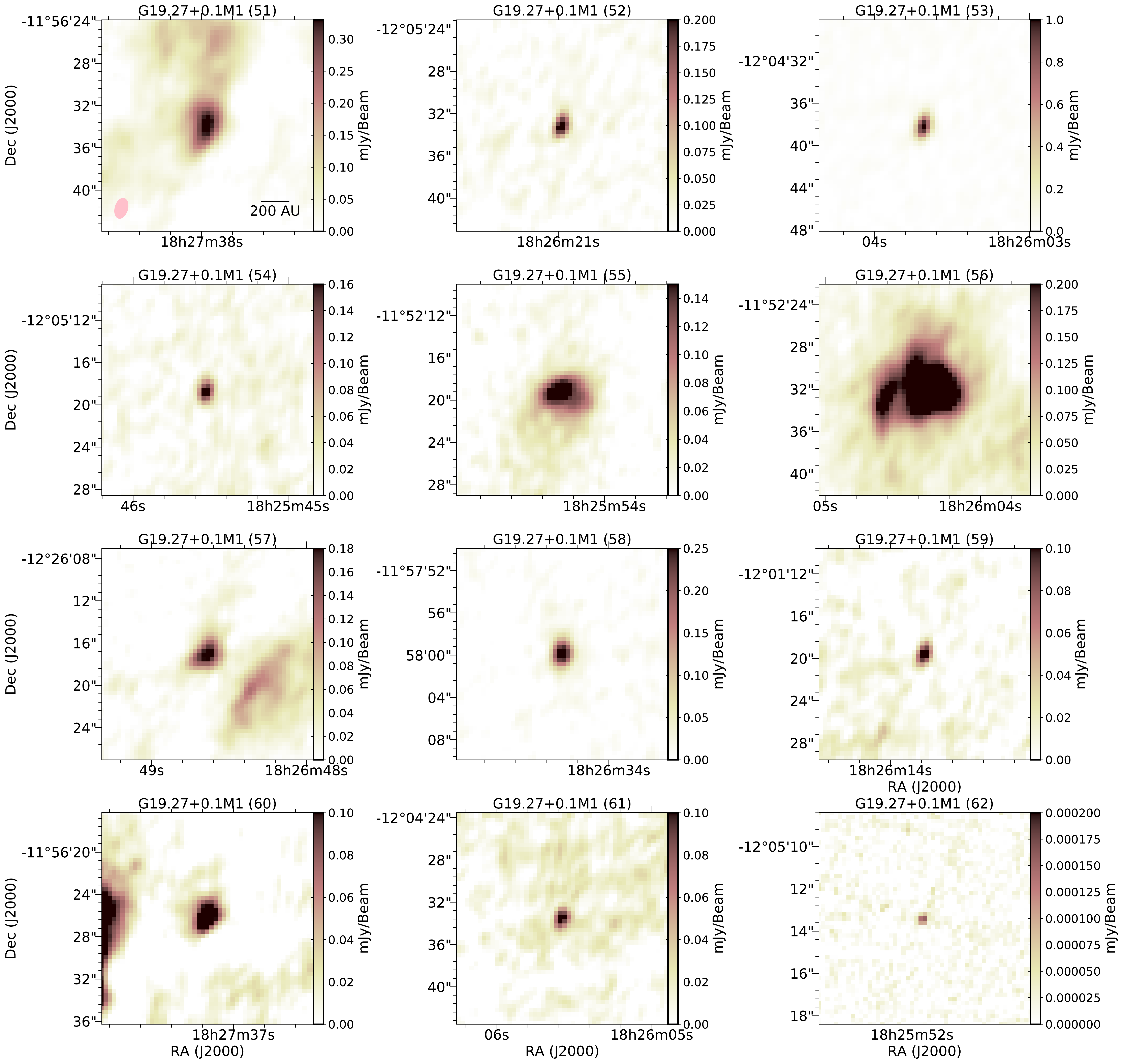}
	\caption{\label{f:stamps07} Close-up views of the C~band (color scale image) and K~band (contours) continuum images for the sources listed in Table~\ref{t:catalogue}. Maps for the sources detected in region G19.27$+$0.1\,M1.}
\end{figure*}

\begin{figure*}[!ht]
	\centering
	\includegraphics[width=\columnwidth]{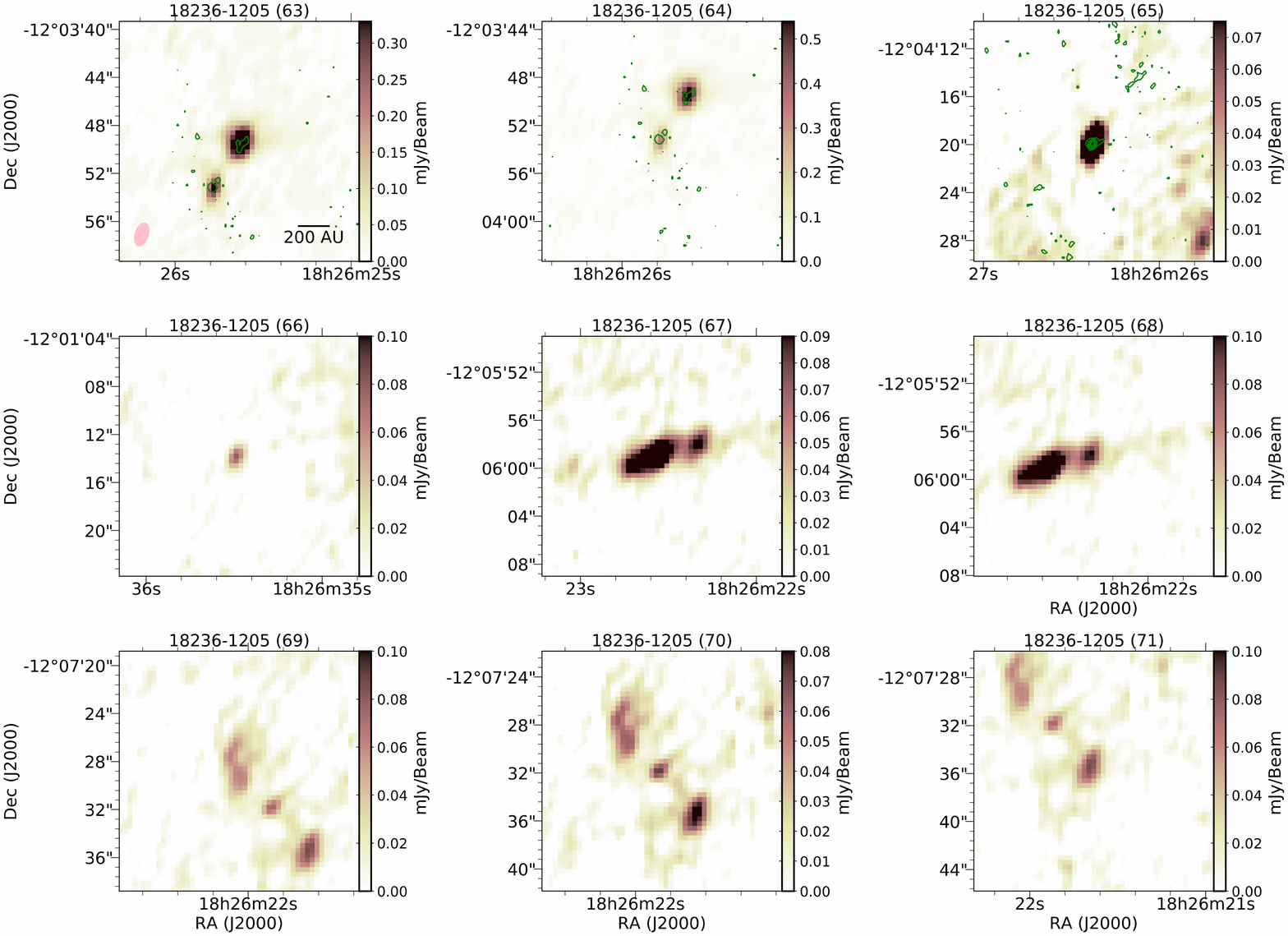}
	\caption{\label{f:stamps09} Close-up views of the C~band (color scale image) and K~band (contours) continuum images for the sources listed in Table~\ref{t:catalogue}. Maps for the sources detected in region IRAS~18236$-$1205.}
\end{figure*}

\begin{figure*}[!ht]
	\centering
	\includegraphics[scale=0.33]{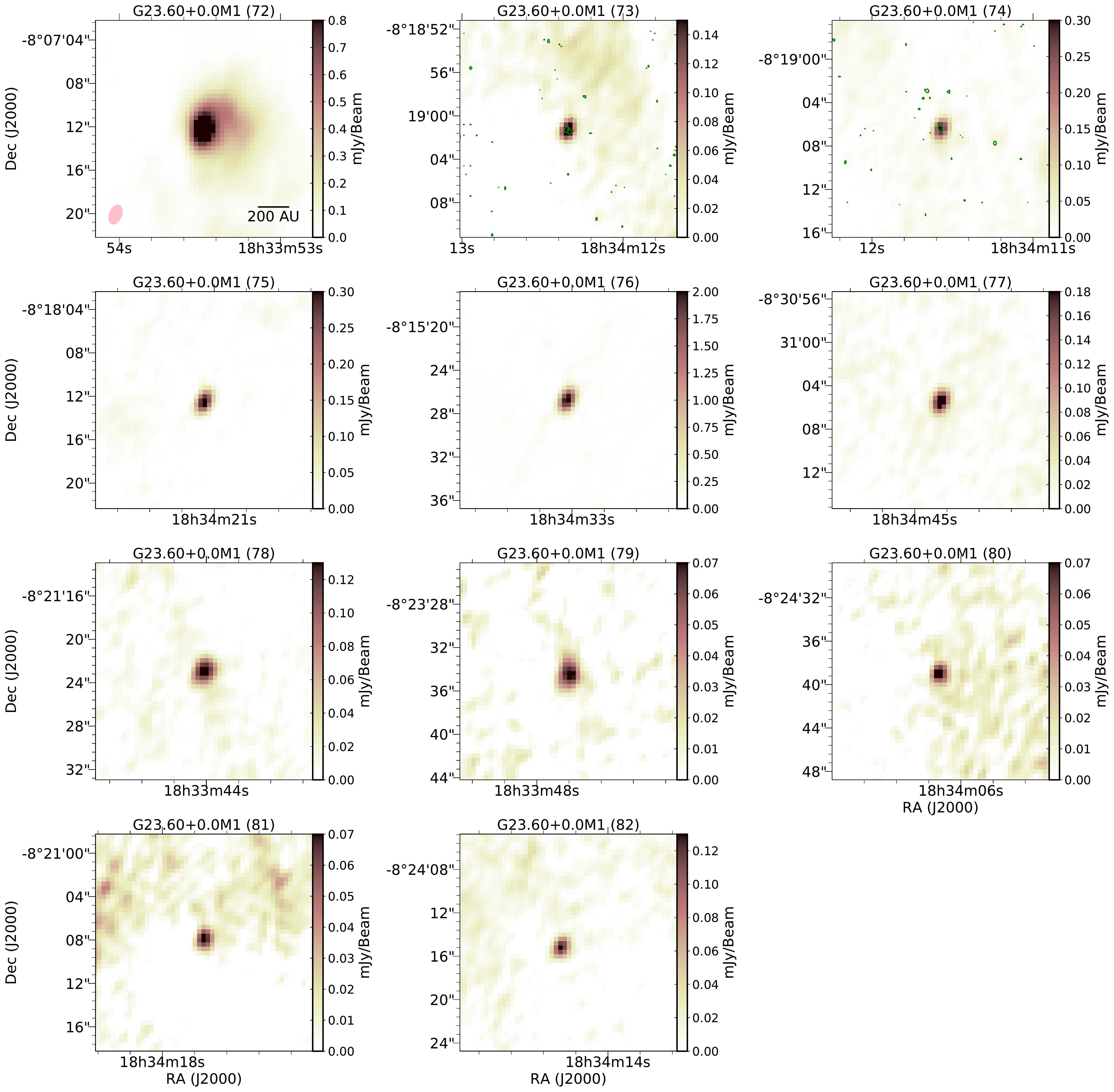}
	\caption{\label{f:stamps10} Close-up views of the C~band (color scale image) and K~band (contours) continuum images for the sources listed in Table~\ref{t:catalogue}. Maps for the sources detected in region G23.60$+$0.0\,M1.}
\end{figure*}

\begin{figure*}[!ht]
	\includegraphics[width=\columnwidth]{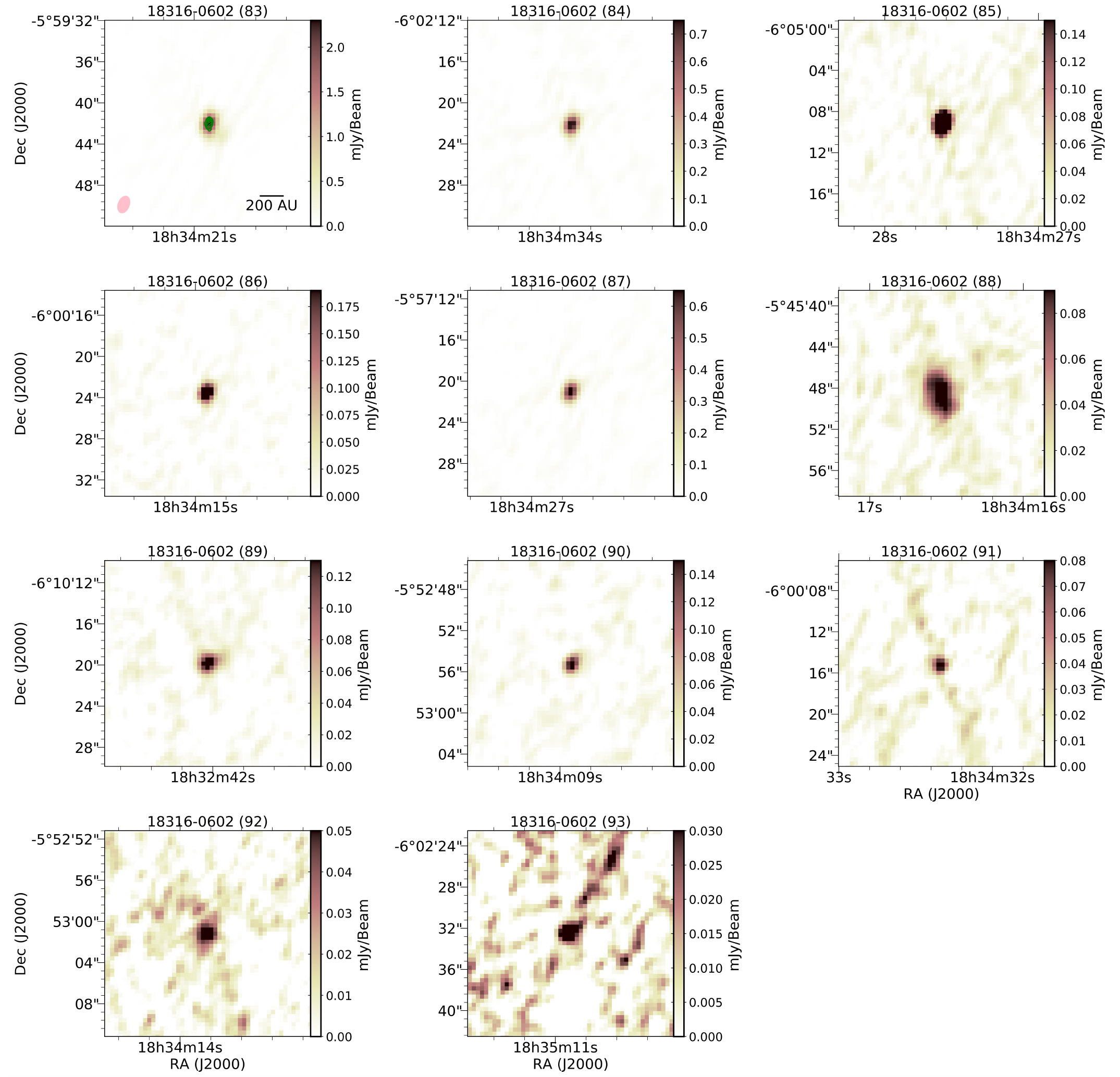}
	\includegraphics[scale=0.33]{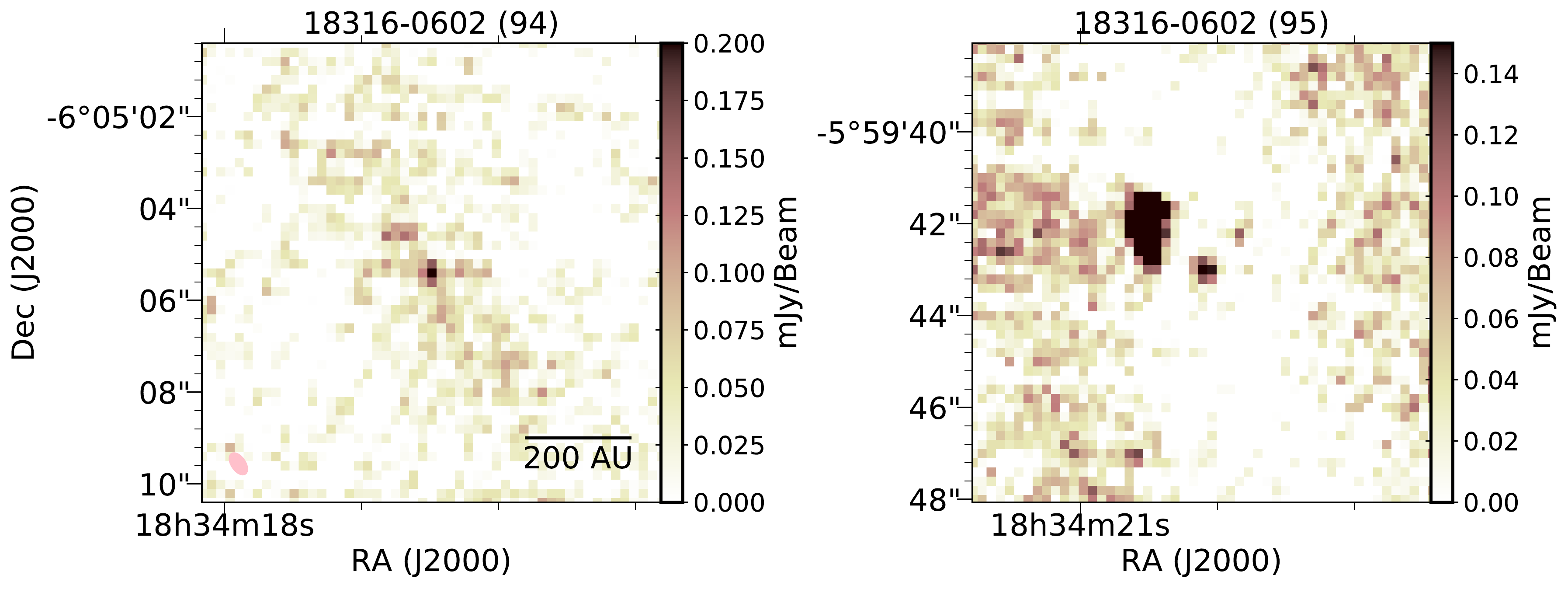}
	\caption{\label{f:stamps11} Close-up views of the C~band (color scale image) and K~band (contours) continuum images for the sources listed in Table~\ref{t:catalogue}. The image of sources \#94 and \#95 correspond to the K~band maps. Maps for the sources detected in region IRAS~18316$-$0602.}
\end{figure*}

\begin{figure*}[!ht]
	\includegraphics[scale=0.33]{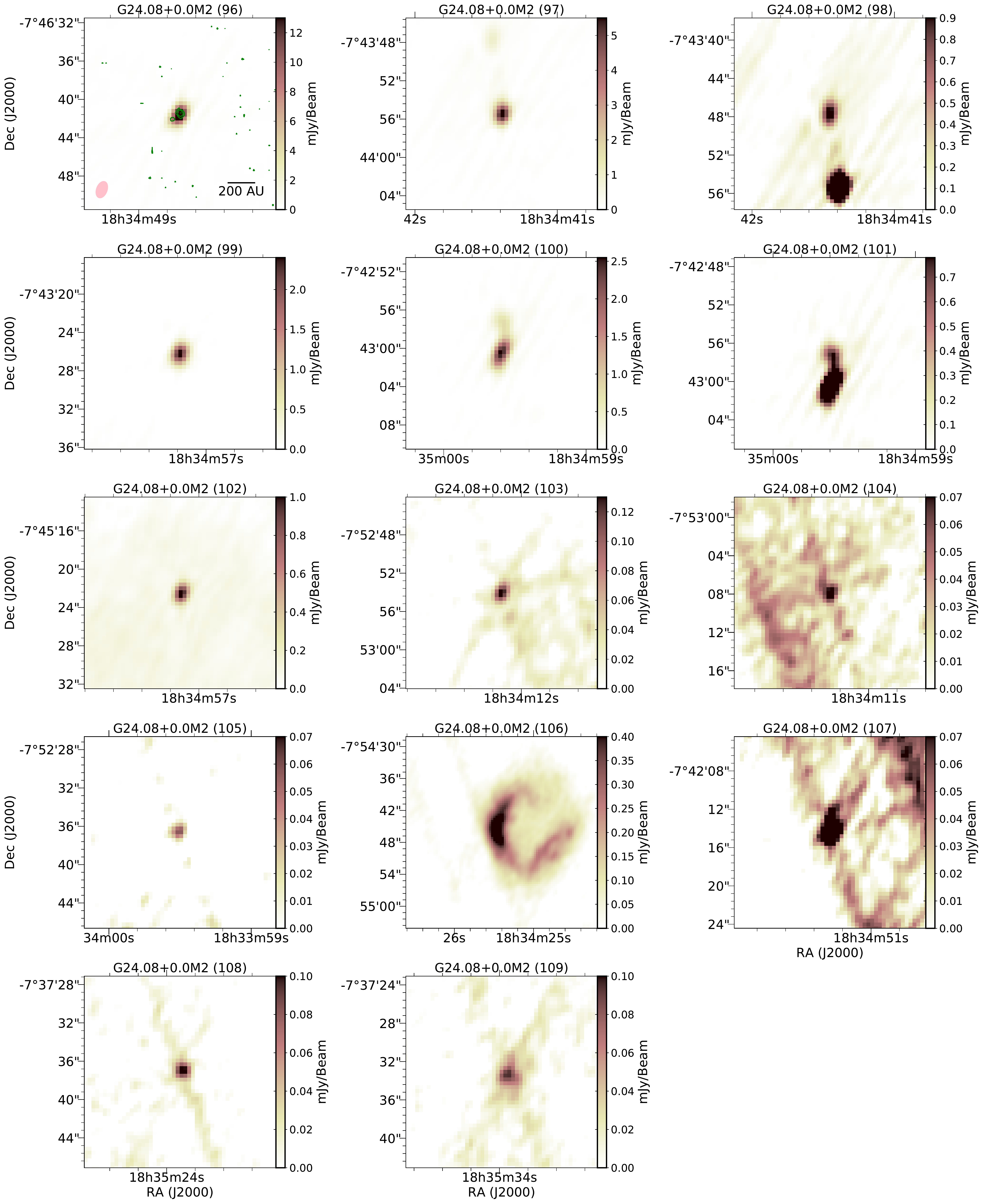}
	\caption{\label{f:stamps12} Close-up views of the C~band (color scale image) and K~band (contours) continuum images for the sources listed in Table~\ref{t:catalogue}. Maps for the sources detected in region G24.08$+$0.0\,M2.}
\end{figure*}

\begin{figure*}[!ht]
	\centering
	\includegraphics[scale=0.33]{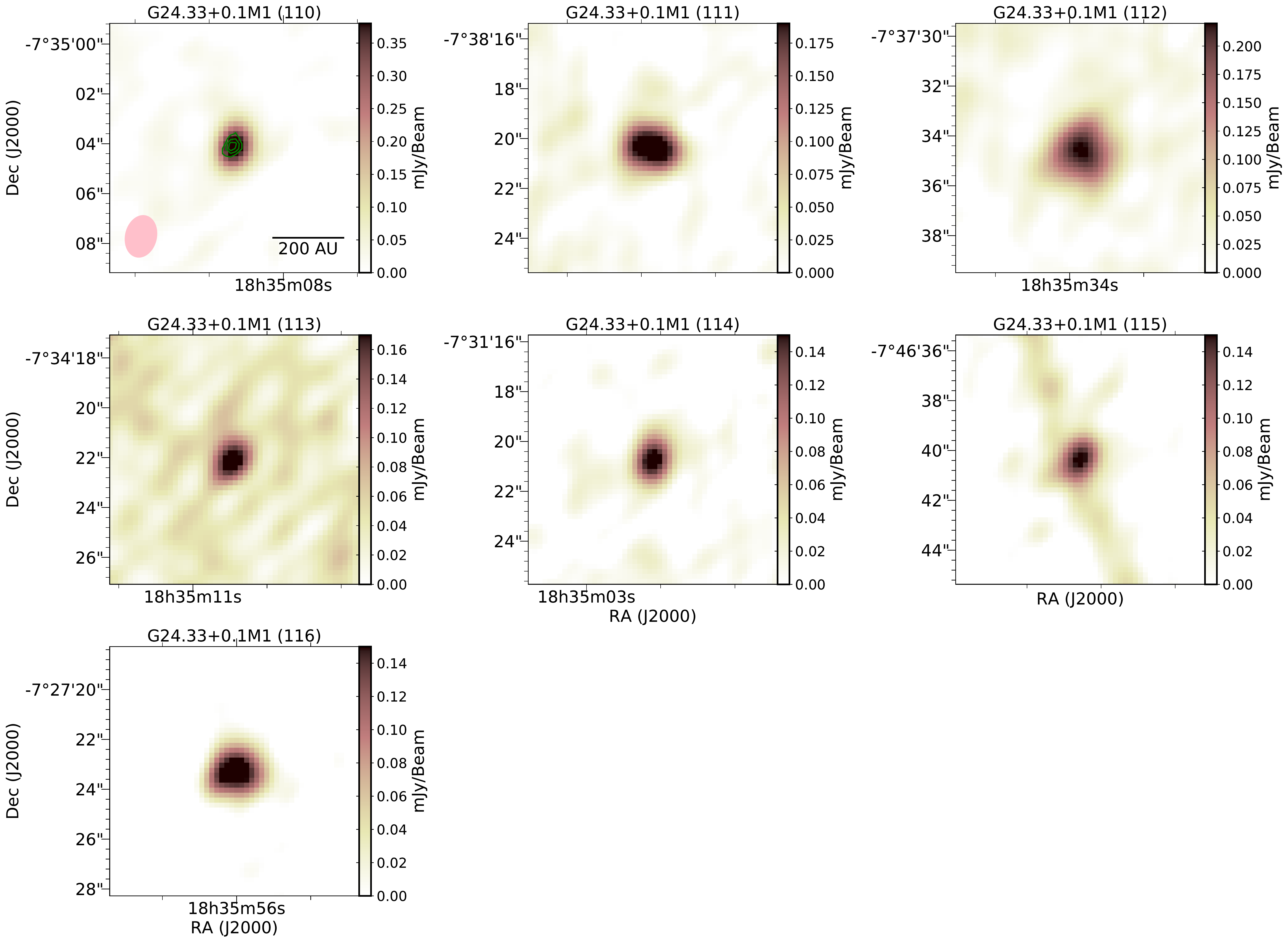}
	\caption{\label{f:stamps13} Close-up views of the C~band (color scale image) and K~band (contours) continuum images for the sources listed in Table~\ref{t:catalogue}. Maps for the sources detected in region G24.33$+$0.1\,M1.}
\end{figure*}

\begin{figure*}[!ht]
	\centering
	\includegraphics[scale=0.33]{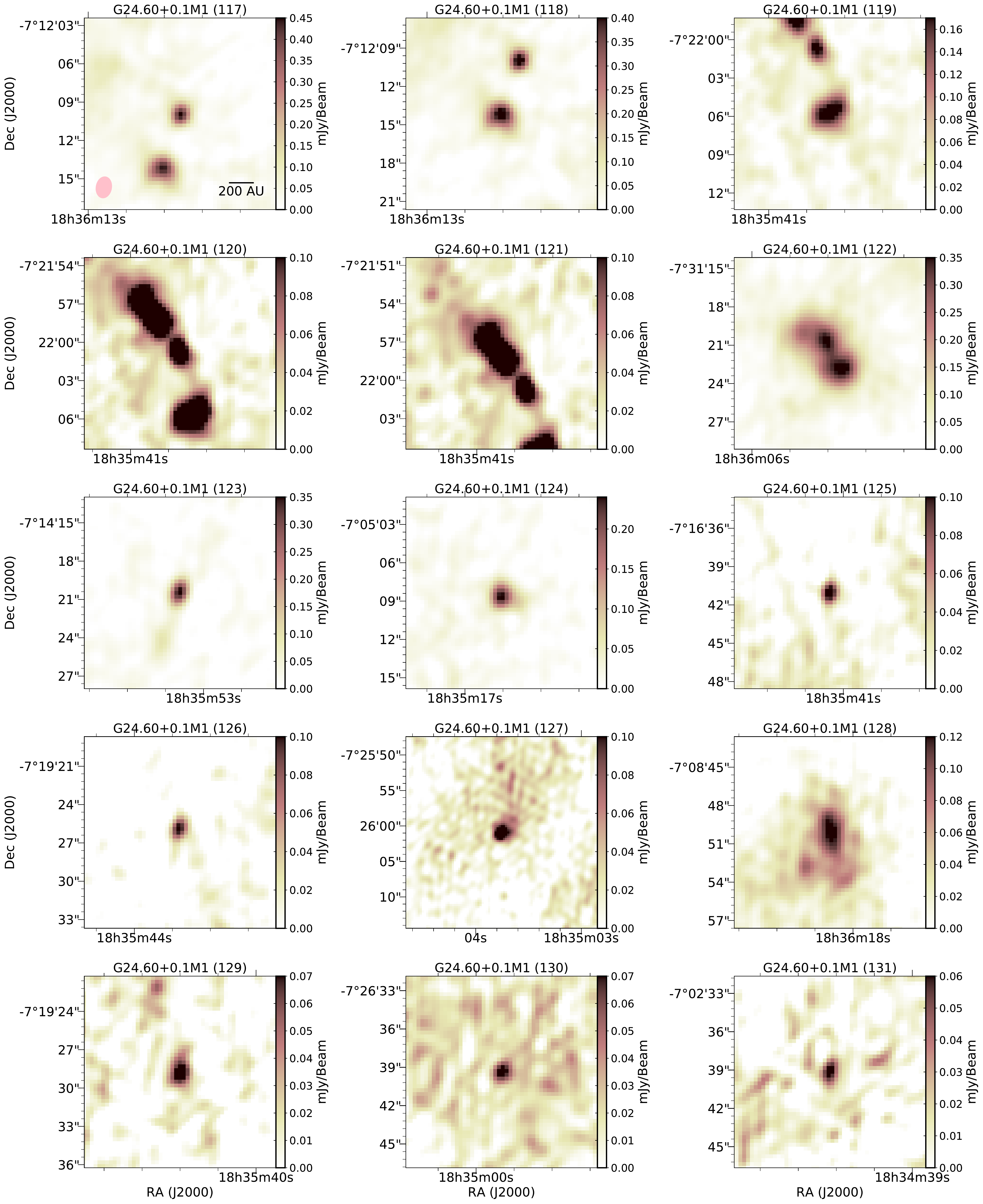}
	\caption{\label{f:stamps14} Close-up views of the C~band (color scale image) and K~band (contours) continuum images for the sources listed in Table~\ref{t:catalogue}. Maps for the sources detected in region G24.60$+$0.1\,M1.}
\end{figure*}

\begin{figure*}[!ht]
	\includegraphics[width=\columnwidth]{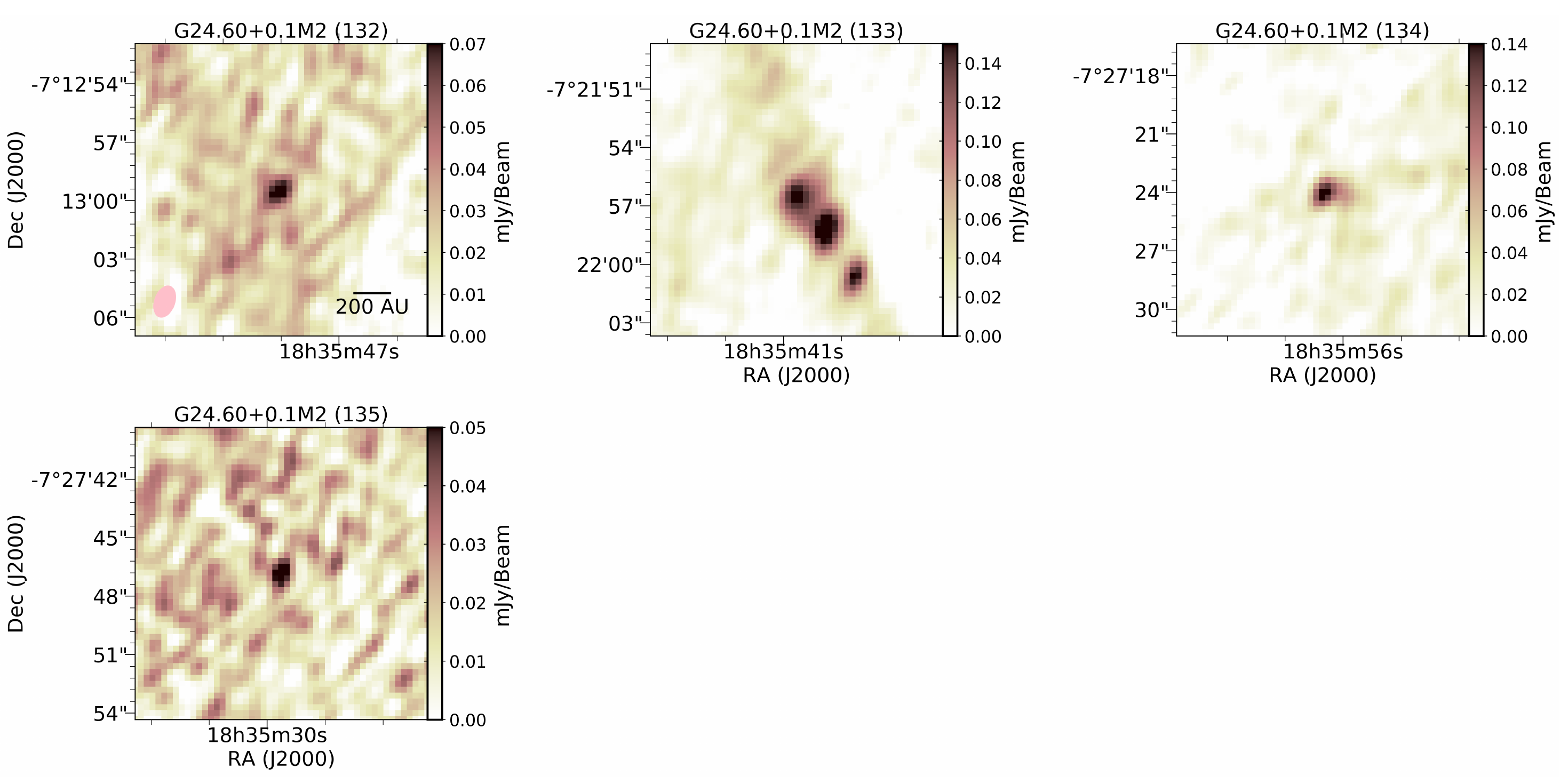} \\
	\includegraphics[scale=0.33]{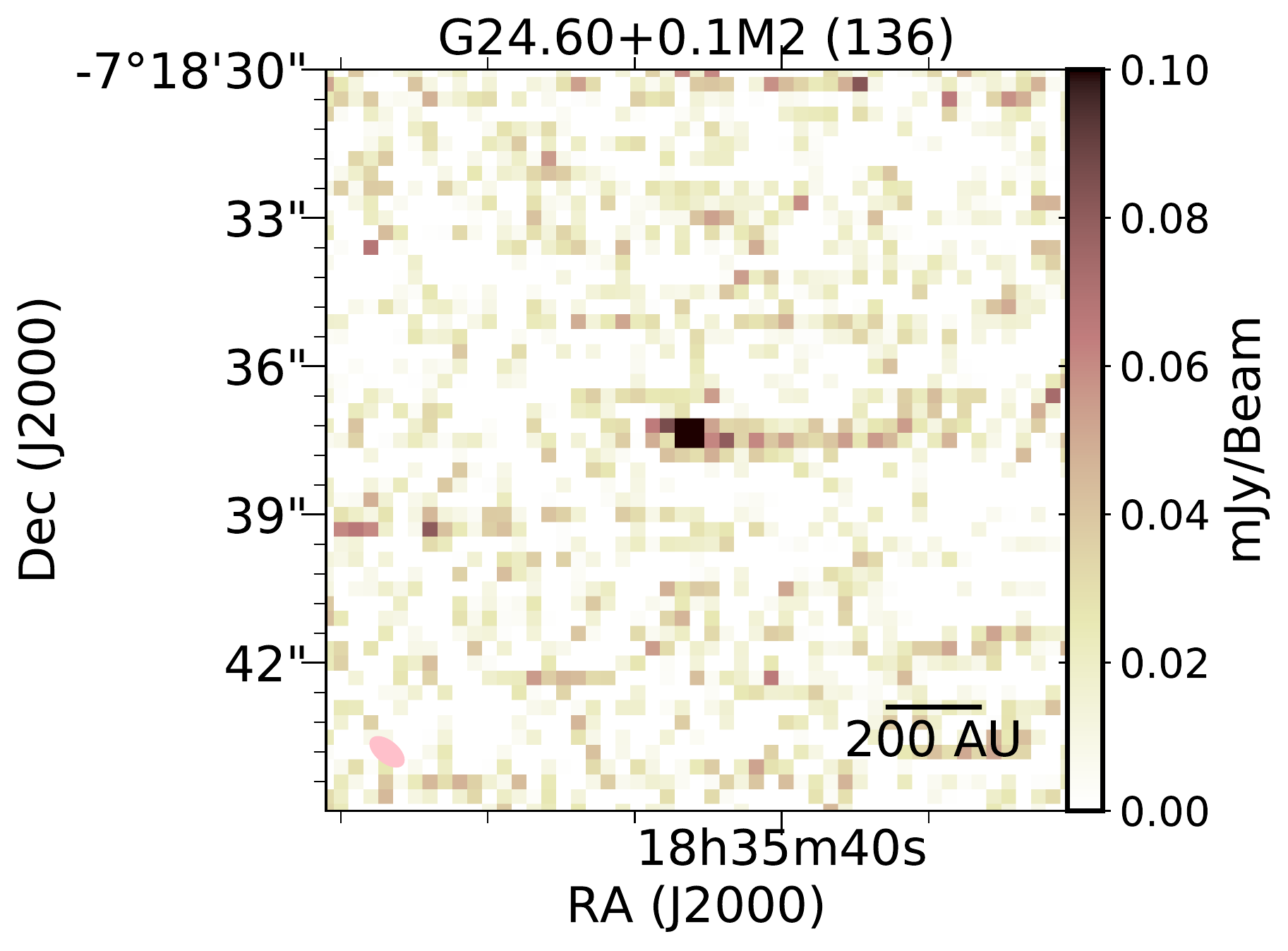}
	\caption{\label{f:stamps15} Close-up views of the C~band (color scale image) and K~band (contours) continuum images for the sources listed in Table~\ref{t:catalogue}. The image of source \#136 corresponds to the K~band map. Maps for the sources detected in region G24.60$+$0.1\,M2.}
\end{figure*}

\begin{figure*}[!ht]
	\centering
	\includegraphics[width=\columnwidth]{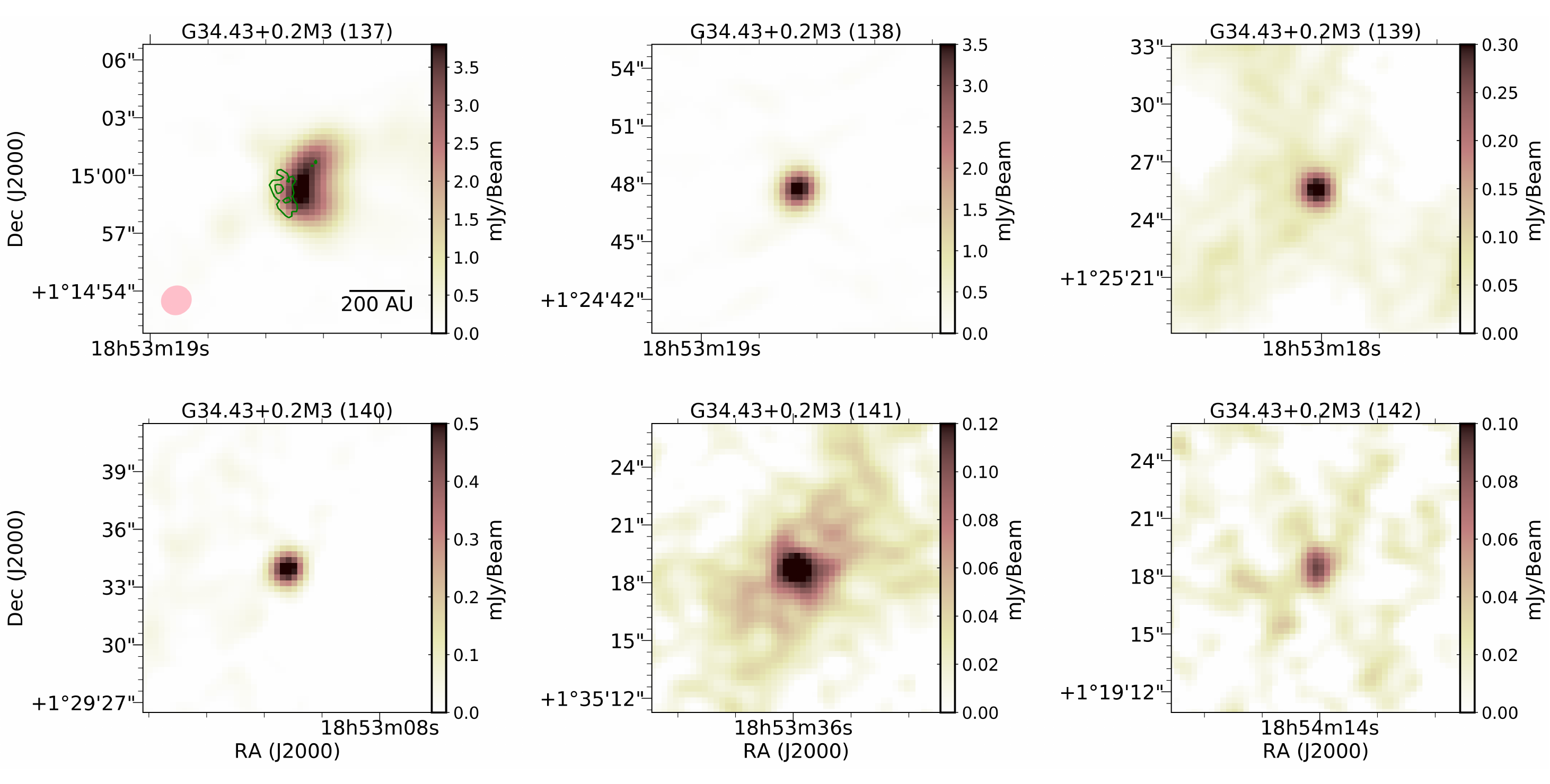}
	\caption{\label{f:stamps16} Close-up views of the C~band (color scale image) and K~band (contours) continuum images for the sources listed in Table~\ref{t:catalogue}. Maps for the sources detected in region G34.43$+$0.2\,M3.}
\end{figure*}

\begin{figure*}[!ht]
	\centering
	\includegraphics[scale=0.33]{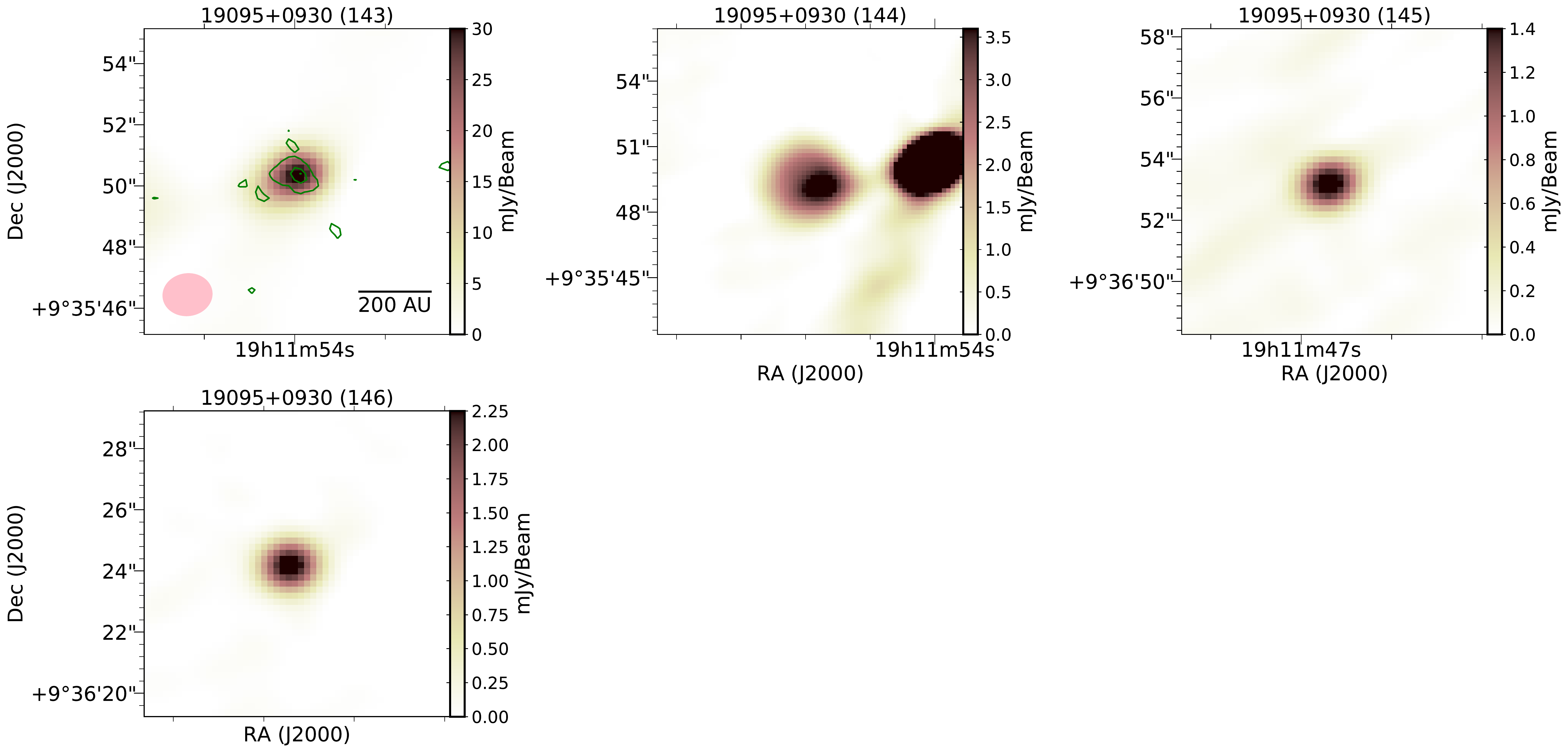}
	\caption{\label{f:stamps17} Close-up views of the C~band (color scale image) and K~band (contours) continuum images for the sources listed in Table~\ref{t:catalogue}. Maps for the sources detected in region IRAS~19095$+$0930.}
\end{figure*}

In the following tables and figures, we provide information on the properties of the radio continuum sources detected in the VLA observations presented in this work. In Table~\ref{t:catalogue}, we list the coordinates of the 146 radio continuum sources together with their flux density, intensity peak and deconvolved size at 6~cm (C~band) and 1.3~cm (K~band). The fluxes and intensities are corrected by the primary beam response of the VLA antennas. For sources outside the C~band primary beam (listed as "oC" in column~(11) of Table~\ref{t:catalogue}) the primary beam correction is not reliable and the flux has to be taken with caution. Similarly, for sources located outside the K~band primary beam (labelled as "oK" in the Table), the K~band flux has to be taken with caution. The last column of the Table lists the spectral index, $\alpha$. For sources with no reliable flux estimate at one of the bands, we do not determine the spectral index. For sources detected at both frequency bands (C and K~bands), the spectral index has been determined using the fluxes determined after creating images with a common \textit{uv}-range (see Sect.~\ref{s:spectralindex} for more details). In Table~\ref{t:sizes}, we list the observed and deconvolved source sizes of all the detected sources. The source sizes are determined as $\sqrt{\theta_\mathrm{major}\times\theta_\mathrm{minor}}$, where $\theta_\mathrm{major}$ and $\theta_\mathrm{minor}$ are listed in Table~\ref{t:sizes}. We transform the angular size of each source into astronomical unit (au) using the distances listed in Table~\ref{t:sample}. We give the source size in Table~\ref{t:catalogue}. Finally, in Table~\ref{t:uvsizes}, we list the intensities, flux densities and sizes determined from the images generated using a common \textit{uv}-range at both C and K bands. In Figs.~\ref{f:stamps01} to \ref{f:stamps17}, we present close-up views of the C and K~band emission for the 146 detected continuum sources.

\onecolumn
\begin{landscape}
\begin{small}
\begin{longtable}{c c c c c c c c c c c c c c}
\caption{\label{t:catalogue} Catalogue of radio continuum sources detected with the VLA.}\\
\hline\hline

&
& \multicolumn{2}{c}{Coordinates}
&
& \multicolumn{3}{c}{C~band\tablefootmark{a}}
&
& \multicolumn{3}{c}{K~band\tablefootmark{b}}
& 
&

\\
\cmidrule{3-4}\cmidrule{6-8}\cmidrule{10-12}

& 
& R.A.(J2000)
& Dec.(J2000)
&
& Intensity
& Flux
& Size 
&
& Intensity 
& Flux
& Size 
&
& Spectral
\\
ID
& Region
& (h:m:s)
& (\degr:\arcmin:\arcsec)
&
& (mJy~beam$^{-1}$)
& (mJy)
& (au)
&
& (mJy~beam$^{-1}$)
& (mJy)
& (au)
&
& index\tablefootmark{c}
\\
\hline
\endfirsthead
\caption{continued.}\\
\hline\hline

&
& \multicolumn{2}{c}{Coordinates}
&
& \multicolumn{3}{c}{C~band\tablefootmark{a}}
&
& \multicolumn{3}{c}{K~band\tablefootmark{b}}
&
& 
\\
\cmidrule{3-4}\cmidrule{6-8}\cmidrule{10-12}

& 
& R.A.(J2000)
& Dec.(J2000)
&
& Intensity
& Flux
& Size 
&
& Intensity 
& Flux
& Size 
&
& Spectral
\\
ID
& Region
& (h:m:s)
& (\degr:\arcmin:\arcsec)
&
& (mJy~beam$^{-1}$)
& (mJy)
& (au)
&
& (mJy~beam$^{-1}$)
& (mJy)
& (au)
&
& index\tablefootmark{c}
\\
\hline
\endhead
\hline
\endfoot
  1 &IRAS\,05358$+$3543 & 05:39:25.838 & $+$35:45:53.861 &&    37.733$\pm$0.272 & \phn38.134$\pm$0.272 &     <2200     && ---                 & ---                  & ---       & iC/oK & --- \\
  2 &                   & 05:39:13.064 & $+$35:45:51.125 && \phn0.387$\pm$0.010 & \phnn0.534$\pm$0.014 &  \phs1300     && ---                 & ---                  & ---       & iC/iK & --- \\
  3 &                   & 05:39:10.056 & $+$35:46:07.731 && \phn0.105$\pm$0.003 & \phnn0.075$\pm$0.002 &     <1800     && ---                 & ---                  & ---       & iC/iK & --- \\
  4 &                   & 05:39:12.839 & $+$35:45:50.854 && \phn0.147$\pm$0.006 & \phnn0.332$\pm$0.015 &  \phs2200     && ---                 & ---                  & ---       & iC/iK & --- \\
  5 &                   & 05:39:09.938 & $+$35:45:17.196 && \phn0.267$\pm$0.008 & \phnn0.403$\pm$0.013 &  \phs1500     && ---                 & ---                  & ---       & iC/iK & --- \\
  6 &                   & 05:39:33.539 & $+$35:48:03.740 && \phn0.282$\pm$0.006 & \phnn0.345$\pm$0.007 &  \phs1000     && ---                 & ---                  & ---       & iC/oK & --- \\
  7 &                   & 05:39:14.367 & $+$35:44:29.634 && \phn0.191$\pm$0.006 & \phnn0.266$\pm$0.009 &  \phs1300     && ---                 & ---                  & ---       & iC/iK & --- \\
  8 &                   & 05:39:12.120 & $+$35:44:02.155 && \phn0.238$\pm$0.006 & \phnn0.262$\pm$0.007 &  \phns500     && ---                 & ---                  & ---       & iC/iK & --- \\
\end{longtable}
\end{small}
\tablefoot{
\tablefoottext{a}{Flux density, intensity peak and deconvolved source size for the sources detected at 6~cm in the C~band images. The fluxes are corrected by the primary beam response of the antennas, except for sources located outside the primary beam (listed as `oC'). Source sizes are calculated as indicated in Appendix~\ref{s:catalogue} and based on the values reported in Table~\ref{t:sizes}. Upper limits in the source size indicate that the source could not be deconvolved (see more details in Table~\ref{t:sizes}).}
\tablefoottext{b}{Flux density, intensity peak and deconvolved source size for the sources detected at 1.3~cm in the K~band images. The fluxes are corrected by the primary beam response of the antennas, except for sources located outside the primary beam (listed either as `iC/oK' or `oC', see Sect.~\ref{s:continuum} for more details about this classification).}
\tablefoottext{c}{Spectral index determined from the fluxes at 6~cm (C~band) and 1.3~cm (K~band). For the sources detected in one band, we use a 5$\sigma$ upper limit for the non-detected band flux. Only for sources located with the primary beam of both images (sources listed as `iC/iK') we can derive reliable fluxes and therefore spectral indices. Sources marked with $^\dagger$ have been re-imaged using the common uv-range (see Table~\ref{t:uvsizes}). More accurate spectral indices, derived using these new images, are listed in Table~\ref{t:candidates}.}}

\end{landscape}

\begin{small}
\begin{longtable}{c c c c c c c c c}
\caption{\label{t:sizes} Observed and deconvolved angular sizes for the radio continuum sources detected in this work (see Table~\ref{t:catalogue})}\\
\hline\hline

&
& \multicolumn{3}{c}{Observed source size}
&
& \multicolumn{3}{c}{Deconvolved diameter}
\\
\cmidrule{3-5}\cmidrule{7-9}    

&
& Major axis
& Minor axis
& PA
&
& Major axis
& Minor axis
& PA
\\
ID
& Region
& (\arcsec)
& (\arcsec)
& (\degr)
&
& (\arcsec)
& (\arcsec)
& (\degr)
\\
\hline
\endfirsthead
\caption{continued.}\\
\hline\hline

&
& \multicolumn{3}{c}{Observed source size}
&
& \multicolumn{3}{c}{Deconvolved diameter}
\\
\cmidrule{3-5}\cmidrule{7-9}    

&
& Major axis
& Minor axis
& PA
&
& Major axis
& Minor axis
& PA
\\
ID
& Region
& (\arcsec)
& (\arcsec)
& (\degr)
&
& (\arcsec)
& (\arcsec)
& (\degr)
\\
\hline
\endhead
\hline
\endfoot
\hline
 & \\
\multicolumn{9}{l}{Observed and deconvolved source sizes for C~band detections} \\
\hline
  1 & IRAS\,05358$+$3543  & \phn1.269$\pm$0.009 & \phn1.239$\pm$0.009 & \phn53.0$\pm$3.8 && \ldots              & \ldots              & \ldots \\
  2 &                     & \phn1.532$\pm$0.034 & \phn1.402$\pm$0.034 &    106.7$\pm$3.4 && \phn0.894$\pm$0.063 & \phn0.633$\pm$0.085 &    114.9$\pm$9.9 \\
  3 &                     & \phn1.113$\pm$0.044 & \phn1.001$\pm$0.043 &    155.4$\pm$5.2 && \ldots              & \ldots              & \ldots \\
  4 &                     & \phn2.411$\pm$0.058 & \phn1.455$\pm$0.057 &    120.4$\pm$1.1 && \phn2.069$\pm$0.069 & \phn0.730$\pm$0.124 &    121.1$\pm$1.2 \\
  5 &                     & \phn1.667$\pm$0.041 & \phn1.409$\pm$0.041 &    104.5$\pm$2.0 && \phn1.105$\pm$0.065 & \phn0.654$\pm$0.098 &    108.5$\pm$3.8 \\
  6 &                     & \phn1.412$\pm$0.028 & \phn1.349$\pm$0.027 & \phnn0.1$\pm$7.5 && \phn0.690$\pm$0.066 & \phn0.472$\pm$0.094 &    169.0$\pm$9.0 \\
\end{longtable}
\end{small}

\onecolumn
\begin{landscape}
\begin{small}
\begin{table*}[h]
\centering
\caption{\label{t:uvsizes}Intensities, fluxes and source sizes for the sources detected at both frequency bands and imaged using the common \textit{uv}-range (see Sect.~\ref{s:spectralindex})}

\begin{tabular}{c c c c c c c c c c c c c c c c}    
\hline\hline

&
&
&
& \multicolumn{3}{c}{Convolved Image Size}
&
& \multicolumn{3}{c}{Deconvolved Image Size}
& 
&
\\
\cmidrule{5-7}\cmidrule{9-11}

&
& Intensity\tablefootmark{a}
& Flux\tablefootmark{a}
& Major Axis
& Minor Axis
& PA
&
& Major Axis
& Minor Axis
& PA
& $\theta_\mathrm{beam}$, PA
& rms
\\
ID
& Region
& (mJy~beam$^{-1}$)
& (mJy)
& (\arcsec)
& (\arcsec)
& (deg)
&
& (\arcsec)
& (\arcsec)
& (deg)
& (\arcsec$\times$\arcsec, \degr)
& (mJy)
\\
\hline
 & \\
\multicolumn{11}{l}{Intensities, fluxes and sizes from the C~band images with common \textit{uv}-range} \\
\hline
 48a & IRAS\,18228$-$1312  & \phn11.50$\pm$0.61 &     55.32$\pm$3.50 & \phn4.60$\pm$0.28 & \phn2.08$\pm$0.10 & \phnn0$\pm$2\phn  && \phn4.24$\pm$0.31 & \phn1.72$\pm$0.13 & \phnn2$\pm$3\phn & 1.88$\pm$1.66, $-$14     & 0.400 \\
 48b &                     & \phnn8.76$\pm$0.99 &     75.41$\pm$9.50 & \phn4.35$\pm$0.50 & \phn3.93$\pm$0.44 & \phn40$\pm$44     && \phn4.15$\pm$0.72 & \phn3.56$\pm$0.65 & \phn52$\pm$54    & 1.88$\pm$1.66, $-$14     & 0.400 \\
 48c &                     & \phn32.31$\pm$1.70 &    129.52$\pm$8.30 & \phn3.21$\pm$0.18 & \phn2.48$\pm$0.12 & \phn61$\pm$8\phn  && \phn3.00$\pm$0.21 & \phn1.73$\pm$0.21 & \phn66$\pm$8\phn & 1.88$\pm$1.66, $-$14     & 0.400 \\
\hline
\end{tabular}
\end{table*}
\end{small}
\tablefoot{
\tablefoottext{a}{The intensities and fluxes are corrected by the primary beam response of the antennas, except for source \#137 which is located outside the primary beam of the antennas and no correction factor can be determined.}}
\end{landscape}


\end{appendix}

\end{document}